\pdfoutput=1
\documentclass[12pt]{emulateapj}
\usepackage{lscape}
\usepackage{graphicx}
\usepackage{multirow}
\usepackage{mathtools}
\usepackage{epsf}
\usepackage{longtable}

\newcommand{\Msun}{$M_{\odot}$}
\newcommand{\degs}{$^{\circ}$ }

\newcommand{\hi}{{\rm H\,}{{\sc i}}}
 
\newcommand{\his}{{\rm H\,}{{\sc i }}}

\newcommand{\mAA}{\AA \,}

\begin{document}	\title{METAL: The Metal Evolution, Transport, and Abundance in the Large Magellanic Cloud Hubble program. I. Overview and Initial Results}

\author{Julia Roman-Duval\altaffilmark{1}, Edward B. Jenkins\altaffilmark{2}, Benjamin Williams\altaffilmark{3}, Kirill Tchernyshyov\altaffilmark{4}, Karl Gordon\altaffilmark{1}, Margaret Meixner\altaffilmark{1,4,6}, Lea Hagen\altaffilmark{1}, Joshua Peek\altaffilmark{1}, Karin Sandstrom\altaffilmark{5}, Jessica Werk\altaffilmark{3}, Petia Yanchulova Merica-Jones\altaffilmark{5}}
\altaffiltext{1}{Space Telescope Science Institute, 3700 San Martin Drive, Baltimore, MD 21218; duval@stsci.edu}
\altaffiltext{2}{Princeton University Observatory, Peyton Hall, Princeton University, Princeton, NJ 08544-1001 USA}
\altaffiltext{3}{Department of Astronomy, Box 351580, University of Washington, Seattle, WA 98195, USA}
\altaffiltext{4}{Department of Physics and Astronomy, The Johns Hopkins University, 3400 North Charles Street, Baltimore, MD 21218, USA}
\altaffiltext{5}{Center for Astrophysics and Space Sciences, Department of Physics, University of California, 9500 Gilman Drive, La Jolla, San Diego, CA 92093, USA}
\altaffiltext{6}{NASA Goddard Space Flight Center, Greenbelt, MD 20771, USA}

\begin{abstract}
Metal Evolution, Transport, and Abundance in the LMC (METAL) is a large Cycle 24 program on the {\it Hubble Space Telescope} aimed at measuring dust extinction properties and interstellar depletions in the Large Magellanic Cloud (LMC) at half-solar metallicity. The 101-orbit program is comprised of COS and STIS spectroscopy toward 33 LMC massive stars between 1150 \mAA and 3180 \mAA, and parallel WFC3 imaging in 7 NUV-NIR filters. The fraction of silicon in the gas-phase (depletion) obtained from the spectroscopy decreases with increasing hydrogen column density. Depletion patterns for silicon differ between the Milky Way, LMC, and SMC, with the silicon depletion level offsetting almost exactly the metallicity differences, leading to constant gas-phase abundances in those galaxies for a given hydrogen column density. The silicon depletion correlates linearly with the absolute-to-selective extinction, $R_V$, indicating a link between gas depletion and dust grain size. Extinction maps are derived from the resolved stellar photometry in the parallel imaging, which can be compared to FIR images from Herschel and Spitzer to estimate the emissivity of dust at LMC metallicity. The full METAL sample of depletions, UV extinction curves, and extinction maps will inform the abundance, size, composition, and optical properties of dust grains in the LMC, comprehensively improve our understanding of dust properties, and the accuracy with which dust-based gas masses, star formation rates and histories in nearby and high-redshift galaxies are estimated. This overview paper describes the goals, design, data reduction, and initial results of the METAL survey.

\end{abstract}

\maketitle

\section{Introduction} \label{introduction}

\begin{figure*}
\includegraphics[width=16cm]{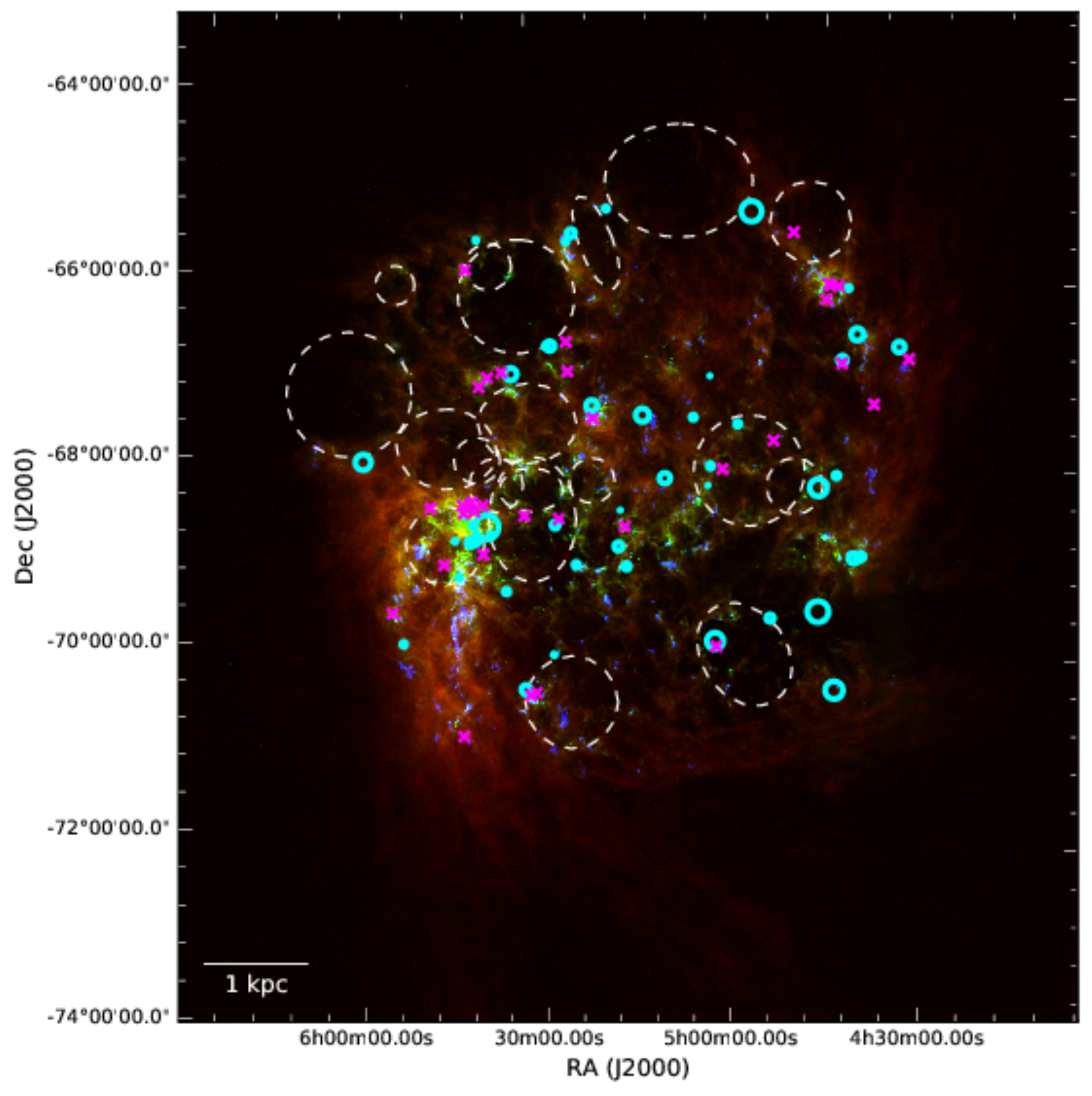}
\caption{Three-color image of the LMC, with the dust column density in red, the atomic gas in green, and molecular gas traced by CO 1-0 emission in blue. The METAL primary targets are show by magenta crosses. The parallel WFC3 fields are $\simeq$ 5 arcmin away from the primary targets and overlap with the magenta crosses on the 8 \degs scale of the LMC. Cyan circles show SN remnants. Dashed line white ellipses show super bubbles in \his}
\label{plot_targets}
\end{figure*}

\indent Recent observations and modeling have shown that interstellar dust can grow and evolve in the interstellar medium (ISM), and that dust growth is critical for galaxy evolution. In both local and high-redshift galaxies, the dust production rates in evolved stars \citep{bladh2012, riebel2012, srinivasan2016} and supernova remnants \citep{matsuura2011} are largely insufficient compared to the dust destruction rates in interstellar shocks \citep{jones1994, jones1996} to explain the dust masses of galaxies over cosmic times \citep{morgan2003, boyer2012,  rowlands2012, rowlands2014,  zhukovska2013}. This so-called Òdust budget crisisÓ can be resolved by a combination of two effects: 1) if dust grains can grow in the ISM by accreting gas-phase metals \citep{zhukovska2008, draine2009, rowlands2014, mckinnon2016}, effectively modifying the relation between dust and gas mass; and/or 2) if dust grains grow in size by coagulation, leading to changes in their optical properties (but not the dust abundance) and therefore in the relation between dust emission and dust mass \citep{rowlands2014}. In the latter case, the increase in emissivity caused by the increase in size can lead to an overestimation of dust masses.Thus, explaining dust masses over cosmic times requires that dust grow in the ISM, and therefore that the dust properties (abundance, composition, size distribution, and optical properties) change significantly with environment, particularly density. \\
\indent Moreover, dust grains absorb stellar light in the UV-optical and re-emit it in the FIR, which represents 30 --- 50\% of the output of a galaxy. This significant alteration to the emitted light implies that our ability to interpret observations of galaxies and trace their stellar, dust, and gas content over cosmic time over the entire spectral range critically relies on our models of how the dust abundance and properties vary with environment, particularly metallicity and density. These models require observational constraints to understand the processes responsible for dust formation and destruction and their timescales.  \\
\indent Since variations in the dust abundance and properties are difficult to observationally characterize in unresolved galaxies, we must measure the micro-astrophysical processes responsible for dust evolution in the ISM in local galaxies, and apply our knowledge to the more distant universe. With distances of 50 kpc \citep{schaefer2008} and 62 kpc \citep{hilditch2005}, the Large and Small Magellanic Clouds (LMC and SMC) are the only galaxies where a large sample of depletions can be obtained in a reasonable time, allowing us to resolve the dust evolution processes at the level of individual clouds while also getting the broad spatial coverage needed to understand the ISM on global scales. \\
\indent This study based on observations with the {\it Hubble Space Telescope} (HST) focuses on observations in the LMC. The SMC is the subject of another {\it Hubble} program and paper \citep[Cycle 23 program GO-13778, PI Jenkins, see][]{jenkins2017} dedicated to measuring depletions toward 18 sightlines with the {\it Space Telescope Imaging Spectrograph} (STIS). The LMC metallicity \citep[1/2 solar][]{russell1992} lies approximately midway between that of the MW (solar) and the SMC \citep[1/5 solar][]{russell1992}, and provides the link between the large differences in dust properties seen beween the MW and SMC, where the dust-to-gas ratio departs from a linear scaling with metallicity \citep{remyruyer2014, RD2014, RD2017}, the PAH fraction is an order of magnitude lower than in the MW \citep{sandstrom2010}, and the UV extinction curves distinctively lack a 2175 \mAA  bump and are steeper in the FUV than in the MW \citep{gordon2003}. In addition, the LMC's gas disk is thinner \citep[120 pc][]{elmegreen2001} and less inclined than the SMC, alleviating confusion in velocity and distance structure along the line-of-sight. Finally, we have considerable ancillary data to trace the LMCÕs stellar and ISM content from Spitzer, Herschel, FUSE, IUE, and ground-based facilities \citep[][and references therein]{meixner2013}.\\
\indent  There have been several investigations of dust evolution in the LMC by measuring its dust and gas content with Spitzer, Herschel \citep{meixner2006, meixner2013}, IRAS, and Planck \citep{RD2017}, in addition to ground-based measurements of CO \citep{wong2011}, \his \citep{kim2003}, and H$\alpha$ \citep{gaustad2001}. These emission-based measurements suggest that the dust abundance and size increase from the diffuse to the dense ISM \citep{RD2014, RD2017}. Because the FIR surface brightness from dust emission is the product of the dust opacity and surface density, however, those quantities are degenerate. The FIR opacity of dust is known to vary with metallicity, density, and temperature \citep{stepnik2003, kohler2012, demyk2017}, but the dependence of the opacity on such physical parameters is not well constrained. As a result, the observed variations of the dust abundance measured from FIR emission could stem from variations of the opacity. Additionally, unaccounted for CO-dark gas and variations of the CO-to-H$_2$ conversion factor can result in an apparent variation of the dust abundance. These large systematic uncertainties \citep[factors 2-3][]{RD2014} and degeneracies between dust optical properties, dust abundance, and the CO-to-H$_2$ conversion factor prevent us from unambiguously characterizing dust growth (via accretion and/or coagulation) and evolution in the ISM, and from quantifying the effects of metallicity and other environmental parameters on dust properties \citep{RD2014}. Absorption-based measurements are required to resolve these degeneracies because, unlike emission-based measurements, they weakly depend on the excitation conditions.\\
\indent  In the MW, depletion and extinction curve measurements obtained from UV spectroscopy of background stars provide evidence for ISM dust growth. First, elemental depletions collectively correlate with density \citep{jenkins2009}, indicating some combination of metal accretion onto dust at high densities and dust destruction by shocks at low densities; second, variations in the extinction curves within the MW \citep{cardelli1989, valencic2004, gordon2009} require the composition (via accretion of gas-phase metals) and size (via coagulation) of dust to vary \citep{weingartner2001, hirashita2012}. Depletions studies in the LMC (and SMC) suggest that ISM dust growth via accretion also occurs at low metalllicity \citep{tchernyshyov2015, jenkins2017}. Additionally, the striking differences in the morphology of the dust UV extinction curves within and between the MW, LMC, and SMC \citep{gordon2003} imply that the dust size, composition, and optical properties change substantially with metallicity. \\
\indent The most comprehensive depletion study in the LMC  attempting to characterize dust evolution timescales comes from \citet{tchernyshyov2015}, who derived P and Fe depletions for 40 sight-lines based on archival {\it Far-ultraviolet Spectroscopic Explorer} (FUSE) spectra. Of these sight-lines, eight were observed with the {\it Cosmic Origins Spectrograph} (COS) onboard HST in the NUV to derive Si, Cr, Zn depletions. The gas-phase abundances of all elements probed in this study decrease with increasing surface density following metallicity-dependent patterns (slope and zero-points), supporting the metallicity-dependence of ISM dust growth processes and timescales. However, most sight-lines have depletion measurements for Fe and P only, and still missing depletion measurements for the critical dust constituents - Fe, Mg, Si, Ni, C and O - required to constrain the dust abundance and composition and the dust growth timescales as a function of environment, particularly metallicity and density.\\
\indent The METAL ("Metal Evolution and TrAnsport in the Large Magellanic Cloud") large HST program (101 orbits, GO-14675) obtained high S/N UV spectra toward 33 massive stars serving as background lights to measure UV extinction curves and elemental depletions for the main constituents of interstellar dust (Fe, Si, Mg, Ni) and for other elements (Cu, Cr, Zn, S) often used as metallicity tracers (e.g., in damped Lyman alpha systems \citep[DLAs][]{rafelski2012, quiret2016}. In addition to UV spectroscopy, METAL obtained parallel {\it Wide Field Camera 3} (WFC3) NUV-NIR imaging in 7 filters to map the dust column density and extinction properties - A$_{\mathrm{V}}$, R$_{\mathrm{V}}$, strength of the 2175 \mAA bump - in the vicinity of the LMC massive stars to derive extinction maps and characterize the FIR emissivity of dust at half-solar metallicity. The overarching goal of the program is to perform a complete and detailed characterization of the dust abundance and properties in the LMC. Given the legacy value of the spectra and photometry, it is expected that this large sample of observations in the LMC will also be used for other purposes, such as characterizing the wind properties of massive stars at half-solar metallicity, and studying the nature and origin of galactic-scale outflows in the LMC. \\
\indent The paper is organized as follows. In Section \ref{design_section}, we provide an overview of the design of the program. In Section \ref{reduction_section}, we describe the METAL data reduction. Section \ref{photometry_section} describes the point source extraction and photometry for the parallel WFC3 imaging data. In Section \ref{results_section}, we present initial results from the METAL data analysis that highlight the overarching science goals of the program. Section \ref{conclusion_section} summarizes this work.

\section{Survey Design}\label{design_section}

Achieving our goal of measuring dust properties in the LMC required careful selection of stellar targets for ISM spectroscopy, instrument set-ups, and exposure times. We describe the rationale for the METAL program design below.

\subsection{Target sample}

\begin{figure}
\centering
\includegraphics[width=8cm]{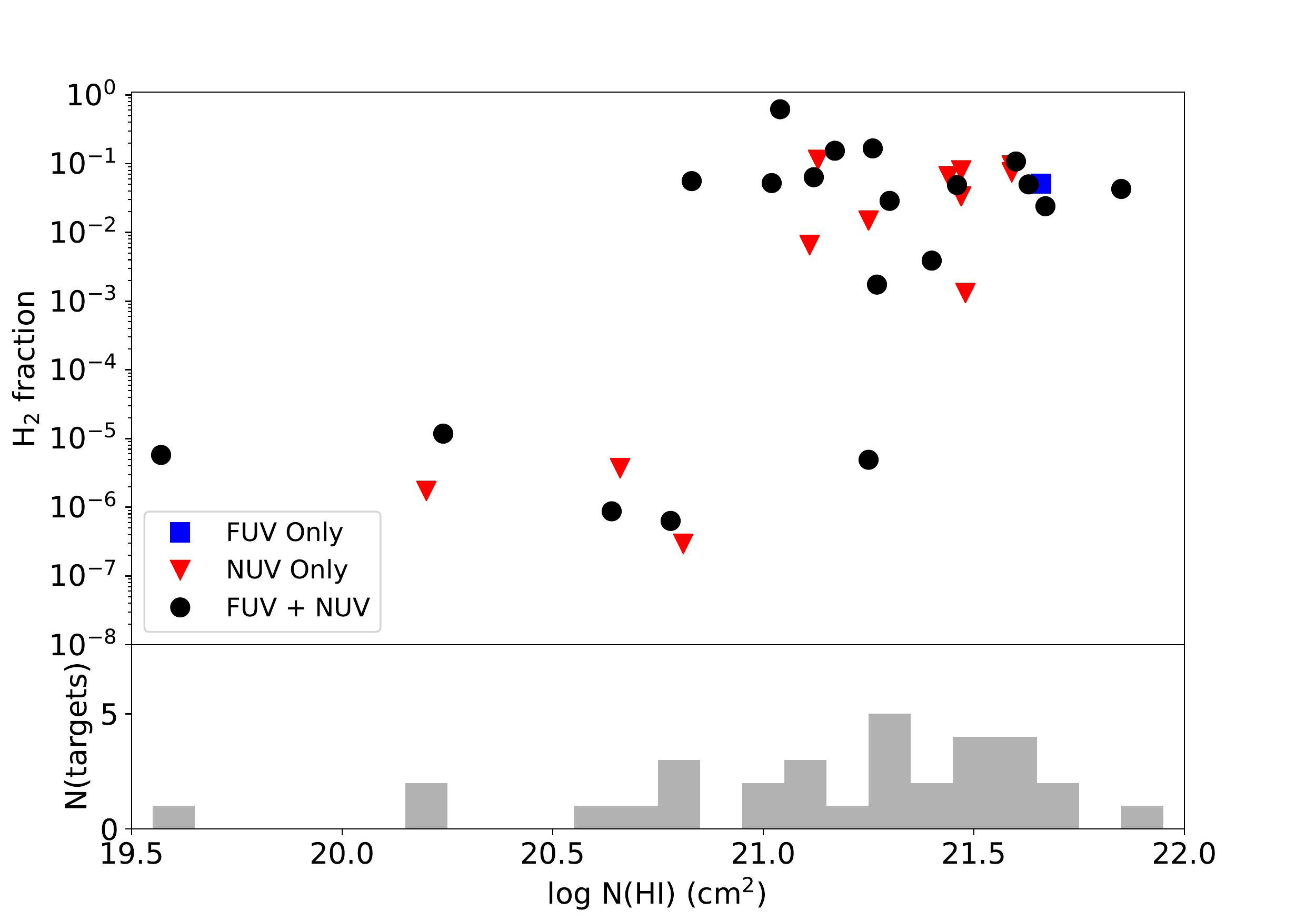}
\caption{\his column density and H$_2$ fraction of our 33 LMC targets for spectroscopy. The different colors indicate whether both FUV and NUV spectra were obtained as part of METAL (black), or, if either FUV or NUV spectra were available in the archive, whether METAL obtained the FUV only (blue) or NUV only (red). The number of targets per 0.1 dex \his column density interval is shown in the bottom panel.}
\label{plot_param_space}
\end{figure}

\begin{deluxetable*}{ccccccc}
\centering
\tabletypesize{\scriptsize}
\tablecolumns{6}
\tablewidth{\textwidth}
\tablecaption{Spectroscopic targets and their stellar parameters}
\tablenum{1}
 
 \tablehead{Target & Ra & Dec & SpT & $v\times sin(i)$  &$V$ }\\
 
 \startdata
 & $h$ & $deg$ & & km s$^{-1}$ & mag \\
 \hline
 &&&&\\
 SK-67 2 & 04h47m04.451s & -67d06m53.12s & B1Ia+ & 143.0 & 11.26 \\

SK-67 5 & 04h50m18.918s & -67d39m38.10s & O9.7Ib & 87.0 & 11.34 \\
SK-69 279 & 04h54m14.256s & -69d15m13.35s & Ofpe/WN9 & 84.0 & 12.78 \\
SK-67 14 & 04h54m31.889s & -67d15m24.58s & B1.5Ia & 88.0 & 11.53 \\
SK-66 19 & 04h55m53.951s & -66d24m59.35s & B0Ia & ... & 12.79 \\
PGMW 3120 & 04h56m46.812s & -66d24m46.72s & O5.5V((f*)) & 111.0 & 12.8 \\
PGMW 3223 & 04h57m00.859s & -66d24m25.12s & O8V & 147.0 & 12.95 \\
SK-66 35 & 04h57m04.440s & -66d34m38.45s & BC1Ia & ... & 11.58 \\
SK-65 22 & 05h01m23.070s & -65d52m33.40s & O6Iaf+ & 66.0 & 12.08 \\
SK-68 26 & 05h01m32.248s & -68d10m42.93s & BC2Ia & ... & 11.65 \\
SK-70 79 & 05h06m37.262s & -70d29m24.16s & B0III & 86.0 & 12.71 \\
SK-68 52 & 05h07m20.423s & -68d32m08.59s & B0Ia & 70.0 & 11.63 \\
SK-69 104 & 05h18m59.501s & -69d12m54.82s & O6Ib(f) & 117.0 & 12.1 \\
SK-68 73 & 05h22m59.781s & -68d01m46.62s & Of/WN & ... & 11.45 \\
SK-67 101 & 05h25m56.221s & -67d30m28.67s & O8II & 103.0 & 12.67 \\
SK-67 105 & 05h26m06.192s & -67d10m56.79s & O4f+O6V & ... & 12.42 \\
BI 173 & 05h27m09.941s & -69d07m56.46s & O8.5II(f) & ... & 12.96 \\
BI 184 & 05h30m30.657s & -71d02m31.60s & B0.5V & ... & 13.84 \\
SK-71 45 & 05h31m15.654s & -71d04m09.69s & O4-5III(f) & 117.0 & 11.54 \\
SK-69 175 & 05h31m25.520s & -69d05m38.59s & WN11h & ... & 11.9 \\
SK-67 191 & 05h33m34.028s & -67d30m19.72s & O8V & 100.0 & 13.46 \\
SK-67 211 & 05h35m13.905s & -67d33m27.51s & O2III(f*) & 173.0 & 12.29 \\
BI 237 & 05h36m14.628s & -67d39m19.18s & O2V & 83.0 & 13.0 \\
SK-68 129 & 05h36m26.768s & -68d57m31.90s & B1I & 96.0 & 12.78 \\\
SK-69 220 & 05h36m43.692s & -69d29m47.47s & LBV & 0.0 & 10.58 \\
SK-66 172 & 05h37m05.394s & -66d21m35.18s & O2III(f*)+OB & 101.0 & 13.13 \\
BI 253 & 05h37m34.461s & -69d01m10.20s & O2V & 155.0 & 13.76 \\
SK-68 135 & 05h37m49.112s & -68d55m01.69s & ON9.7Ia+ & 70.0 & 11.36 \\
SK-69 246 & 05h38m53.384s & -69d02m00.93s & WN6h & ... & 11.1 \\
SK-68 140 & 05h38m57.18s & -68d56m53.1s & B0Ia & ... & 12.79 \\
SK-71 50 & 05h40m43.192s & -71d29m00.65s & O6.5II & 296.0 & 13.44 \\
SK-68 155 & 05h42m54.93s & -68d56m54.5s & O8Ia & 96.0 & 12.75 \\
SK-70 115 & 05h48m49.654s & -70d03m57.82s & O6.5Iaf & 164.0 & 12.24 \\

\enddata

\end{deluxetable*}

\begin{deluxetable*}{ccccccccc}
\centering
\tabletypesize{\scriptsize}
\tablecolumns{8}
\tablewidth{\textwidth}
\tablecaption{Spectroscopic targets and their interstellar parameters}
\tablenum{2}
 
 \tablehead{Target & Ra & Dec & $E(B-V)$  & $\log N($\hi)$_{\mathrm{LMC}}$\tablenotemark{a}& $\log{N(\mathrm{H}_2})$\tablenotemark{b}&  $\log N($\hi)$_{\mathrm{MW}}$\tablenotemark{a} & nearby \his shell?}\\
 
 \startdata
 & $h$ & $deg$ &   mag & cm$^{-2}$ &  cm$^{-2}$ & cm$^{-2}$ & \\
 \hline
 &&&&&&\\
SK-67 2 & 04h47m04.451s & -67d06m53.12s & 0.26 & 21.04 $\pm$  0.12 & 20.95 & 20.75 $\pm$ 0.28 & n \\
SK-67 5 & 04h50m18.918s & -67d39m38.10s & 0.14 & 21.02 $\pm$ 0.04 & 19.46 & 20.06 $\pm$ 0.33 & n \\
SK-69 279 & 04h54m14.256s & -69d15m13.35s & 0.21 & 21.59 $\pm$ 0.05 & 20.31 & 18.5 $\pm$ 0.53 & n \\
SK-67 14 & 04h54m31.889s & -67d15m24.58s & 0.08 & 20.24 $\pm$ 0.06 & 15.01 & 20.56 $\pm$ 0.03 & n \\
SK-66 19 & 04h55m53.951s & -66d24m59.35s & 0.36 & 21.85 $\pm$ 0.07 & 20.2 & 20.74 $\pm$ 1.35 & y \\
PGMW 3120 & 04h56m46.812s & -66d24m46.72s & 0.25 & 21.48 $\pm$ 0.03 & 18.3 & 18.5 $\pm$ 0.72 & n \\
PGMW 3223 & 04h57m00.859s & -66d24m25.12s & 0.19 & 21.4 $\pm$ 0.06 & 18.69 & 20.45 $\pm$ 1.14 & n \\
SK-66 35 & 04h57m04.440s & -66d34m38.45s & 0.11 & 20.83 $\pm$ 0.04 & 19.3 & 20.75 $\pm$ 0.06 & n \\
SK-65 22 & 05h01m23.070s & -65d52m33.40s & 0.11 & 20.66 $\pm$ 0.03 & 14.93 & 20.3 $\pm$ 0.03 & y \\
SK-68 26 & 05h01m32.248s & -68d10m42.93s & 0.29 & 21.6 $\pm$ 0.06 & 20.38 & 19.68 $\pm$ 1.02 & n \\
SK-70 79 & 05h06m37.262s & -70d29m24.16s & 0.24 & 21.26 $\pm$ 0.04 & 20.26 & 20.08 $\pm$ 0.98 & y \\
SK-68 52 & 05h07m20.423s & -68d32m08.59s & 0.18 & 21.3 $\pm$ 0.06 & 19.47 & 19.42 $\pm$ 0.89 & n \\
SK-69 104 & 05h18m59.501s & -69d12m54.82s & 0.1 & 19.57 $\pm$ 0.68 & 14.03 & 20.73 $\pm$ 0.04 & n \\
SK-68 73 & 05h22m59.781s & -68d01m46.62s & 0.4 & 21.66 $\pm$ 0.02 & 20.09 & 18.5 $\pm$ 0.48 & n \\
SK-67 101 & 05h25m56.221s & -67d30m28.67s & 0.08 & 20.2 $\pm$ 0.04 & 14.14 & 20.7 $\pm$ 0.01 & n \\
SK-67 105 & 05h26m06.192s & -67d10m56.79s & 0.17 & 21.25 $\pm$ 0.04 & 19.13 & 20.65 $\pm$ 0.2 & n \\
BI 173 & 05h27m09.941s & -69d07m56.46s & 0.15 & 21.25 $\pm$ 0.05 & 15.64 & 20.77 $\pm$ 0.19 & n \\
BI 184 & 05h30m30.657s & -71d02m31.60s & 0.2 & 21.12 $\pm$ 0.04 & 19.65 & 20.87 $\pm$ 0.05 & y \\
SK-71 45 & 05h31m15.654s & -71d04m09.69s & 0.16 & 21.11 $\pm$ 0.03 & 18.63 & 20.63 $\pm$ 0.06 & y \\
SK-69 175 & 05h31m25.520s & -69d05m38.59s & 0.17 & 20.64 $\pm$ 0.03 & 14.28 & 20.67 $\pm$ 0.02 & y \\
SK-67 191 & 05h33m34.028s & -67d30m19.72s & 0.1 & 20.78 $\pm$ 0.03 & 14.28 & 20.73 $\pm$ 0.02 & n \\
SK-67 211 & 05h35m13.905s & -67d33m27.51s & 0.1 & 20.81 $\pm$ 0.04 & 13.98 & 20.63 $\pm$ 0.04 & n \\
BI 237 & 05h36m14.628s & -67d39m19.18s & 0.2 & 21.63 $\pm$ 0.03 & 20.05 & 19.51 $\pm$ 0.91 & n \\
SK-68 129 & 05h36m26.768s & -68d57m31.90s & 0.22 & 21.59 $\pm$ 0.14 & 20.2 & 21.26 $\pm$ 0.38 & y \\
SK-69 220 & 05h36m43.692s & -69d29m47.47s & ... & 21.28 & 19.04 & ... & y \\
SK-66 172 & 05h37m05.394s & -66d21m35.18s & 0.18 & 21.27 $\pm$ 0.03 & 18.21 & 20.18 $\pm$ 0.13 & y \\
BI 253 & 05h37m34.461s & -69d01m10.20s & 0.23 & 21.67 $\pm$ 0.03 & 19.76 & 19.83 $\pm$ 0.9 & y \\
SK-68 135 & 05h37m49.112s & -68d55m01.69s & 0.26 & 21.46 $\pm$ 0.02 & 19.87 & 18.5 $\pm$ 0.54 & y \\
SK-69 246 & 05h38m53.384s & -69d02m00.93s & 0.18 & 21.47 $\pm$ 0.02 & 19.71 & 19.03 $\pm$ 0.61 & y \\
SK-68 140 & 05h38m57.18s & -68d56m53.1s & 0.29 & 21.47 $\pm$ 0.11 & 20.11 & 21.49 $\pm$ 0.12 & y \\
SK-71 50 & 05h40m43.192s & -71d29m00.65s & 0.2 & 21.17 $\pm$ 0.05 & 20.13 & 20.76 $\pm$ 0.14 & n \\
SK-68 155 & 05h42m54.93s & -68d56m54.5s & 0.28 & 21.44 $\pm$ 0.09 & 19.99 & 21.49 $\pm$ 0.08 & y \\
SK-70 115 & 05h48m49.654s & -70d03m57.82s & 0.2 & 21.13 $\pm$ 0.08 & 19.94 & 20.95 $\pm$ 0.13 & n \\

\enddata
\tablenotetext{a}{The \his column densities are from this work (see Section \ref{nhi_fitting}), except for SK-69 220, which is from \citet{welty2012}}
\tablenotetext{b}{The H$_2$ column densities are from \citet{welty2012}} 
\end{deluxetable*}

\indent We selected our targets from the \citet{welty2012} catalog of massive stars in the LMC that have FUSE archival data and spectroscopically determined \his and H$_2$ column densities (for abundance determinations). Amongst the targets that could be observed in 5 orbits or less, we selected at least 3 when available, but preferably 5, stars per 0.1 dex in N(\hi), some with and some without nearby expanding shells \citep{kim1999} and supernova remnants \citep{temim2015} nearby, in order to account for the variety of dynamic conditions and density structures (traced by the H$_2$ fraction) for a given hydrogen column density. We preferentially selected targets with archival COS/STIS medium-resolution FUV or NUV data when available to minimize the observational cost of the program. The METAL program obtained full UV coverage (FUV $+$ NUV) spectra of 20 targets. FUV spectra were available in the archive for 11 targets, and in those cases, only the NUV spectra were obtained as part of METAL. In only one case was a NUV archival spectrum available and did METAL obtain the FUV counterpart. \\
\indent  Figure \ref{plot_targets} shows the target sample with dust, atomic and molecular gas tracers in color-scale. Supernova remnants from \citet{temim2015} and \his shells from \citet{kim1999} are shown as well, since they can alter the dynamic conditions near our targets due to the presence of nearby interstellar shocks. The targets  uniformly cover the \his disk of the LMC, with a typical separation of 1$^{\circ}$ ($\simeq$ 900 pc) between targets. Our sample's \his log column density spans 20.5---22, with H$_2$ column densities spanning 14---21 (or H$_2$ fractions ranging from 10$^{-6}$ to 0.6). Tables 1 and 2 list the primary targets and their stellar and ISM parameters, such as \his and H$_2$ column density, respectively. The ISM parameter space (\hi, H$_2$ column densities, presence of a nearby \his shell) probed by this sample is shown in Figure \ref{plot_param_space}. We note that, while the sample was initially built based on the \his column density information from \citet{welty2012}, \his column densities were re-derived from the better-quality METAL spectra (Section \ref{nhi_fitting}). The \his column density values and associated uncertainties from this work are listed in Table 2 and shown in Figure \ref{plot_targets}. The H$_2$ column densities are taken directly from \citet{welty2012}. \\
\indent The massive stars are the targets for the primary single-object spectroscopic observations. The WFC3 imaging was obtained in parallel to the spectroscopy with no constraints on the telescope's roll angle. Therefore, the parallel WFC3 fields are randomly distributed in annuli about 5 arcmin ($\sim$ 70 pc) away from the massive stars.

\begin{figure*}
\includegraphics[width=\textwidth]{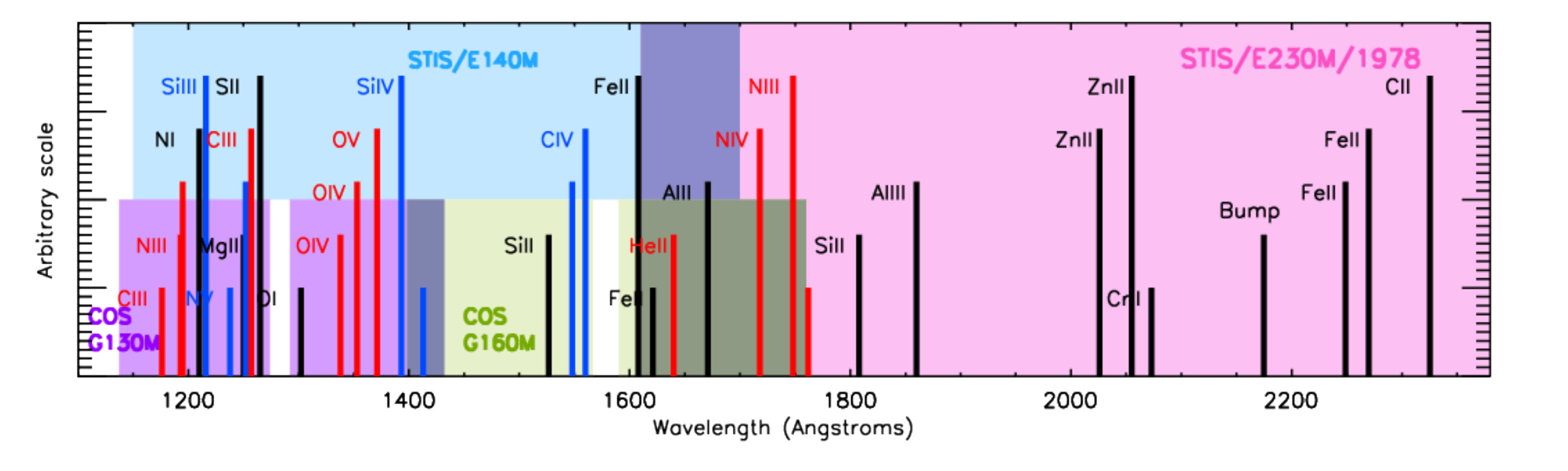}
\caption{Spectral lines required for characterizing interstellar elemental depletions (black), as well as spectral lines of interest for legacy studies of stellar winds (red) and the circum-galactic medium (blue). The approximate wavelength ranges of the COS G130M/1291 (purple) and G160M/1589 (green) settings, as well as the STIS/E140M (blue) and E230M/1978 (pink) settings are indicated by the shaded regions. The y-scale is arbitrary.}
\label{plot_lines}
\end{figure*}

\subsection{STIS and COS spectroscopy}

\begin{figure*}
\centering
\includegraphics[width=12cm]{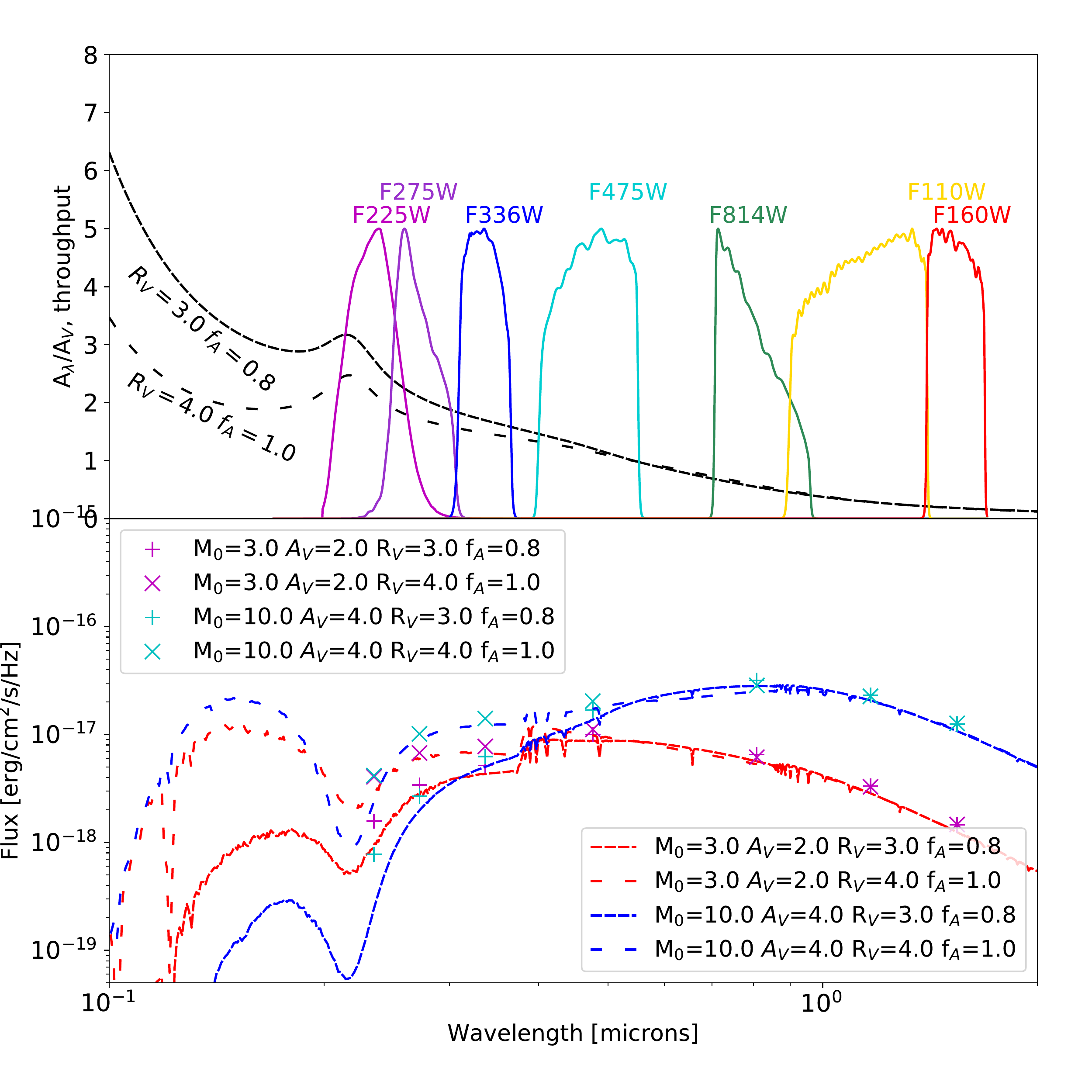}
\caption{(Top) Transmission function of the WFC3 filters used in the METAL program (colors), compared to the shape of dust UV extinction curves (black) with different parameters ($R_V$ = 3 and f$_A$ = 0.8 in solid; $R_V$ = 4 and $f_A$ = 1 in dashed). (Bottom) Model Spectra (lines) and synthetic SEDs (points) for a 10 million year old O star ($M$ $=$ 10 \Msun) reddened by $A_V$ $=$ 4 of dust (blue spectra, teal points) and a 10 million year old A star ($M$ $=$ 3 \Msun) reddened by $A_V$ $=$ 4 of dust (red spectra, purple points). For each star, we show the spectra and SEDs for two sets of dust properties: $R_V$ $=$ 3 and $f_A$ = 0.8 (solid spectra, crosses), and $R_V$ $=$ 4 and $f_A$ = 1 (dashed spectra, x's). The model spectra and corresponding SEDs are taken from the BEAST tool \citep{gordon2016}. The combination of the UV, optical and IR resolves the degeneracies between reddening and stellar mass, and discriminates between different dust properties.}
\label{plot_filters}
\end{figure*}

\begin{figure*}
\centering
\includegraphics[width=15cm]{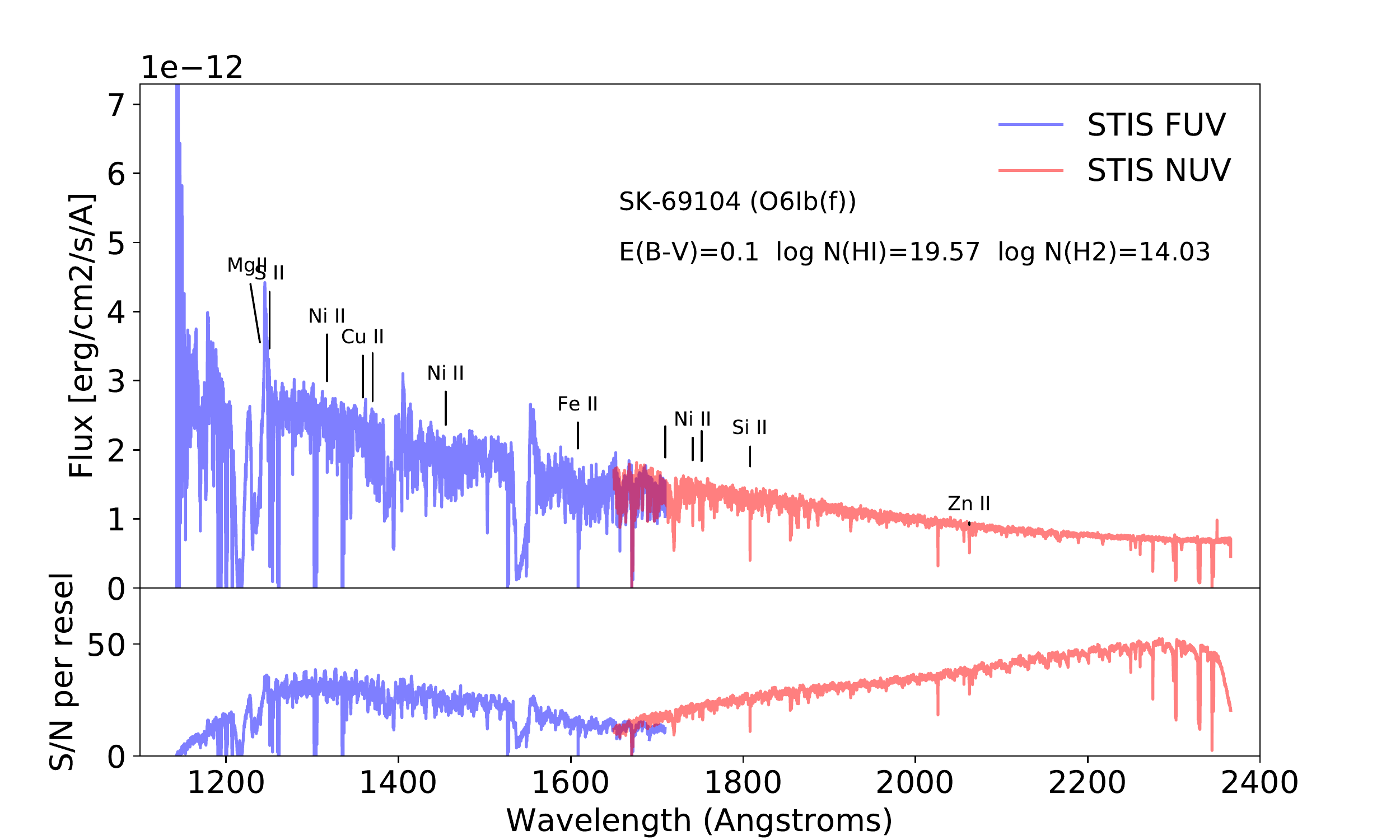}
\includegraphics[width=15cm]{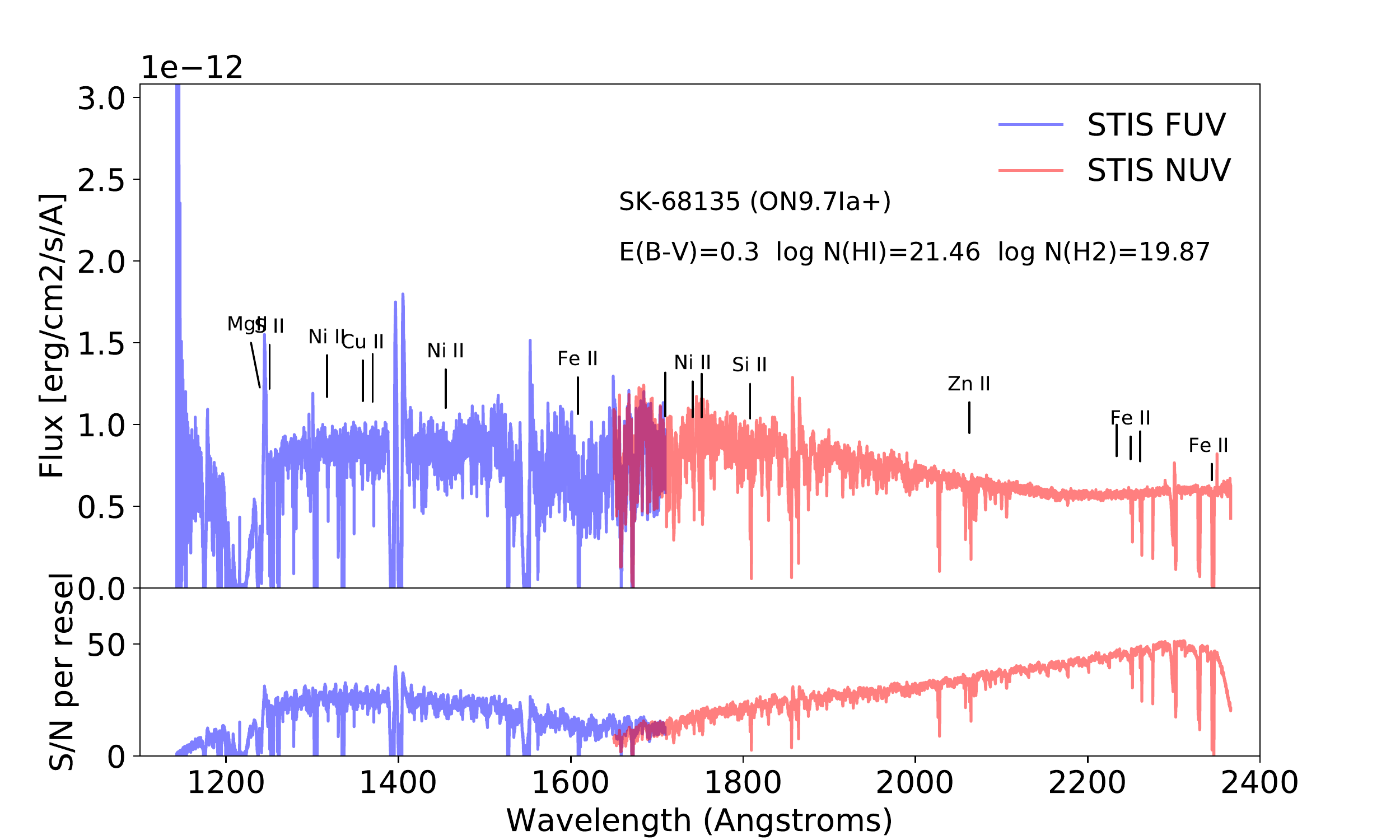}
\end{figure*}
\begin{figure*}
\centering
\includegraphics[width=15cm]{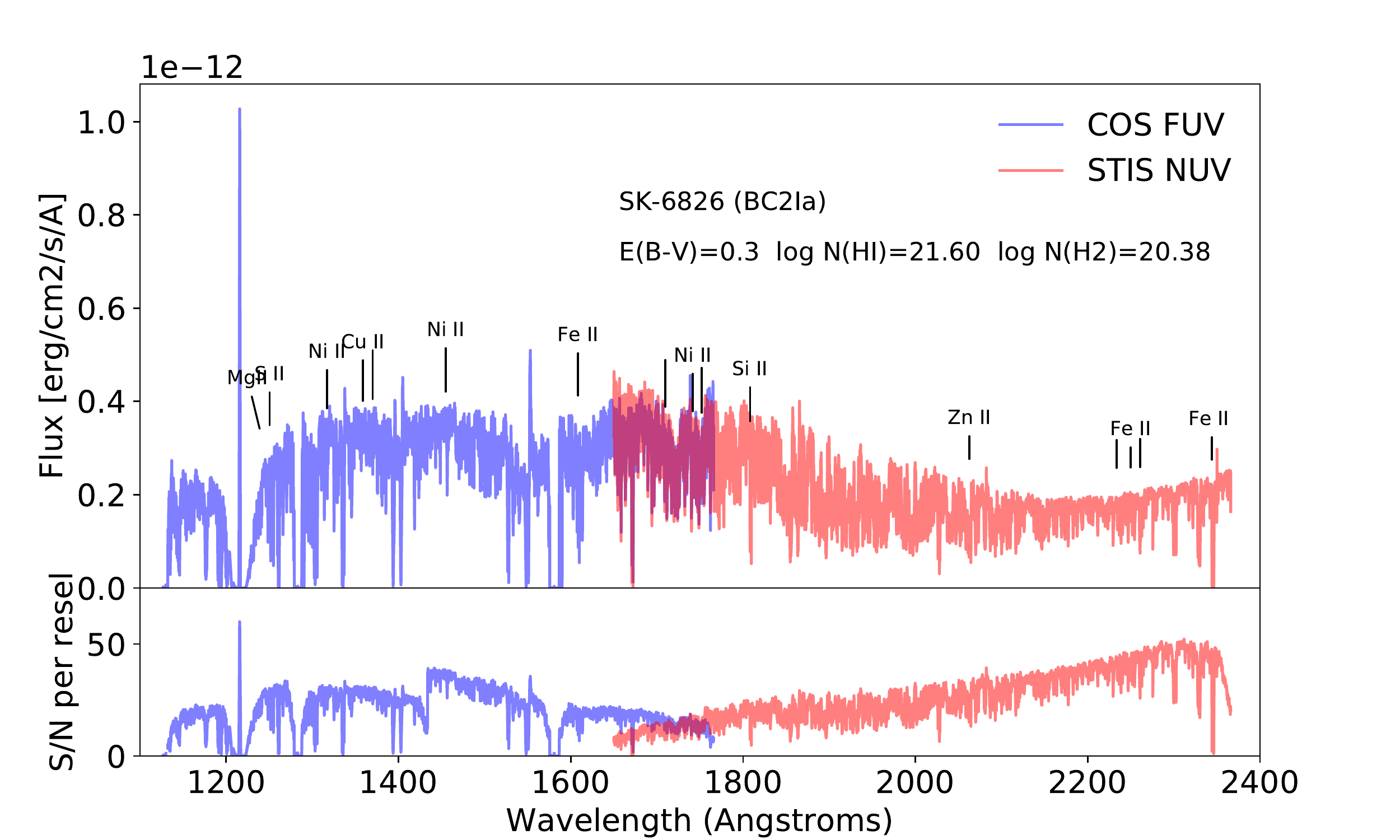}
\includegraphics[width=15cm]{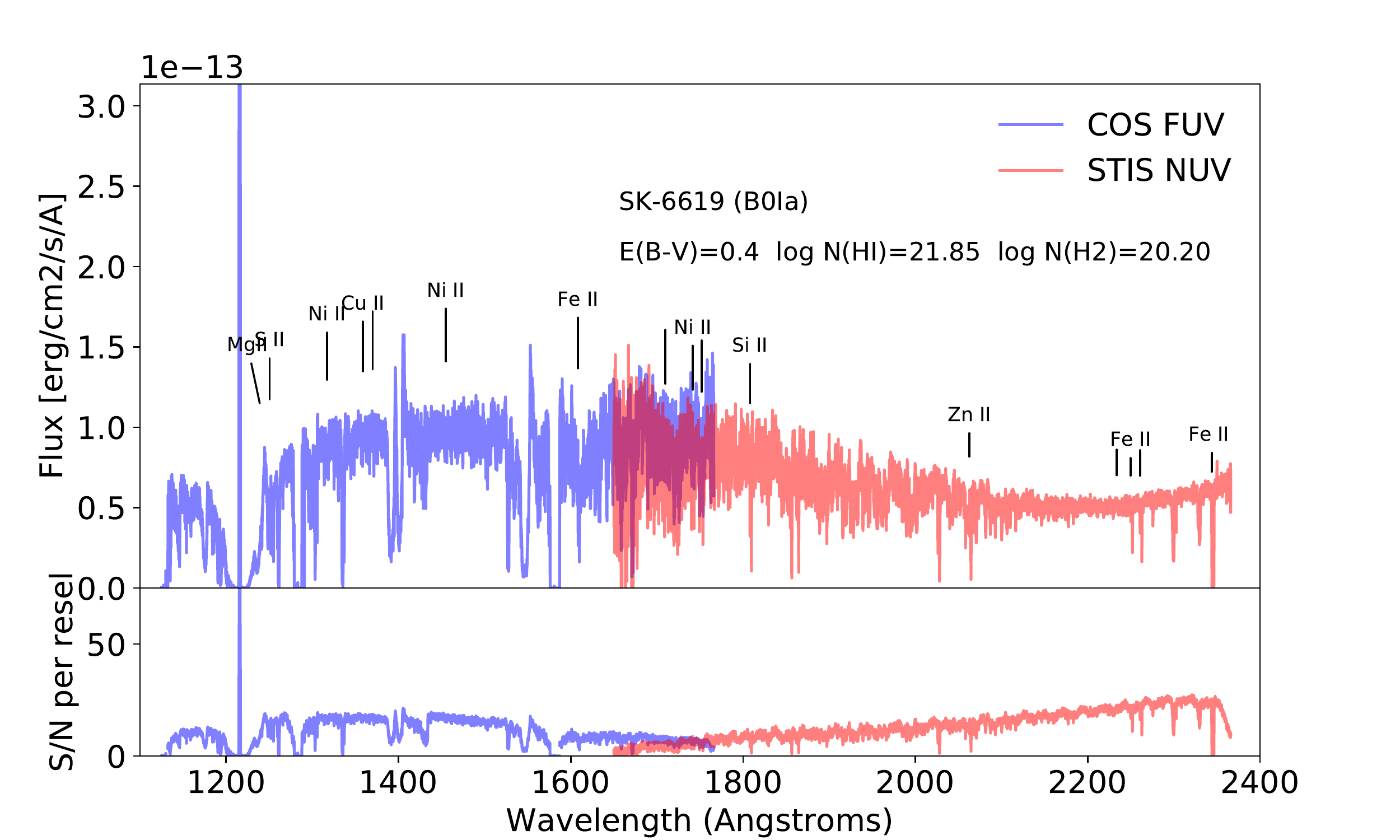}
\caption{Full FUV (blue) + NUV (red) spectra for SK-69 104, SK-68 135, SK-68 26, and SK-66 19 (from top to bottom). The first two stars were observed with STIS in the FUV, while the last two targets were observed with COS in the FUV. The spectra shown here correspond to sight-lines with a range of \his column densities.}
\label{show_spectra}
\end{figure*}

\begin{deluxetable}{cccc}
\centering
\tabletypesize{\scriptsize}
\tablecolumns{4}
\tablewidth{8cm}
\tablecaption{Spectral lines used for depletion measurements}
\tablenum{3}
\tablehead{Element/ion & 12 + log(X/H)$_{\mathrm{LMC, tot}}$ &  Wavelength & log $\lambda f_{\lambda}$} \\
 
 \startdata
& & \AA & \AA  \\
\hline
& &&\\
Mg II& 7.26$\pm$0.08 & 1239.925 & -0.106  \\
 & & 1240.395  & -0.355 \\
 &&&\\
Si II & 7.35 $\pm$0.10 & 1808.013 & 0.575 \\
&&&\\
P II  & 5.10$\pm$0.1 & 1152.818 & 2.451 \\
&&&\\
S II & 6.94 $\pm$0.04& 1250.578 &0.809 \\
 & & 1253.805& 1.113\\
 &&&\\
Cr II & 5.37$\pm$0.07&  2056.254 & 2.326 \\
& &  2066.161 & 2.024 \\
&&&\\
Fe II &  7.32$\pm$0.08 & 1608.451 & 1.968 \\
&& 1611.201 & 0.347\\
& &  2260.780 & 0.742\\
& & 2249.877 & 0.612 \\
&&&\\
Ni II& 5.92$\pm$0.07 & 1370.132 &  1.906  \\
&& 1317.217  & 1.876  \\
 && 1741.549 &  1.871  \\
& & 1709.600 &  1.743  \\
&& 1751.910 &   1.686  \\        
&& 1454.842   & 1.672 \\
&&&\\
Zn II & 4.31$\pm$0.15 & 2062.664 & 2.804 \\
&&&\\
Cu II & 3.89$\pm$0.04 & 1358.773 & 2.569 \\
\enddata
\tablecomments{Mg, Si, P, Cr, Fe, Ni, Zn LMC abundances are from \citet{tchernyshyov2015}; S, Cu abundances are from \citet{asplund2009} scaled by a factor 0.5.}
\end{deluxetable}

\indent Line-widths in the neutral ISM are typically $<$ 10 km s$^{-1}$ \citep[e.g.,][]{welty2003}. Measuring column densities from interstellar lines would therefore ideally require high-resolution (R$\simeq$100,000), which can only be achieved with the STIS E140H and E230H onboard HST. However, exposure times for extragalactic stars would be prohibitively long in the context of a survey of a large number of sight-lines with full FUV+NUV coverage. Therefore, we opt for medium-resolution with COS and STIS for the METAL program. While we do not resolve the individual components of interstellar absorption, we can still get a relatively accurate estimate of the metal-ion column densities. Nonetheless, as we will see in Section \ref{depletion_section}, two of the METAL targets have archival E230H spectra that we use for benchmarking our column density estimates with the E230M.\\
\indent The spectral lines required for characterizing ISM depletions are shown in Figure \ref{plot_lines} and listed in Table 3. Those lines can be covered by COS G130M/1291 and G160M/1589 (at resolving power R $\simeq$ 20,000 corresponding to 15 km s$^{-1}$) or STIS/E140M (at R $\simeq$ 45,000 corresponding to 7 km s$^{-1}$) in the FUV, in combination with STIS/E230M/1978 (at R $\simeq$ 30,000 corresponding to 10 km s$^{-1}$) in the NUV. STIS/E230M is the most efficient way to cover the NUV at medium resolution because it covers all NUV lines with one setting (the striping of COS/NUV would require multiple settings and more time than STIS). \\
\indent In the FUV, targets were preferentially observed with STIS/E140M to maximize spectral resolution, which is critical in ISM studies due to the low velocity dispersions. For most targets, the difference in orbits between COS/FUV+STIS/NUV and STIS/FUV+NUV was less than an orbit due to the overhead in changing instruments, and in those instances STIS/E140M was used. All observations using STIS E140M and E230M used the 0.2''x0.2'' aperture. The targets that could not be observed in less than 5 orbits (FUV+NUV) with STIS, or for which STIS/E140M was significantly more expensive than COS, were observed with COS in the FUV. Some of our targets already had COS/FUV, or STIS/E140M, or STIS/E230M archival observations. We verified that the S/N for those archival data was sufficient to achieve our science objectives, and did not repeat those observations. The targets observed with STIS E140M (FUV) and E230M/11978 (NUV) are listed in Table A1 (in the Appendix), along with the list of observations, settings, exposure times, and S/N per pixel (a STIS/E140M resolution element contains 1.4 pixel and the E230M resel is 1.9 pixels). Similarly, the targets observed with COS/G130M and G160M in the FUV and STIS in the NUV are listed in Table A2, along with the observations, exposure times and S/N per pixel (a COS/FUV resolution element contains 6-8 pixels).\\
\indent Extinction curves require high spectro-photometric accuracy ($<$ 5\%) and wide NUV wavelength coverage in the NUV to cover the 2175 \mAA bump. For this purpose, archival IUE or STIS low-resolution spectra were available for most targets. For the 7 stars for which they were not, we inserted short (100s) STIS/G140L and/or G230L exposures with S/N=100 at R=10 resolution. Table A3 (see Appendix) lists all the low-resolution STIS observations obtained as part of METAL to complement the IUE and HST/STIS low-resolution archives. The S/N listed in Table A3 is computed per pixel. \\
\indent  In the FUV, we expected the weakest line to be Mg II (1240 \AA). Assuming LMC abundances from \citet{russell1992}, and conservative depletions, the minimum equivalent width (EW) we expected for this line was 10 m\AA. This can be achieved (at 4$\sigma$) with S/N=25 per resolution element with COS/1291, and S/N=15 per resolution element with STIS/E140M. For COS/G160M, we targeted S/N = 15 per resolution element to probe the strong Fe II and Si II lines. In the NUV, we targeted S/N = 15 per resolution element with the E230M on the Si II (1808 \AA) and Fe II (2249 \mAA --- 2260 \AA) lines. This S/N was also sufficient to derive Cr, Zn column densities. For the COS observations, we used spectral dithering with 4 FPPOS to minimize fixed pattern noise. COS observations were taken in TIME-TAG mode, while STIS observations were taken in ACCUM mode.\\
\indent Straightforward COS ACQ/IMAGE target acquisitions were efficient with $<$ 100 s exposure times based on NUV magnitudes. In some cases, bright object limit concerns with field objects necessitated spectroscopic target acquisitions with COS. For the STIS observations, exposure times for target acquisitions with the 50CCD were $<$ 1s. Altogether, the program was completed in 101 orbits.

\subsection{WFC3 Imaging}

\indent In parallel with the spectroscopic observations, we obtained WFC3 exposures of fields adjacent (about 5 arcmin away) to the primary targets. Due to frequent buffer dumps required by COS exposures, we did not attach parallels to the one-orbit COS observations. The large nature of the program precluded timing or orientation constraints, and as a result, the fields were observed at the positions given by the roll of the telescope at the time of the observations. Dithering was not possible since the pointing was imposed by the primary spectroscopic observations. However, when possible several exposures were obtained in the same filter to allow for cosmic ray rejection with the UVIS detector (in the IR, this can be done as part of the up-the ramp data reduction). \\
\indent The photometric parallel observations follow the same wavelength-coverage strategy as the Panchromatic Hubble Andromeda Treasury (PHAT) survey \citep{dalcanton2012} and the SMIDGE (Small Magellanic Cloud Investigation of Dust and Gas Evolution) survey \citep{yanchulova2017}. The filter selection was motivated by the need to 1) obtain as deep a color-magnitude diagram (CMD) as possible, and 2) resolve the degeneracy between stellar effective temperature and dust extinction for both hot and cool stars (Figure \ref{plot_filters}). The best way to achieve this is to observe with F225W, F275W, F336W, F475W, F814W, F110W, and F160W. This is demonstrated in Figure \ref{plot_filters}, which shows model spectra and synthetic WFC3 SEDs for stars of different masses and with different levels and types of reddening. The IR bands break the degeneracy between stellar mass and reddening, while the UV-optical bands allow us to discriminate between different types of dust. More details on the filter selection can be found in \citet{dalcanton2012}.\\
 \indent  The PHAT survey did not include F225W, but given the benefits of the F225W in constraining the strength of the 2175 \mAA extinction feature, we added one or more exposures in this filter when possible. Targets observed with 1 STIS orbit (5/33 targets) allowed for WFC3 observations with F275W, F336W, F475W, and F814W. Targets observed with 2 or more orbits of STIS allowed for WFC3 exposures with the full suite of filters (F225W, F275W, F336W, F475W, F814W, F110W, and F160W). The more orbits, the more exposures were taken to optimize cosmic ray rejection and exposure time. The list of WFC3 exposures, the orientation of each field (PA), the coordinates of the field centers, the exposure times, and durations of the post-flash implemented to reduce CTE (charge transfer efficiency) effects, are listed in the Appendix (Table A4). \\

\section{Data Reduction}\label{reduction_section}
\indent In the following section, we describe in detail the data reduction for the different components (medium-resolution COS and STIS spectroscopy, WFC3 imaging) of the METAL observations (doi:10.17909/T90Q30).

\begin{figure*}
\centering
\includegraphics[width=8cm]{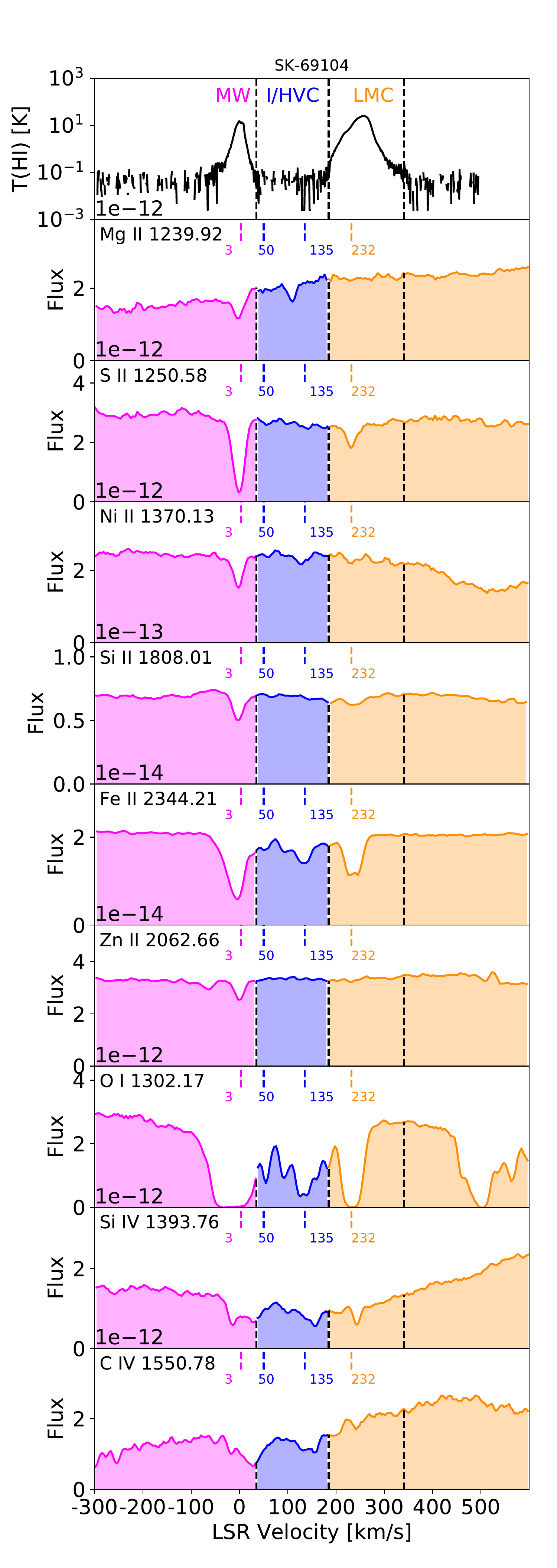}
\includegraphics[width=8cm]{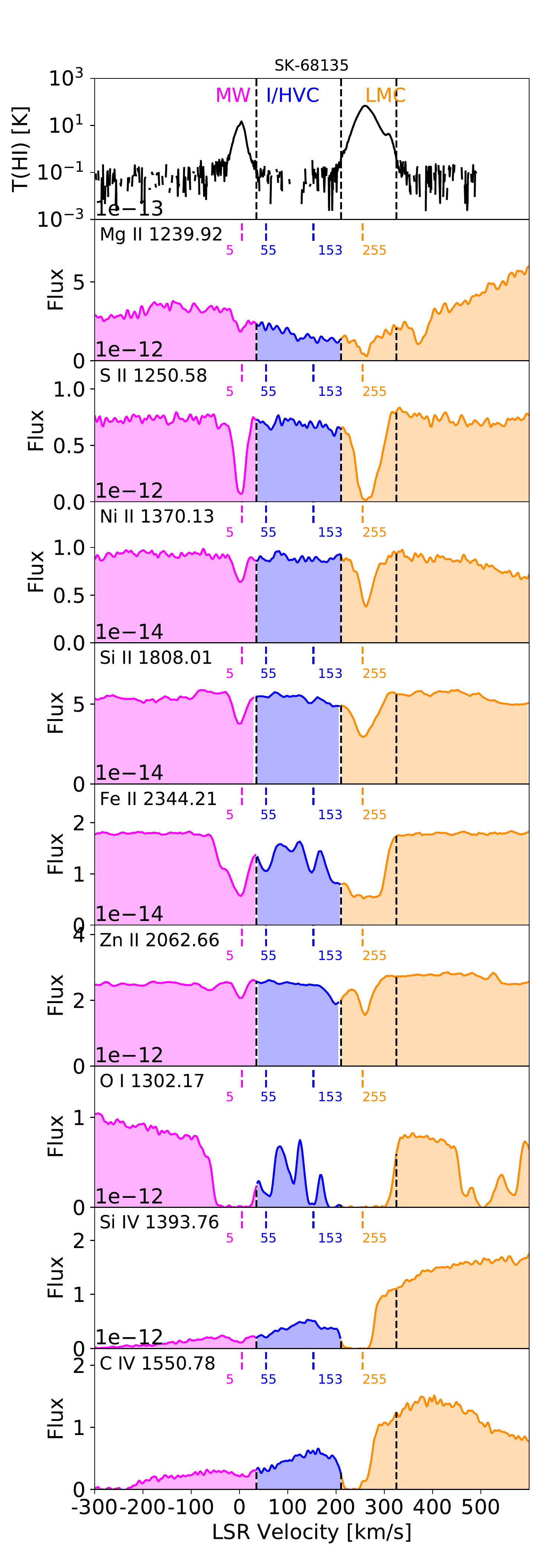}
\end{figure*}

\begin{figure*}
\centering
\includegraphics[width=8cm]{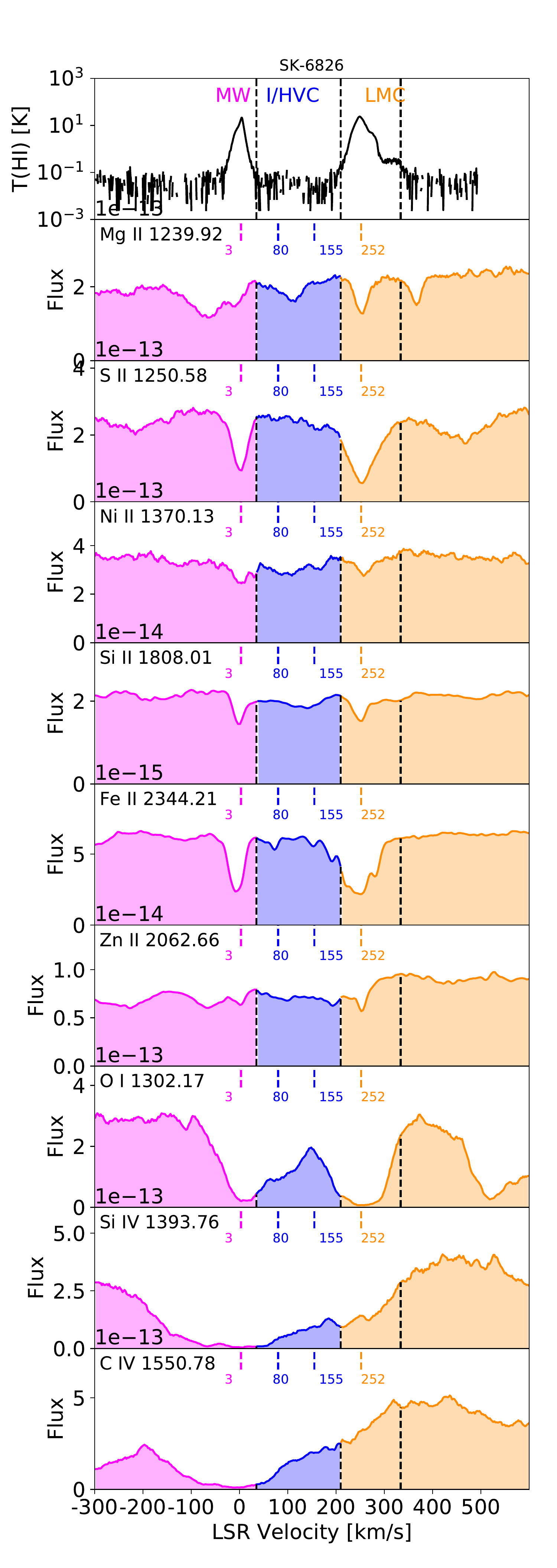}
\includegraphics[width=8cm]{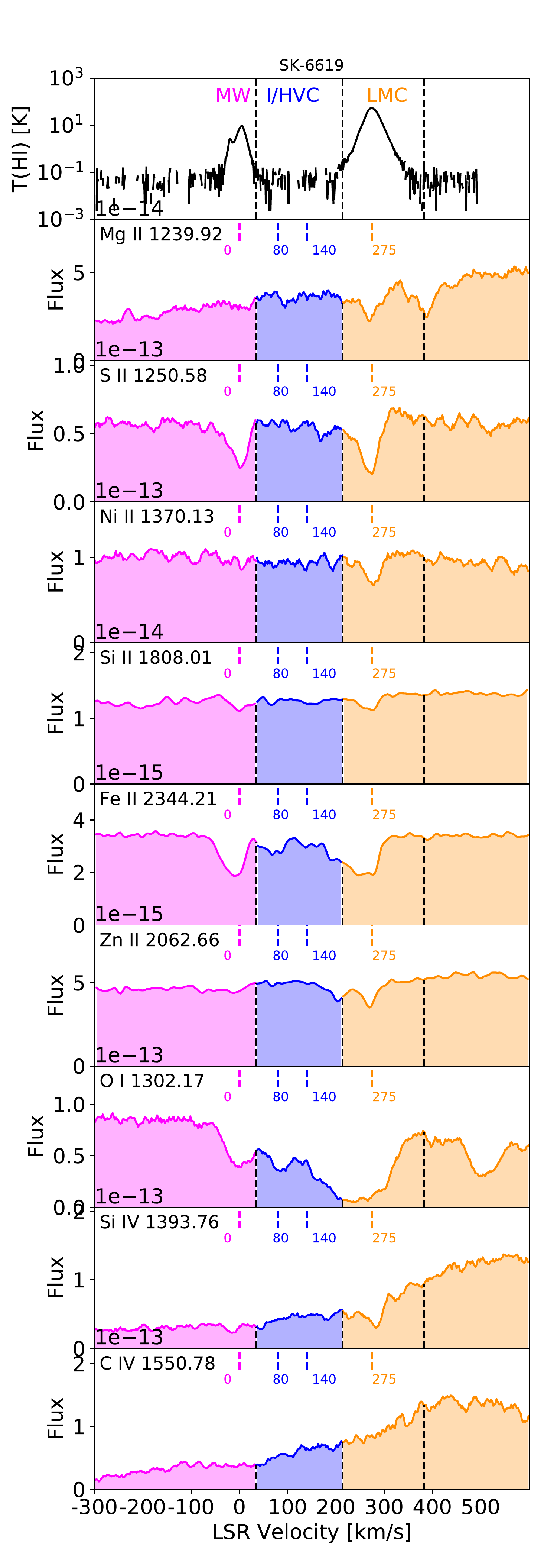}
\caption{Blow-ups of ISM spectral lines of interest (as labeled in each panel) for the same sub-sample of targets as in Figure \ref{show_spectra}. The spectra are shown as a function of LSR velocity. The top panels show the \his emission in the GASS survey \citep{mccluregriffiths2009}, which is used to determine the velocity ranges for the Silicon column density determination (Section \ref{depletion_section}) and for the intermediate velocity gas (Section \ref{ivg_section}). The velocity ranges associated with the MW and LMC are determined from the boundaries of the \his emission peaks, where emission drops below 4$\sigma$  (dashed vertical lines). The magenta-filled area corresponds to MW gas, the blue-filled area corresponds to the intermediate and high velocity gas, while the orange d-filled area is associated with LMC gas. The velocity components identified in the Fe II 2344 \mAA line (one component for the MW, one or two components for the intermediate velocity gas, and one component for the LMC) are shown by colored short vertical dashed lines.}
\label{show_zooms}
\end{figure*}

\subsection{STIS spectroscopy}

\indent The STIS observations, listed in Tables 3, 4, and 5, were taken between September 2016 and July 2017, and retrieved from the {\it Mikulski archive for Space Telescopes} (MAST) Archive between September 2016 and July 2017. The corresponding version of the STIS calibration pipeline, CalSTIS, evolved from 3.4 (November 2013) to 3.4.1 (April 2017). No changes were made in the processing of medium-resolution echelle spectra in the 2017 pipeline version. \\
\indent  The different spectral orders in an exposure and different exposures with the same central wavelength setting were interpolated on a grid of wavelengths that has approximately the same wavelength spacing as the pixel pitch (which varies slightly with wavelength).  All spectral orders and exposures were then co-added onto this common grid.  Weightings of the spectra were done in accord with the inverse squares of the local intensity errors in the different exposures.  However, these weights are measured with S/N smoothed over a 10 pixel interval centered on each wavelength.  This smoothing prevents a bias where individual measurements with chance positive intensity noise excursions are favored over those with excursions in the opposite direction. A subset of the spectra is shown as a function of wavelength in Figure \ref{show_spectra} (SK-69104 and SK-68135). Figure \ref{show_zooms} shows zooms on spectral lines of interest as a function of LSR velocity. The LMC absorption is visible between 200 km/s and 350 km/s.

\begin{figure*}
\centering
\includegraphics[width=8cm]{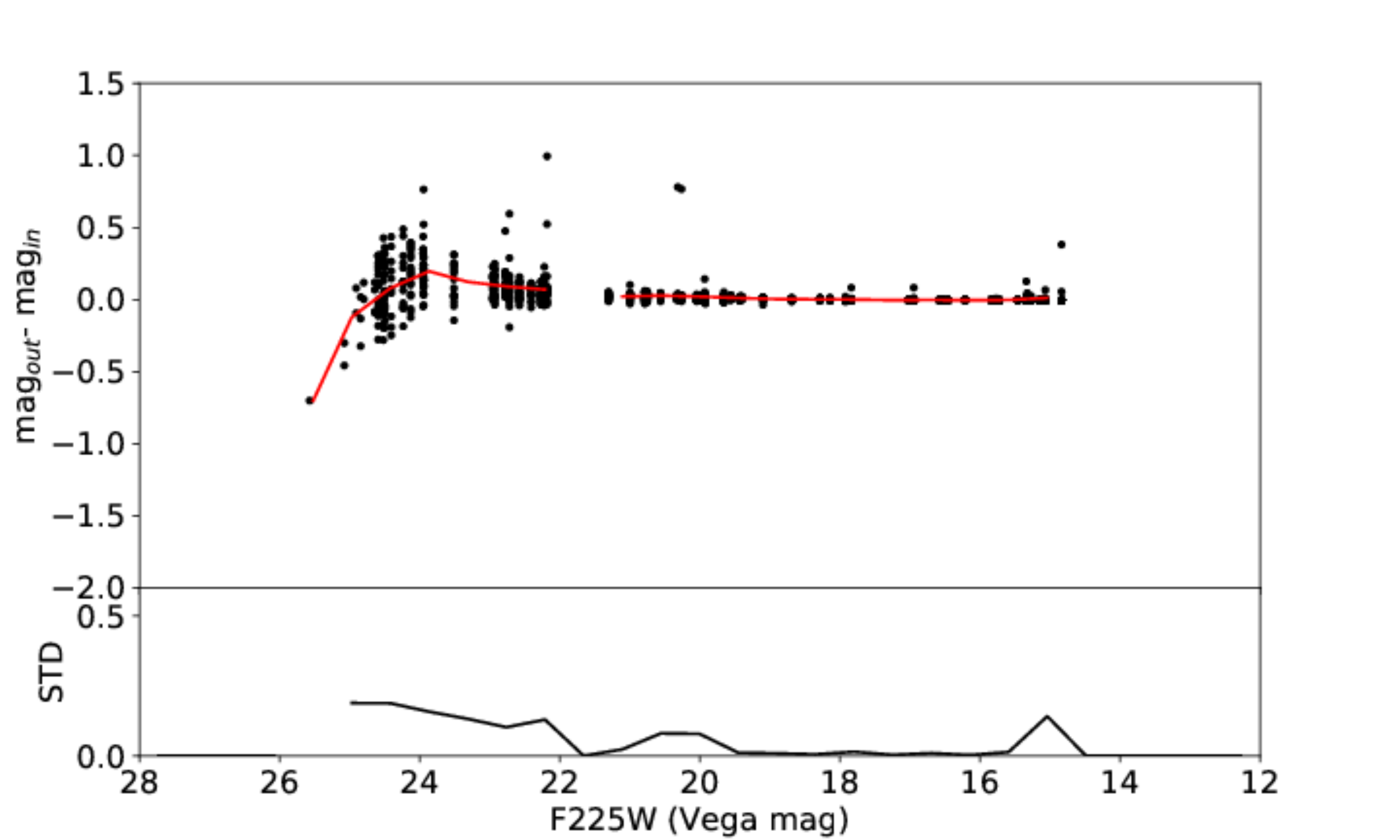}
\includegraphics[width=8cm]{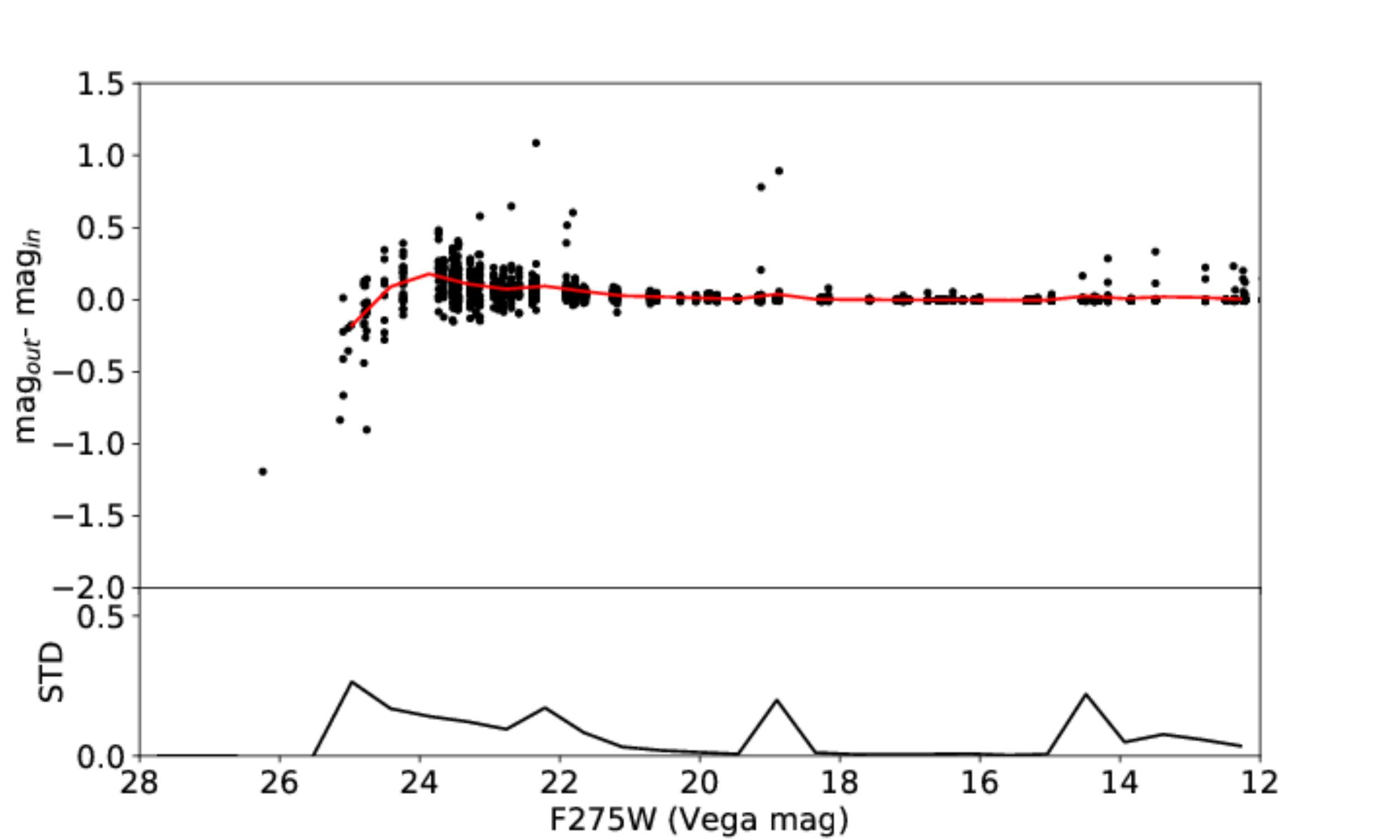}
\includegraphics[width=8cm]{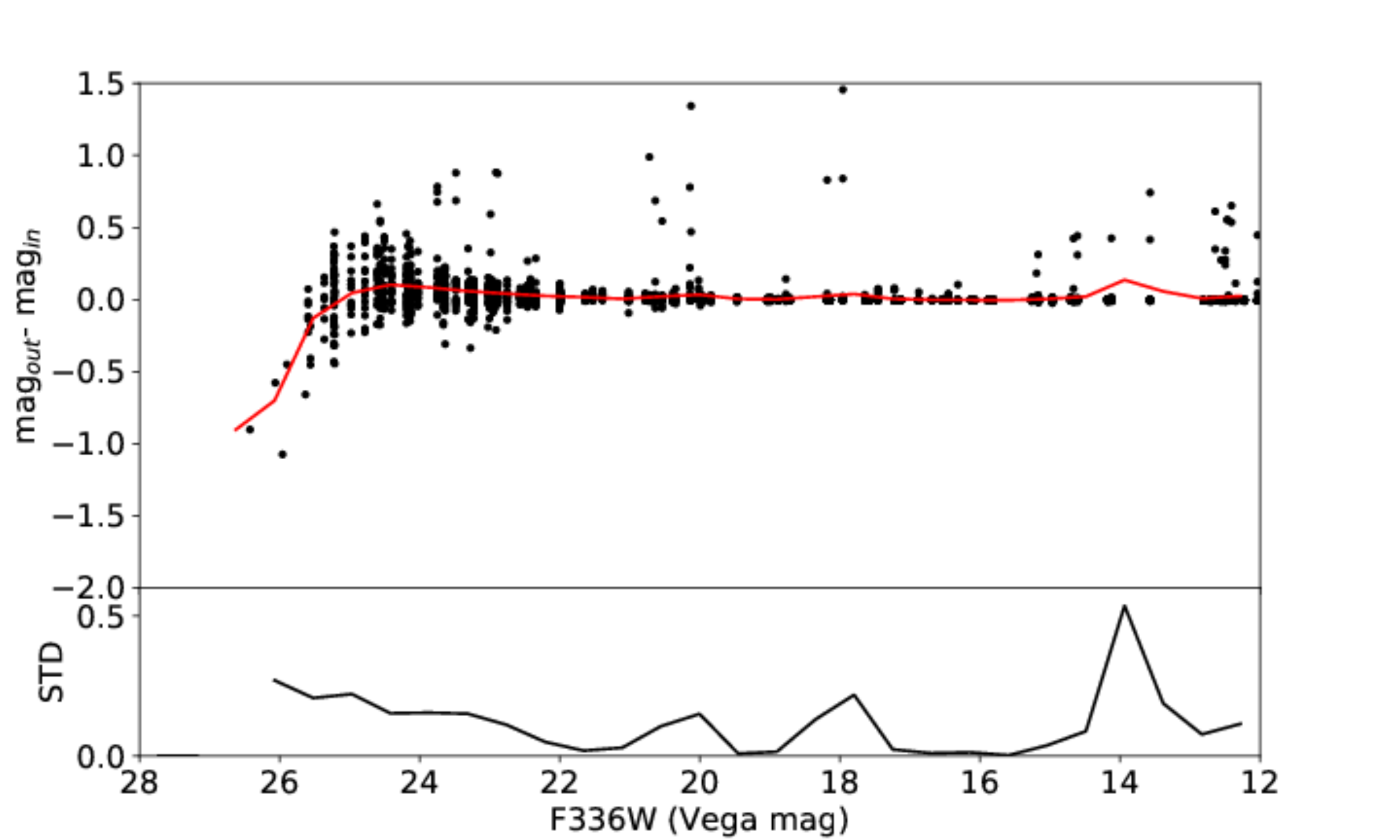}
\includegraphics[width=8cm]{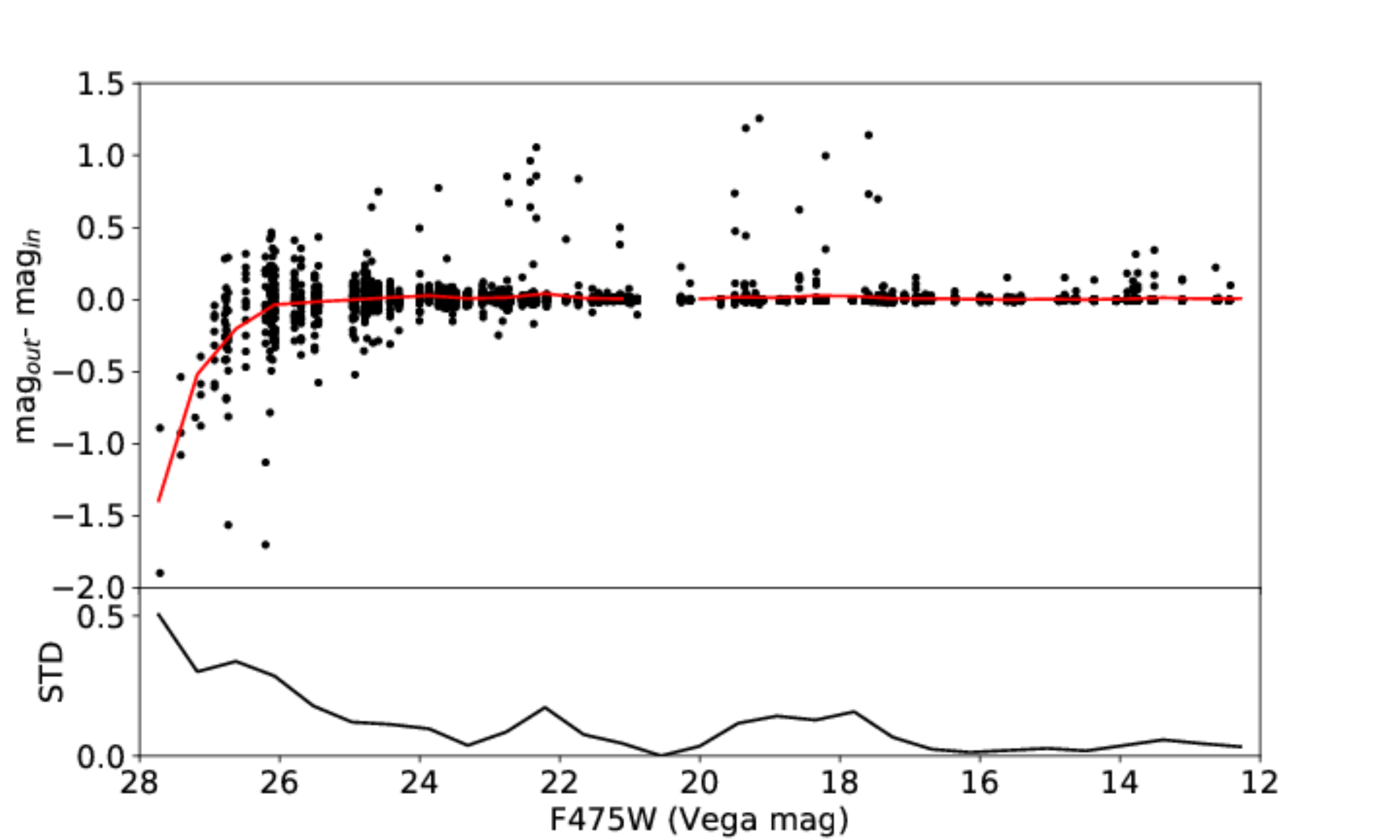}
\includegraphics[width=8cm]{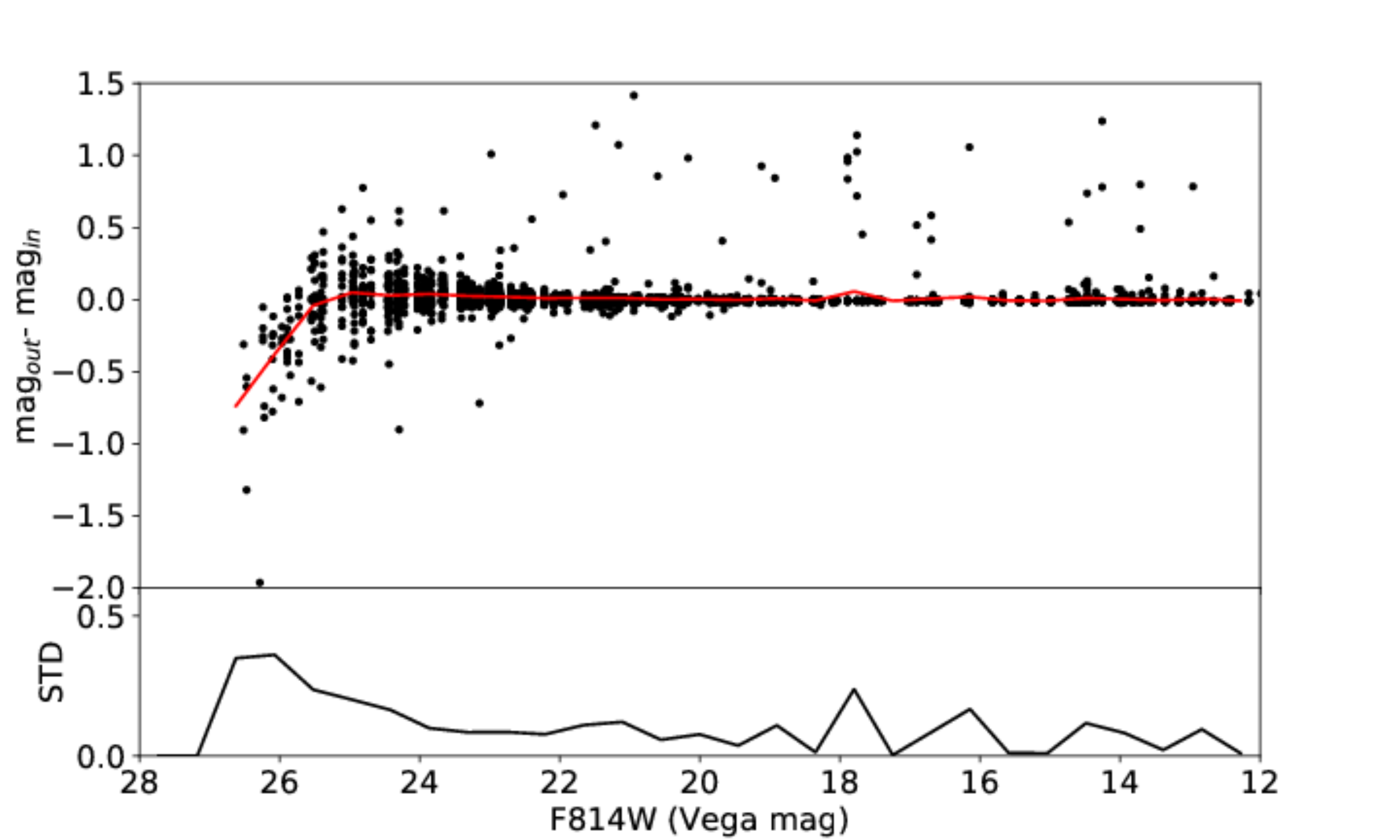}
\includegraphics[width=8cm]{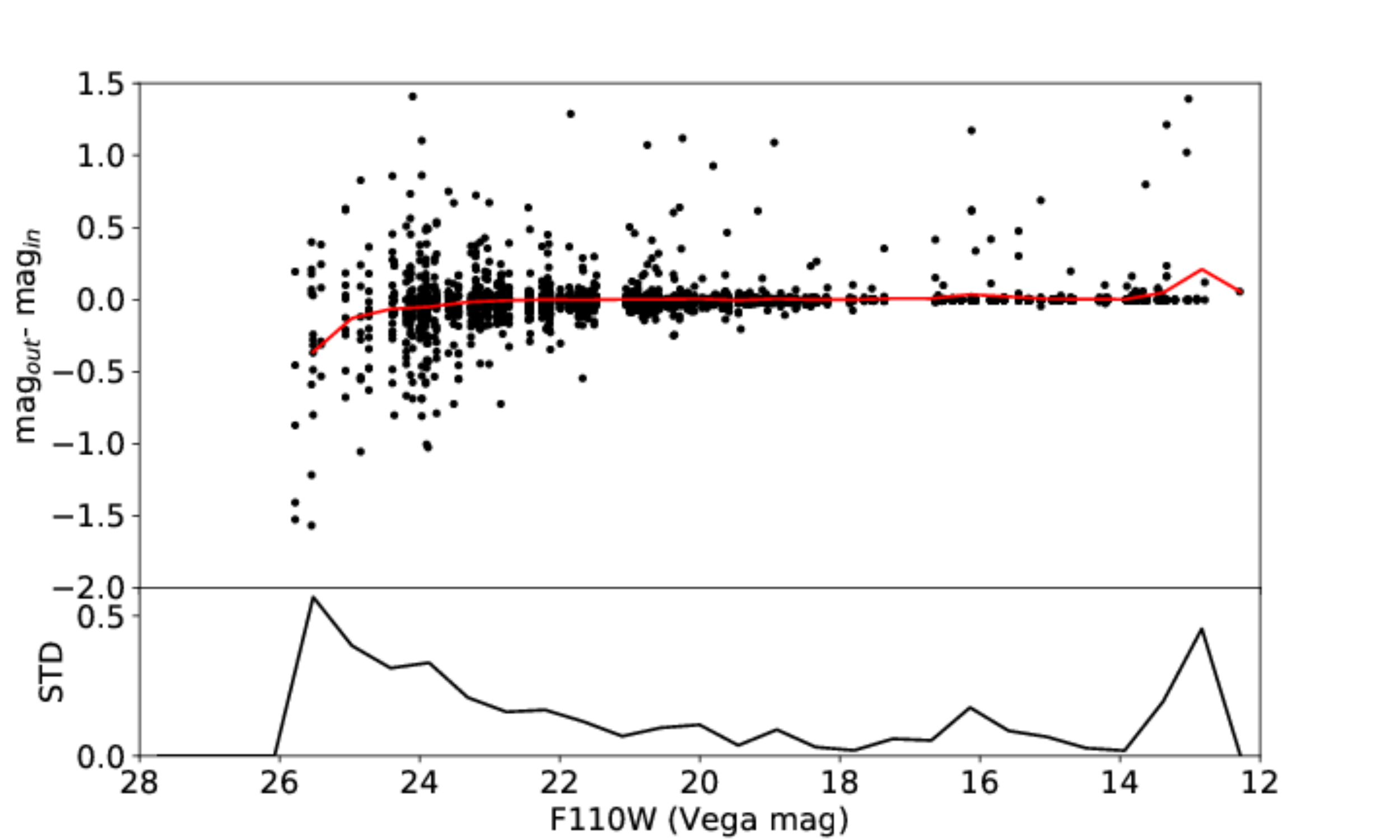}
\includegraphics[width=8cm]{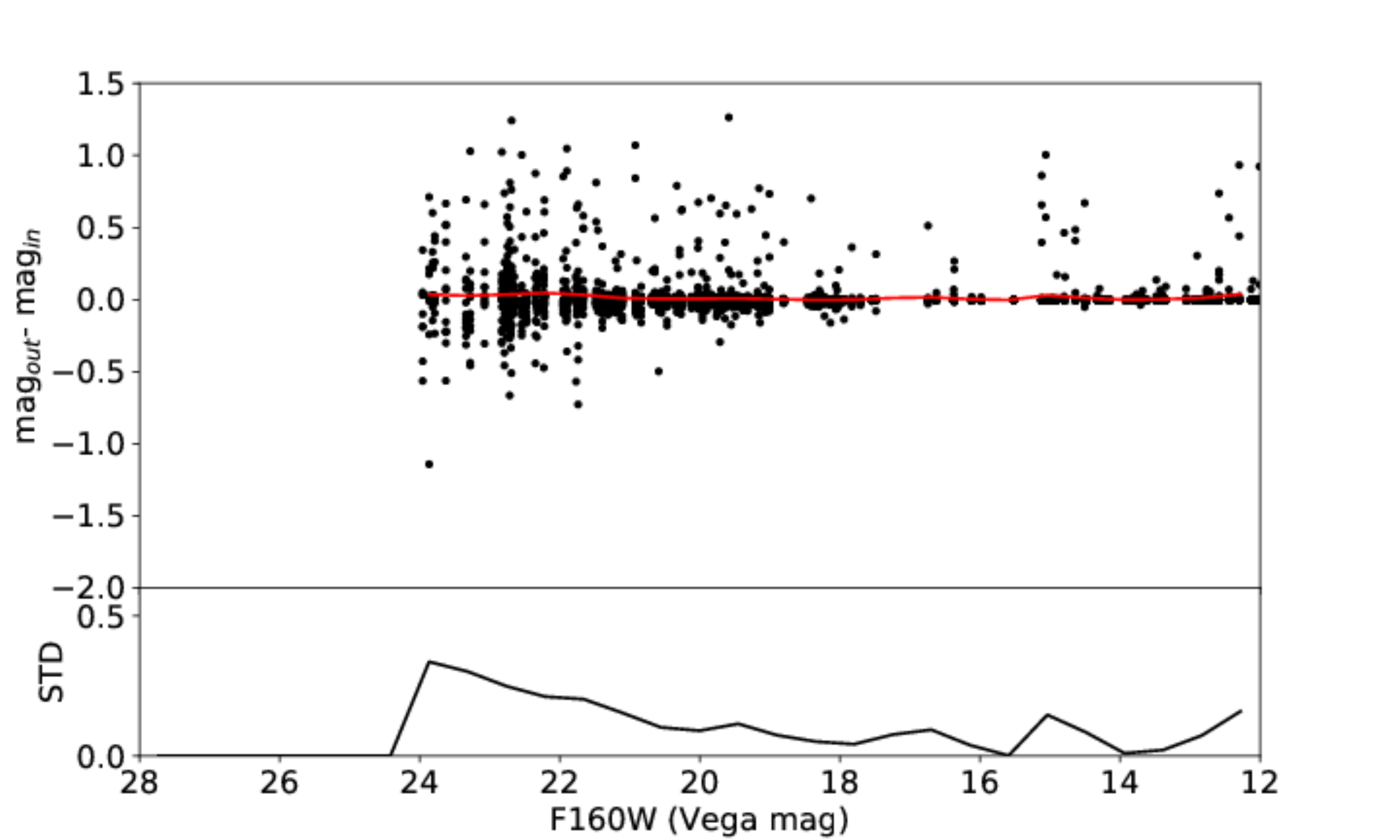}

\caption{Recovered photometry versus input photometry for the artificial star tests in field LMC-5665ne-12232, taken in parallel with the SK-68140 spectroscopy. The top panels show the individual measurements (black points) of the difference between input and output magnitude, and the binned mean (red), corresponding to the bias. The bottom panels show the standard deviation (noise) of the difference between input and output magnitude. Each panel corresponds to a filter, indicated in the x label.}
\label{plot_asts}
\end{figure*}

\begin{figure*}
\centering
\includegraphics[width=8cm]{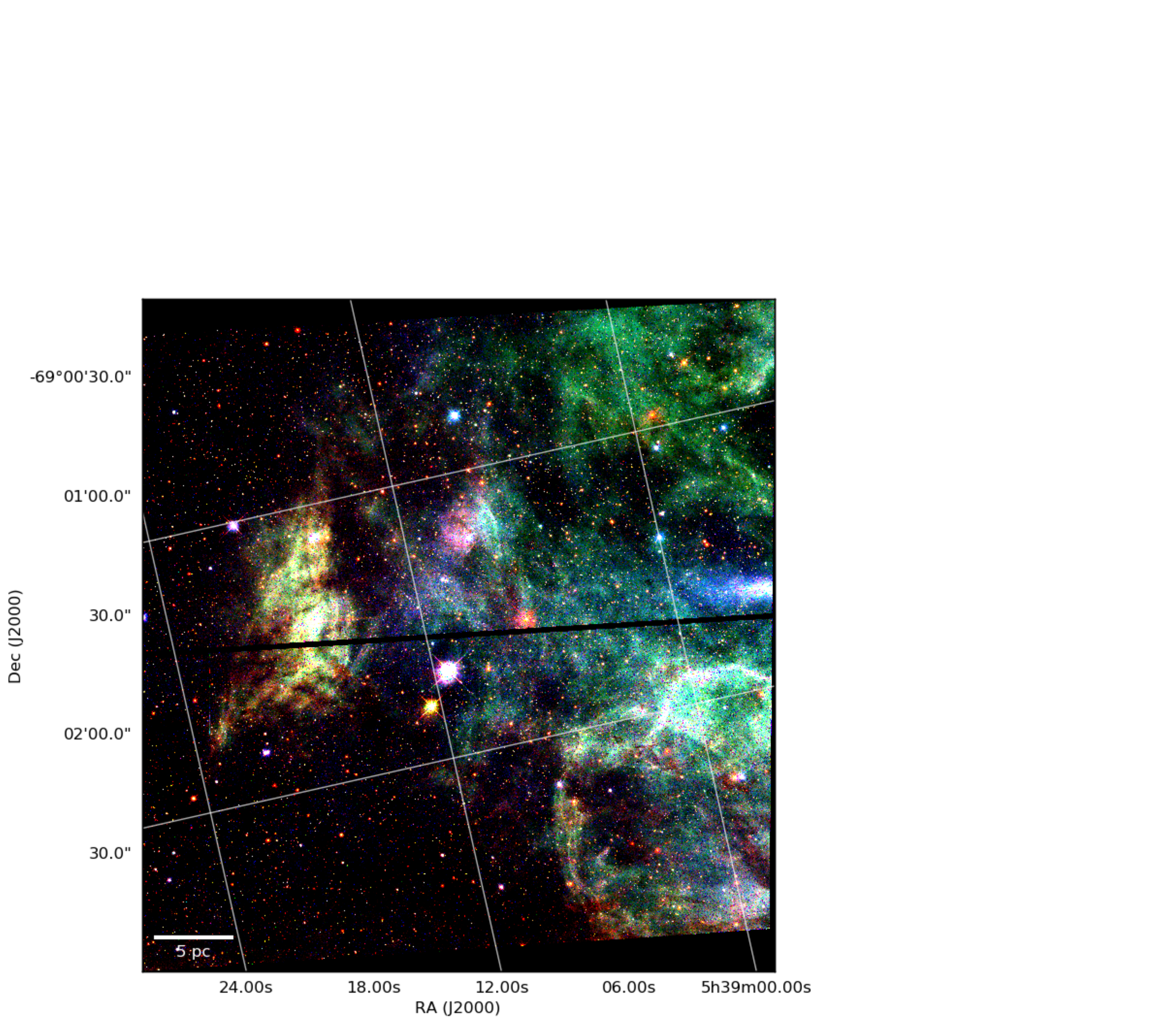}
\includegraphics[width=8cm]{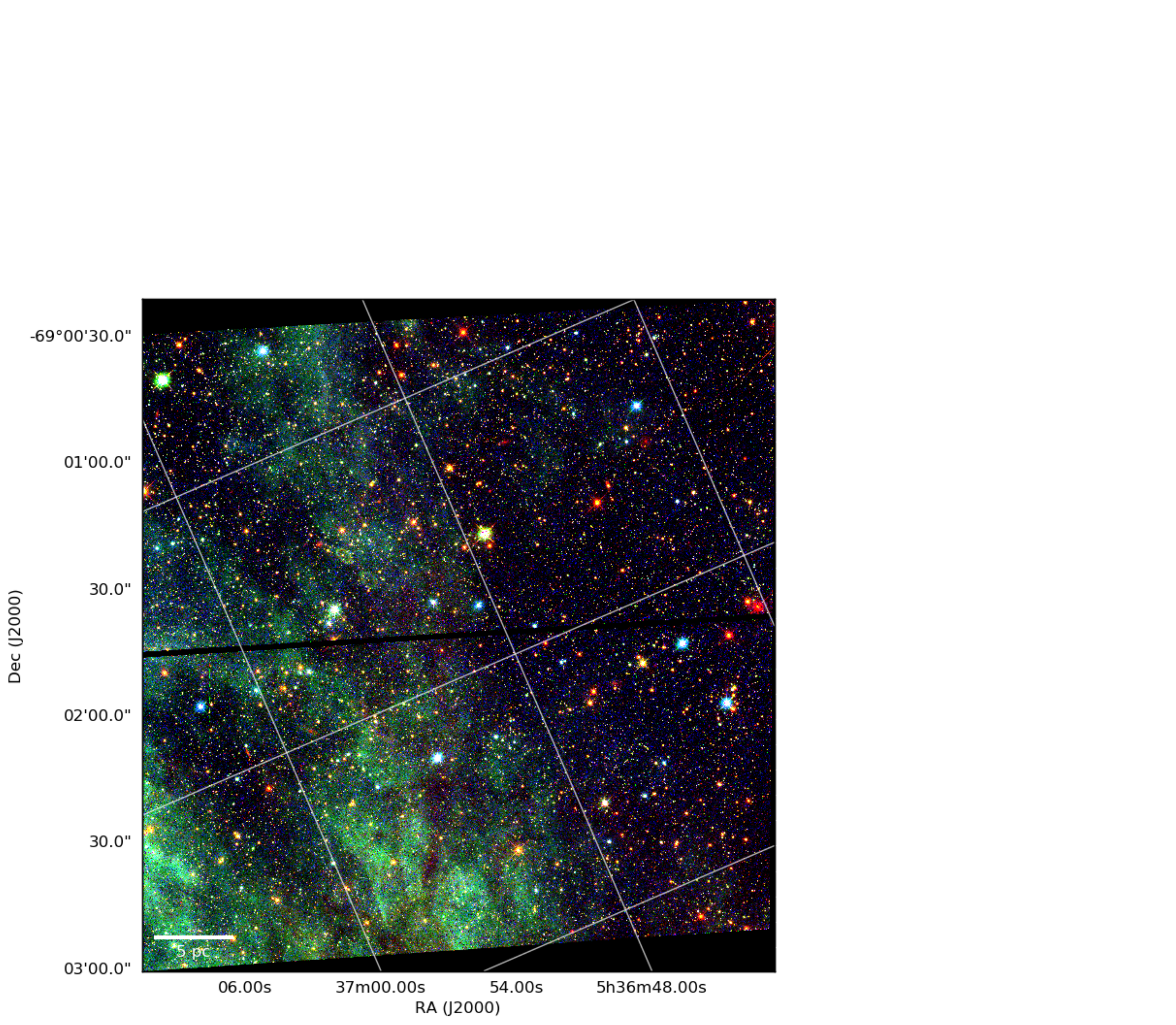}
\includegraphics[width=8cm]{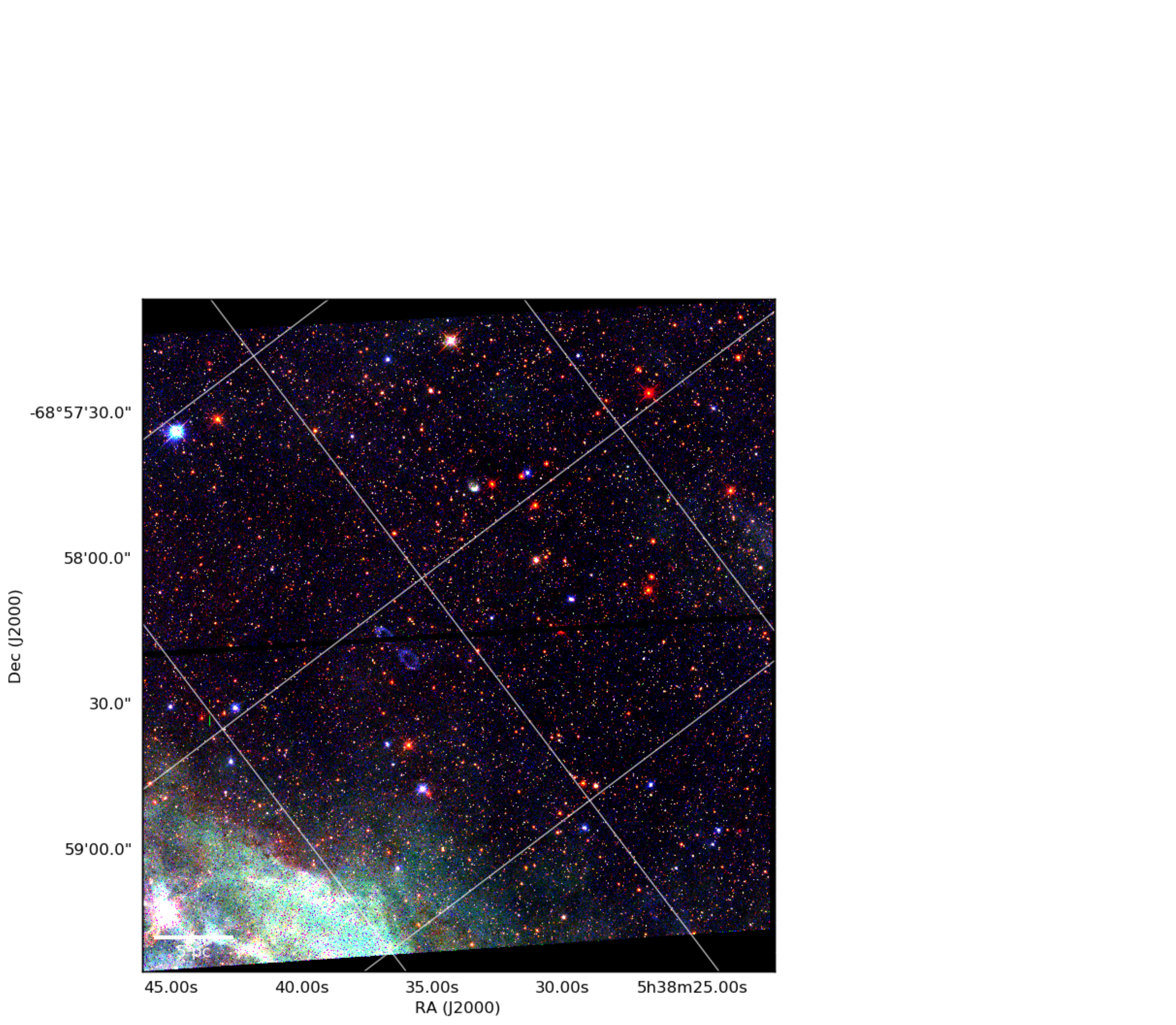}
\includegraphics[width=8cm]{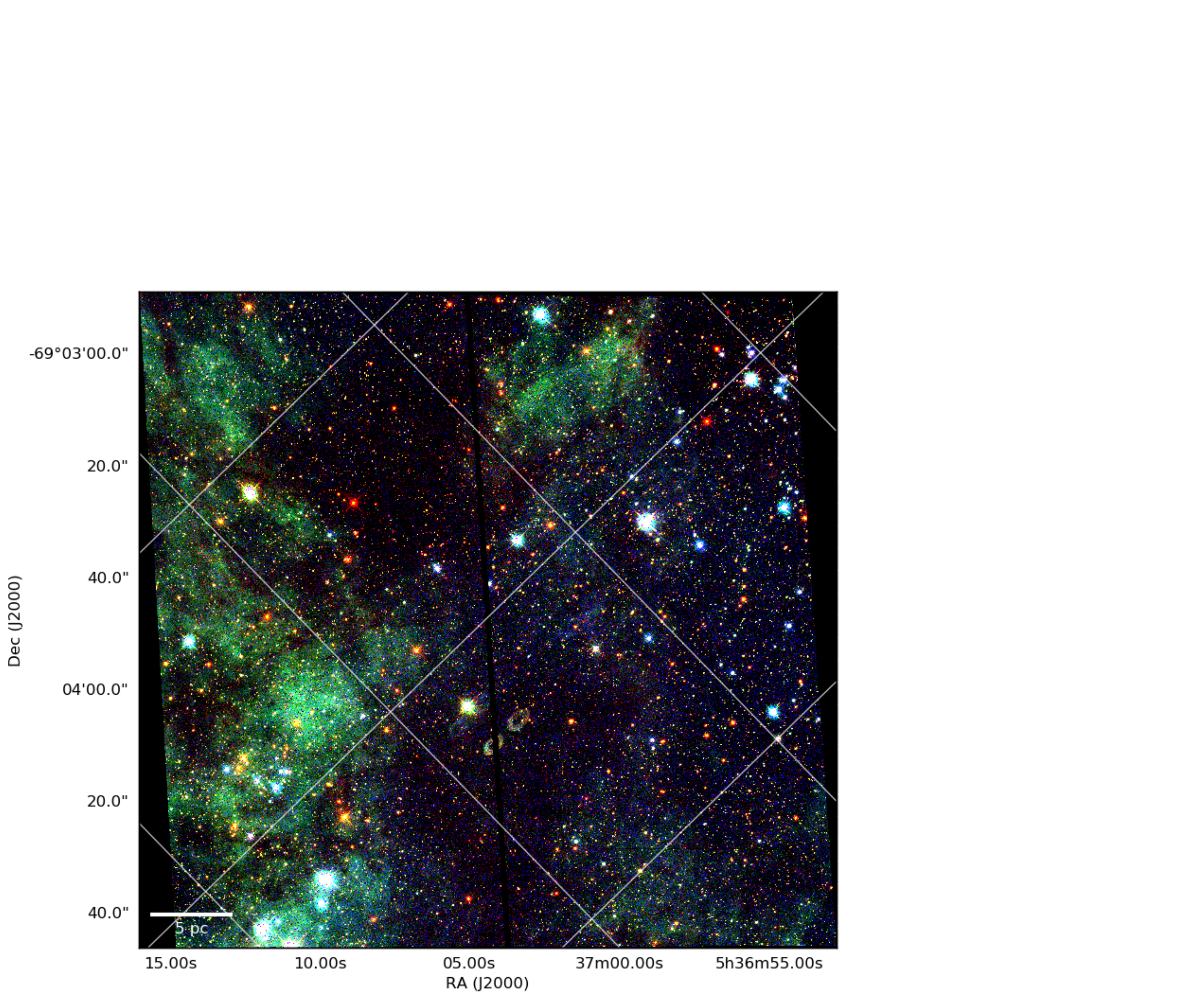}
\includegraphics[width=8cm]{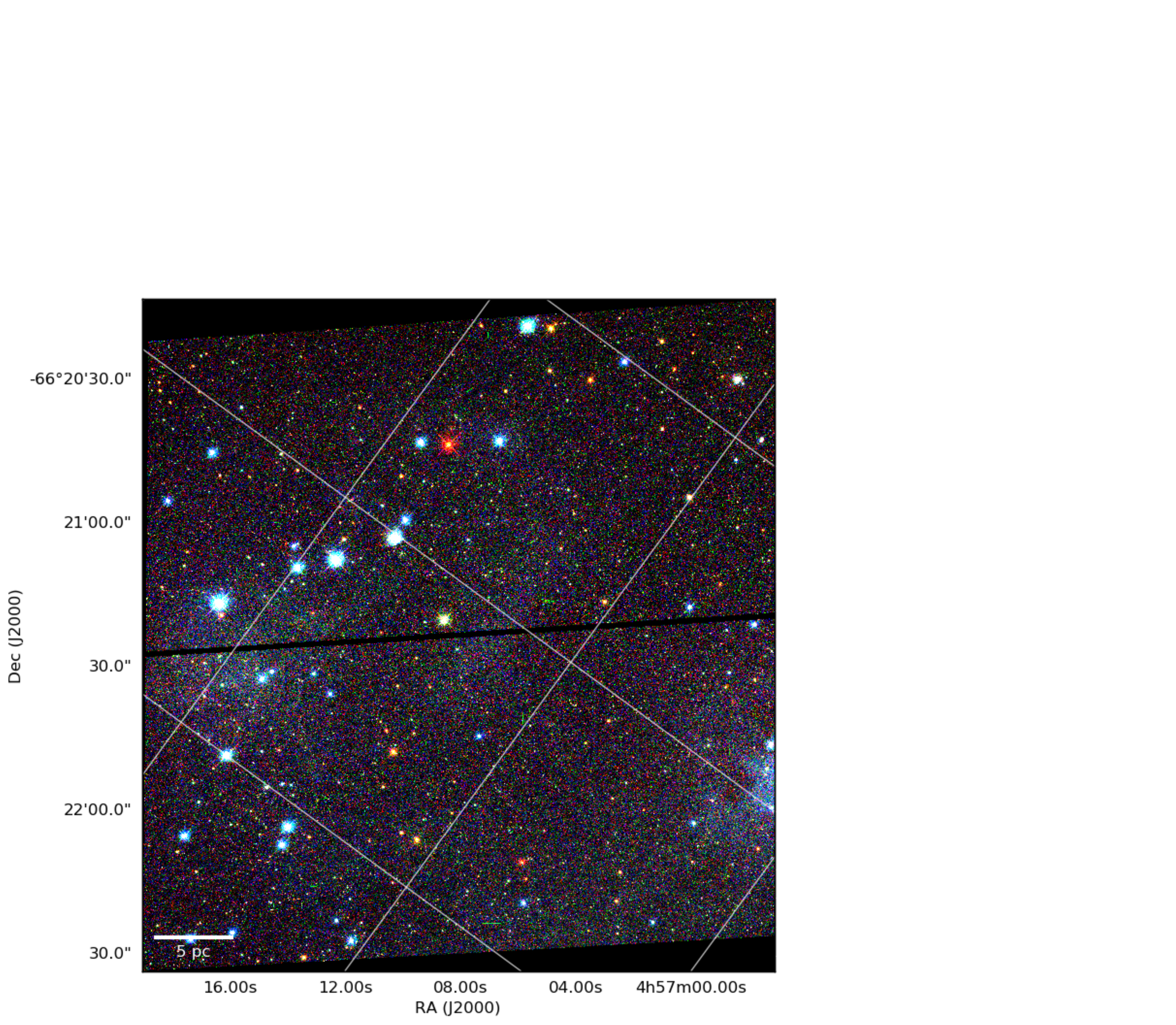}
\includegraphics[width=8cm]{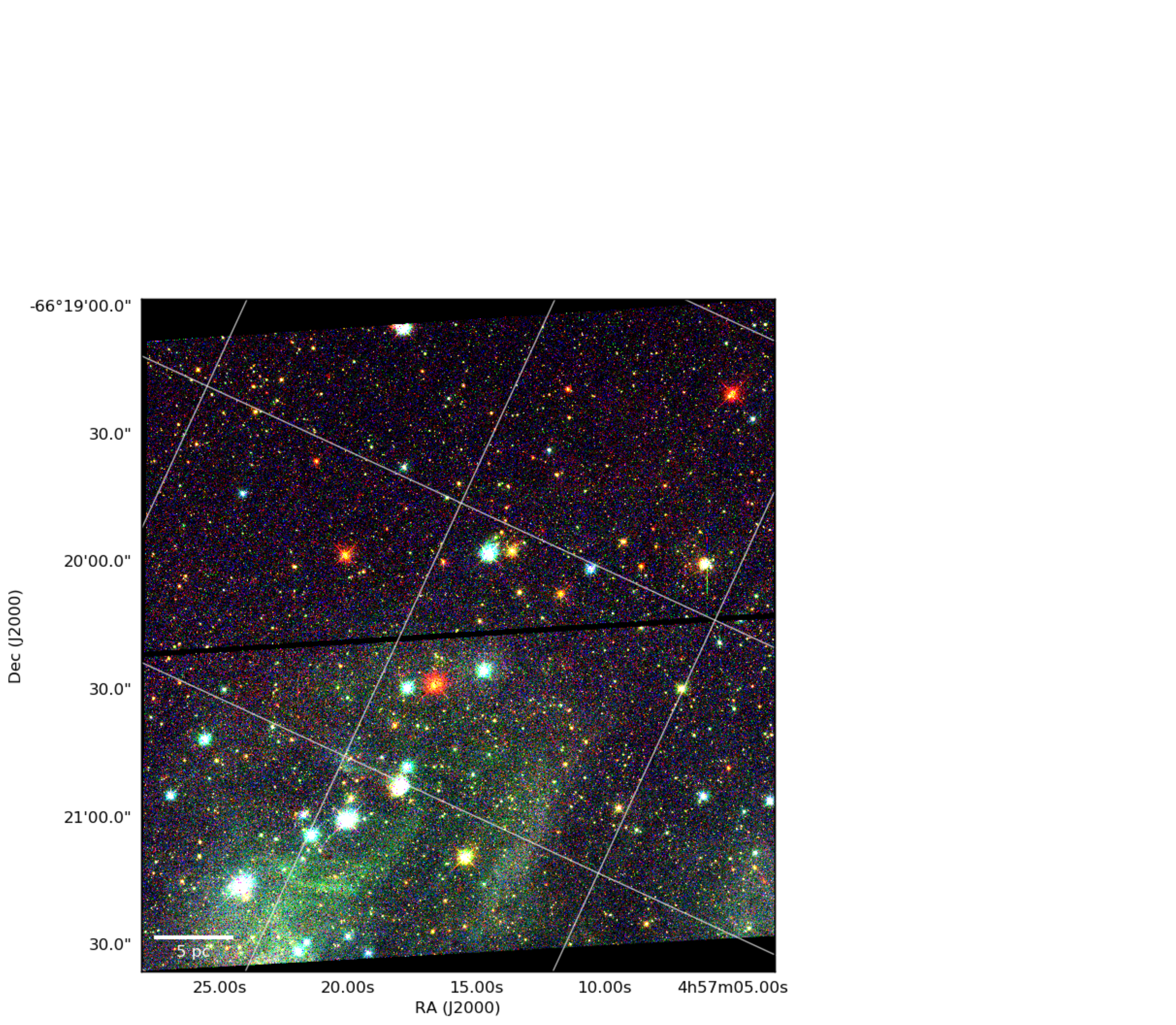}

\caption{Three-color images of select fields, with the F336W in blue, the F475W in green, and the F814W in red. From left to right, top to bottom, the fields were taken as parallel observations for the following spectroscopic targets: SK-68140, SK-68129, SK-68135, BI253, PGMW3120, PGMW3223. The nebular gas emission in F475W is due to [OIII] $\lambda$5007, [OIII] $\lambda$4959, H$\beta$, and H$\gamma$. The F814W includes contributions from nebular [SIII] $\lambda$9069, while F336W includes nebular continuum, Balmer continuum, and He II $\lambda$3203. 
}
\label{wfc3_3col}
\end{figure*}

\begin{figure*}
\centering
\includegraphics[width=8cm]{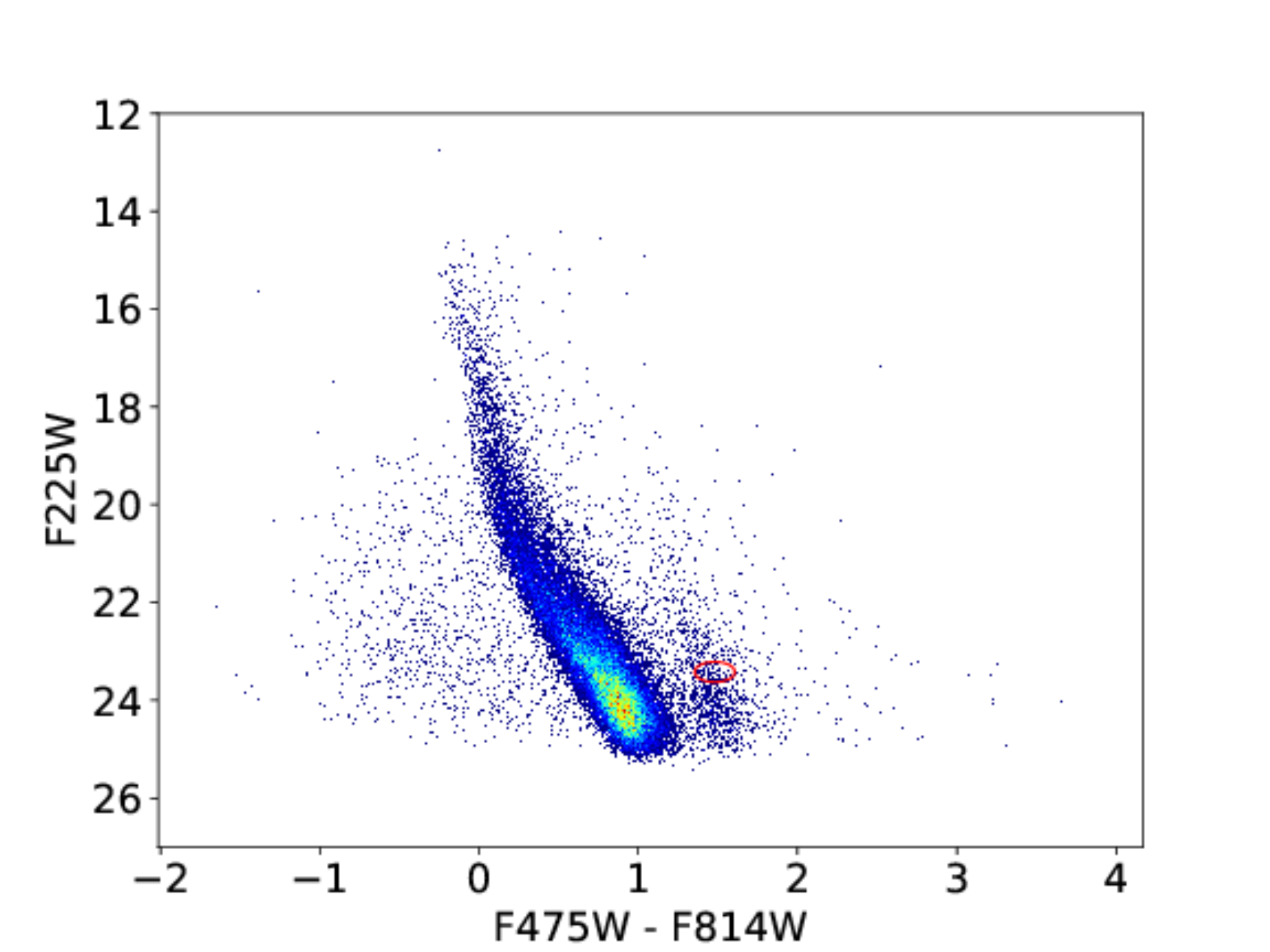}
\includegraphics[width=8cm]{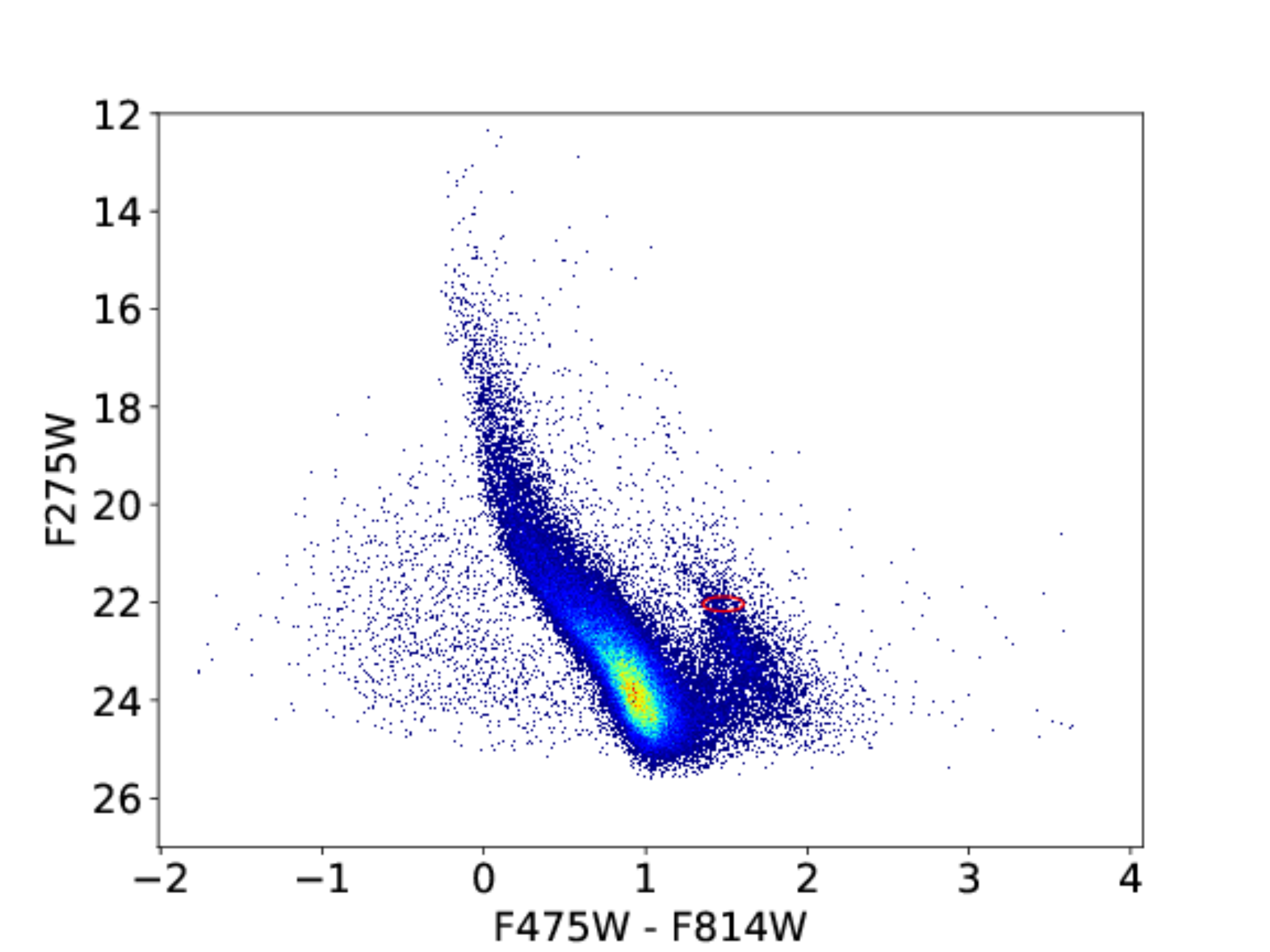}
\includegraphics[width=8cm]{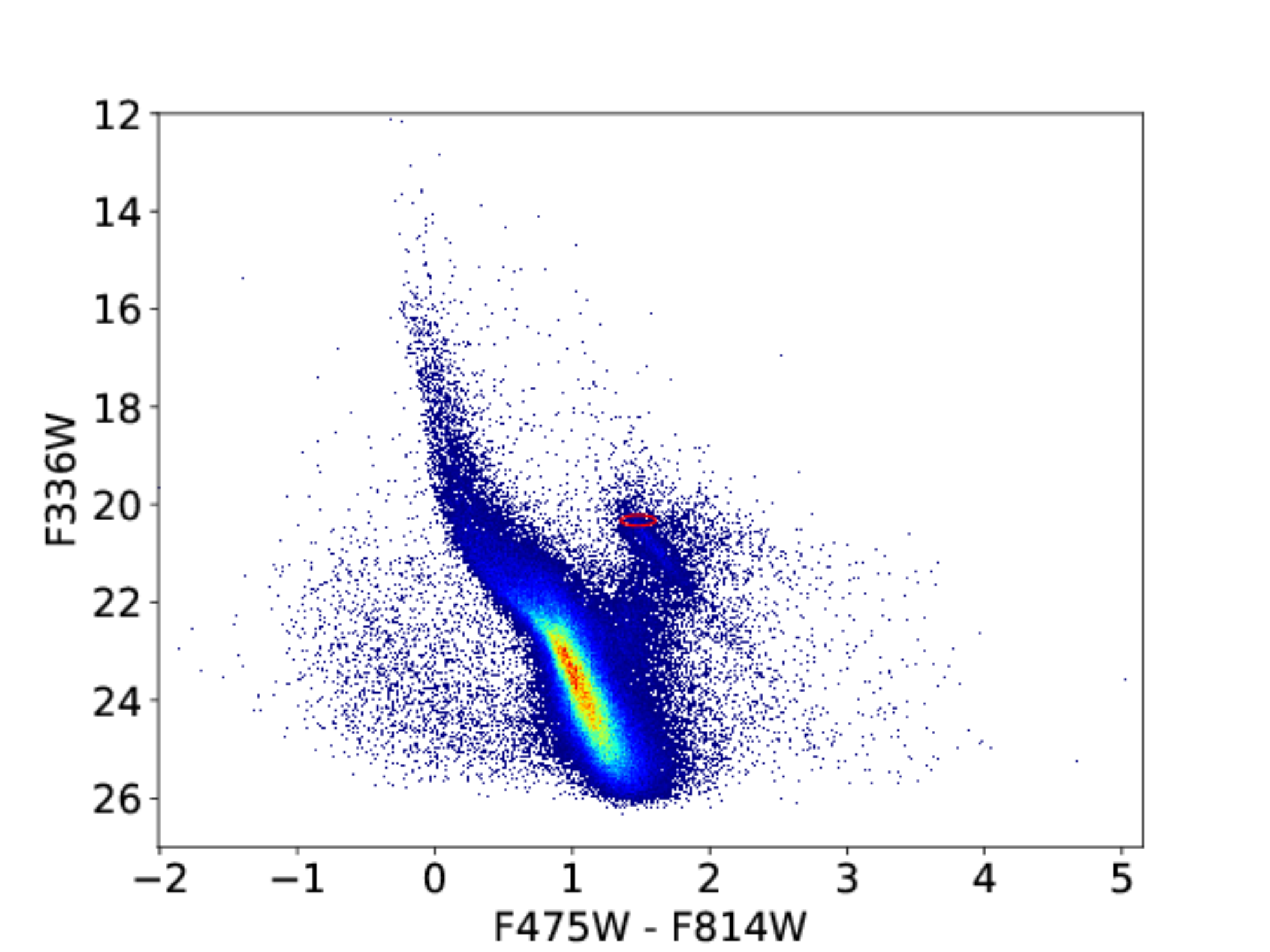}
\includegraphics[width=8cm]{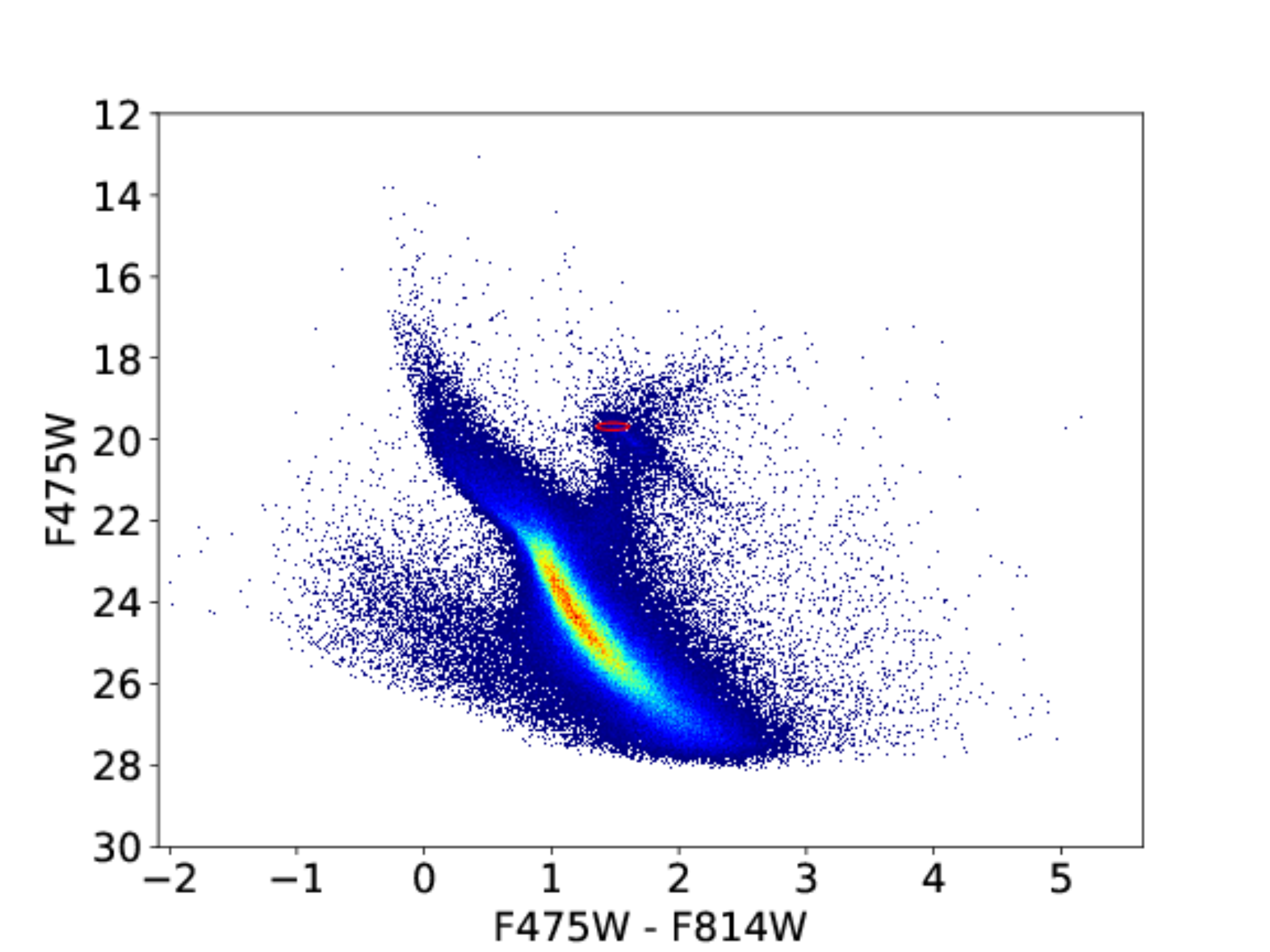}
\includegraphics[width=8cm]{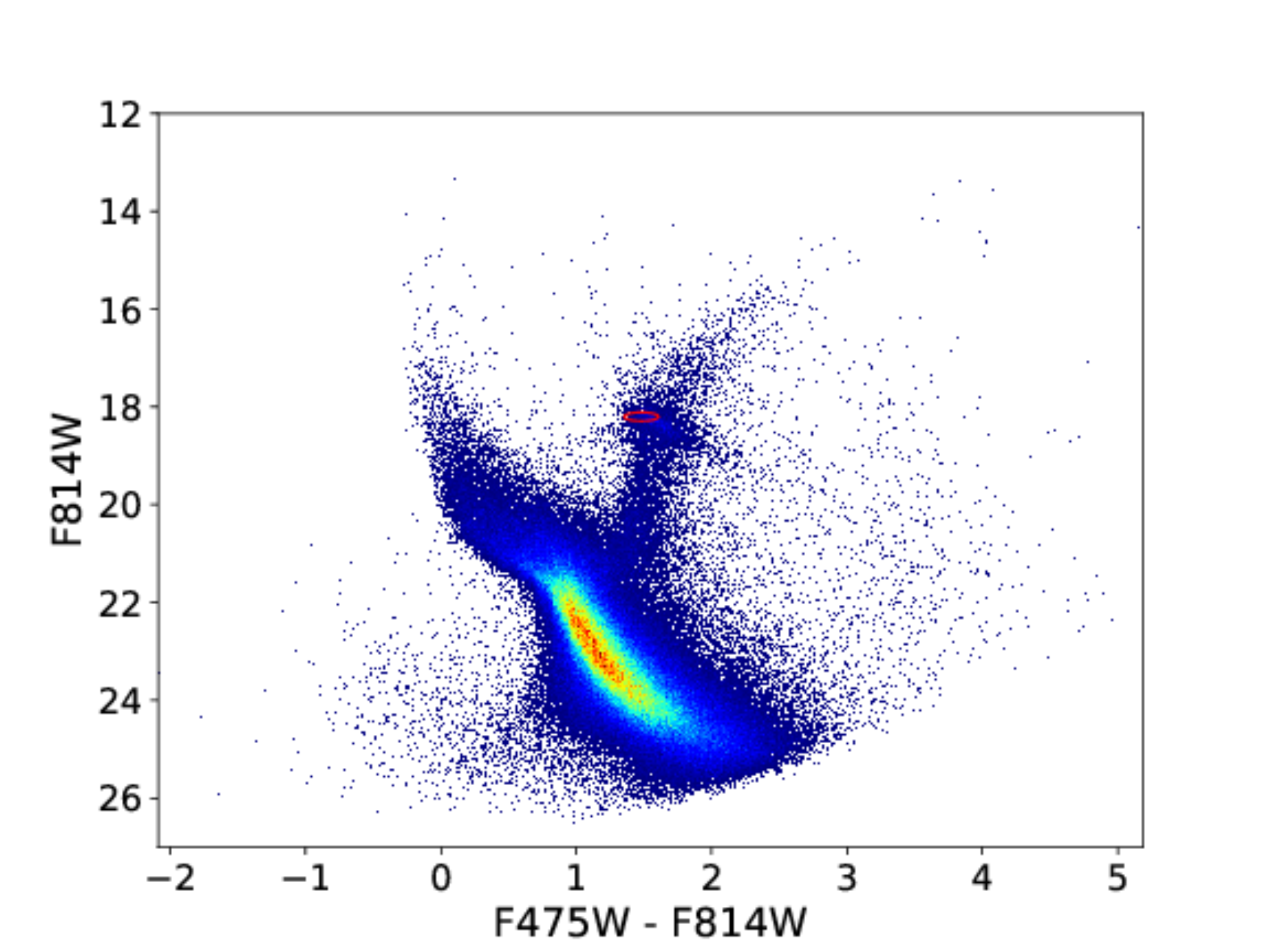}
\includegraphics[width=8cm]{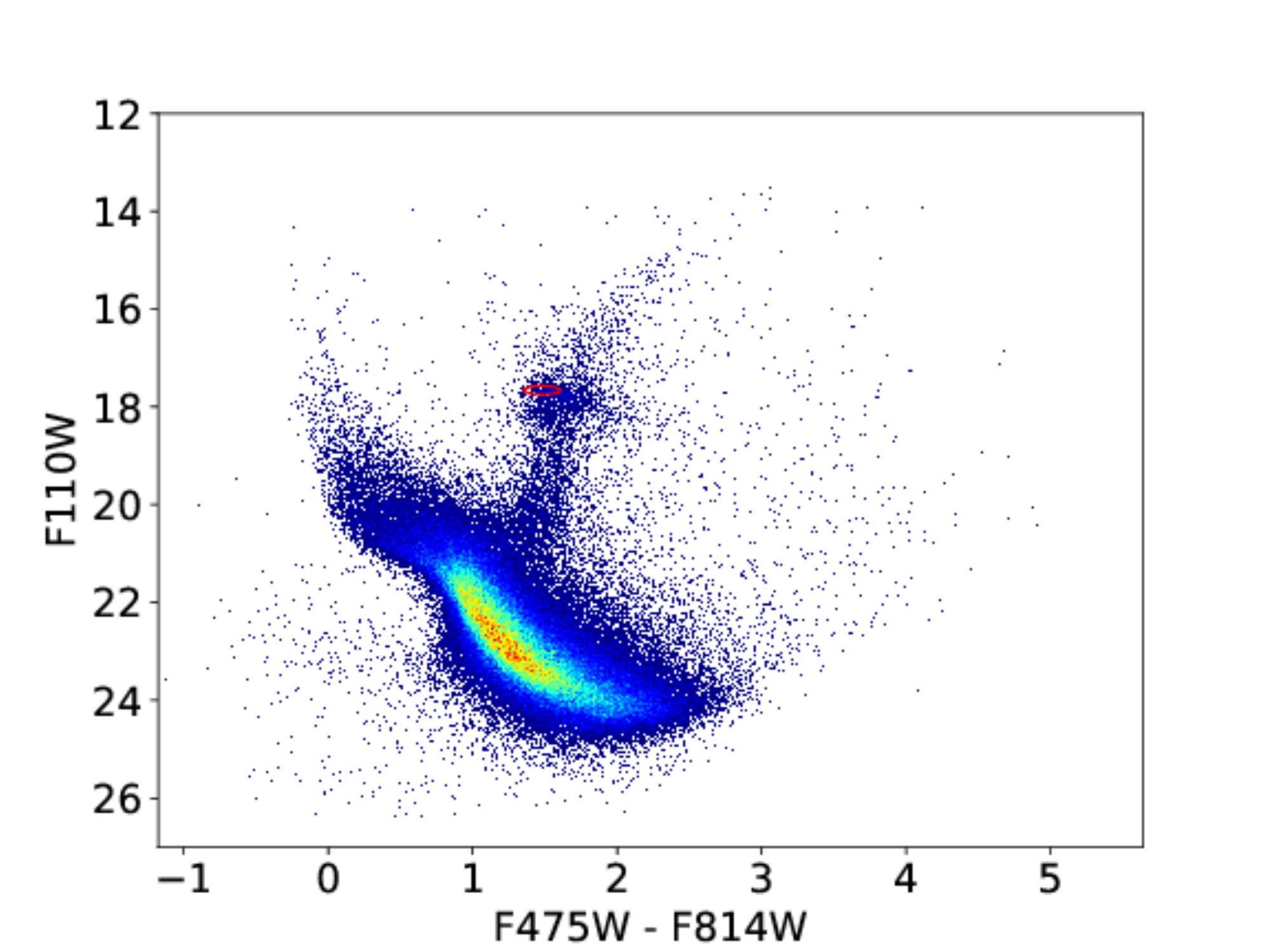}
\includegraphics[width=8cm]{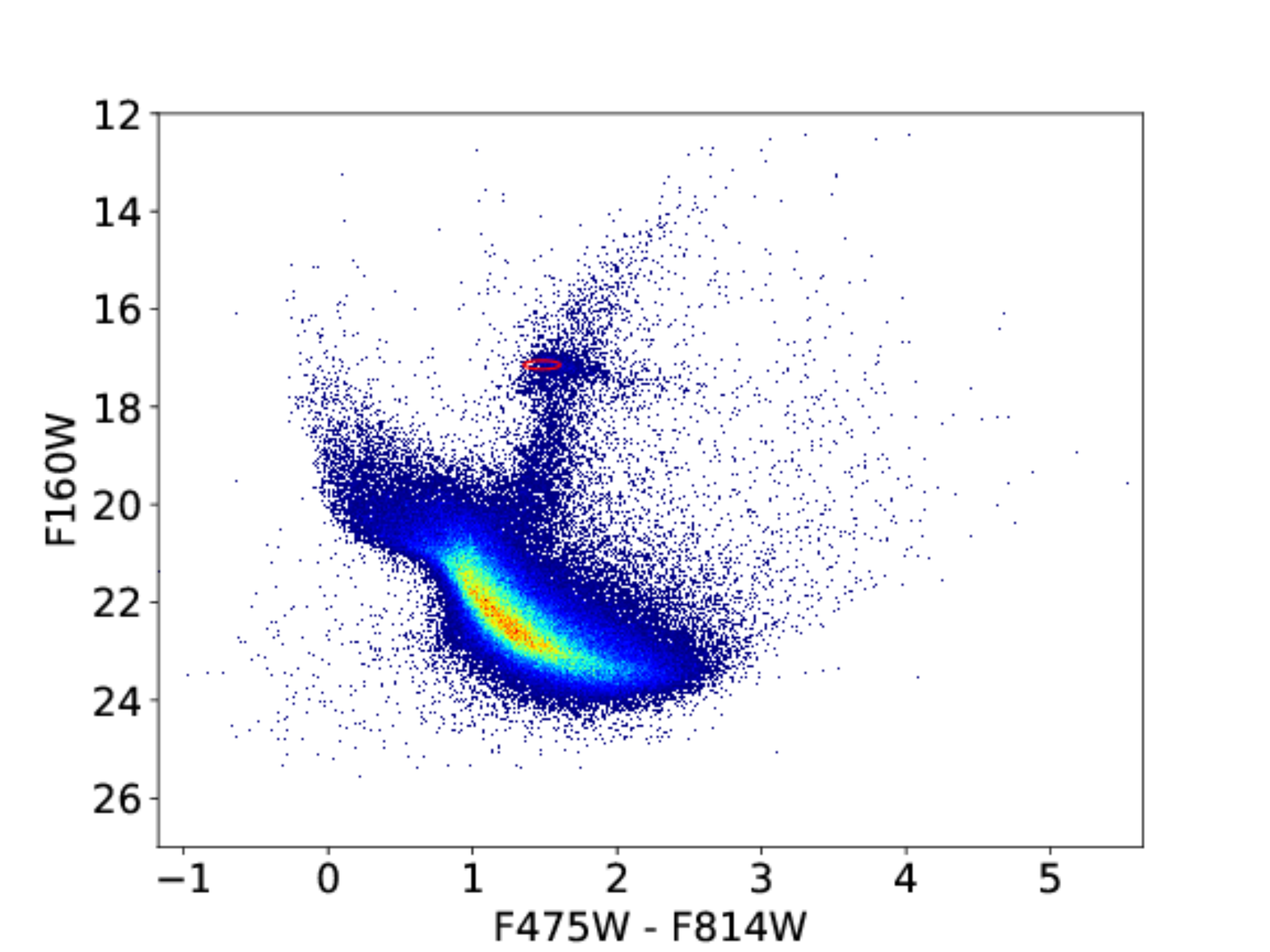}

\caption{Color-magnitude diagrams of sources detected in 3 filters in all the parallel WFC3 fields combined. The point source lists are taken from the "gst" file \citep{williams2014}, which applies a cut on S/N ($>$4). The red ellipse indicates the approximate location of the un-reddened red clump. The cloud of blue and faint magnitudes (e.g., for the F814W vs F475W-F814W CMD, the points with Vega(F814W) $>$ 23, Vega(F475W)-Vega(F814W) $<$ 0.75) stem from contamination by gas emission, most notably in the F475W and F814W filters.}
\label{plot_cmds}
\end{figure*}

\subsection{COS spectroscopy}
\indent The COS observations taken as part of program 14675 (METAL) and listed in Table A2 (see Appendix) were observed between September 2016 and February 2017 using the G130M/1291 and G160M/1589. The METAL COS spectra complemented with archival COS/1327 and G160M/1589 spectra from program 12581. While program 12581 also obtained G160M/1577 and 1600 spectra, the S/N of the G160M/1589 exposures alone matches the S/N of the G160M exposures taken as part of METAL, and therefore, we only use the G160M/1589 observations from GO-12581 for consistency. \\
\indent The COS spectra were retrieved from the MAST Archive and processed with the 3.1.8 version of the COS calibration pipeline, CalCOS. To maximize S/N and spectral coverage, all 4 spectral dithers (FP-POS positions) were used. Since we only observed one central wavelength setting per grating, and only one exposure was taken for each FP-POS setting, no co-addition other than the pipeline-generated {\it x1dsum} combination of the 4 FP-POS was necessary. While wavelength offsets as large as 30 m\mAA ($\sim$10 km/s) are sometimes observed in COS spectra between different cenwave settings, the different FP-POS exposures of a given cenwave obtained in a single visit are observed to be well aligned. Therefore, no specific wavelength alignment between the different FP-POS was performed. In addition, the different COS gratings exhibit different line spread functions. Therefore, spectra taken with different gratings were not co-added. Examples of COS/G130M and G160M FUV spectra are included In Figures \ref{show_spectra} and \ref{show_zooms} (SK-6826 and SK-6619).\\

\subsection{WFC3 imaging reduction and point-source extraction and photometry}\label{photometry_section}

\indent Calibrated images were retrieved from the MAST Archive between September 2016 and July 2017, corresponding to CalWF3 version 3.3 (January 2016) and 3.4.1 (April 2017). \\
\indent The photometry for this project was all performed using the DOLPHOT package \citep[updated HSTphot][]{dolphin2000,dolphin2016}, through the same photometry pipeline used for the PHAT survey, as described in \citet{williams2014}.  There have been a few updates to the photometry routine which we outline here. First, we now use the images that have been corrected for charge transfer efficiency effects ({\tt flc} images) as inputs for our photometry instead of correcting our photometry catalogs for CTE effect after point spread function fitting.  Second, we now use the TinyTim PSF libraries in DOLPHOT instead of the Anderson libraries, as rigorous testing found the TinyTim libraries to provide better spatial uniformity in DOLPHOT. The merged point source lists include magnitude, errors, S/N, roundness and sharpness \citep[see][for details]{williams2014} for all filters observed in a particular field. \\
\indent Using the same package, we also inserted and recovered artificial stars. These artificial star tests (ASTs) are performed by adding a star with known properties to the original images, and then re-running the photometry on the synthetic images to determine if the star is recovered, and how close the measured magnitude is to the true magnitude.  We chose the input photometric properties using the Bayesian Extinction and Stellar Tool \citep[BEAST,][]{gordon2016}  to sample the stellar model space to which our data were sensitive.  By running many thousands of such tests, we can accurately model the photometric bias and error in the model space used by the BEAST for fitting the multi-band spectral energy distributions of the resolved stars. Figure \ref{plot_asts} shows examples of the bias and standard deviation in the recovered photometry compared to the input photometry for the field taken in parallel to the SK-68140 spectroscopy, and all 7 filters. We note that the bias and noise estimated by the ASTs are not used to correct the photometry. Rather, when the stellar SEDs are modeled with the BEAST at a later stage (Hagen et al., in prep, see also Section \ref{ext_map_section}), the bias will be included in the forward-modeling.\\
\indent Three-color images for the WFC3 exposures and the color-magnitude diagrams (CMDs) for the stars detected in 3 filters are shown in Figures \ref{wfc3_3col} and \ref{plot_cmds}, respectively. The total exposure times and depths (Vega mag corresponding to the 50\% completeness limit obtained from the artificial star tests) of the fields observed in each filter are listed in the Appendix (Table A5). The CMDs are contaminated by spurious detections in regions of relatively bright gas emission (e.g., the cloud of blue and faint points with, for the F814W vs F475W-F814W CMD as an example, Vega(F814W) $>$ 23, Vega(F475W)-Vega(F814W) $<$ 0.75). While it is difficult to filter out this contamination based on the photometry (magnitude, error, sharpness or roundness), we expect that fits to reddened stars with the BEAST software will yield poor $\chi^2$ values due to the non-stellar colors of these spurious sources.

\section{Initial Results}\label{results_section}

\indent In this section, we present initial results from METAL that highlight the science objectives of this program.

\subsection{Hydrogen Column Densities}\label{nhi_fitting}

\begin{figure*}
\centering
\includegraphics[width=8.3cm]{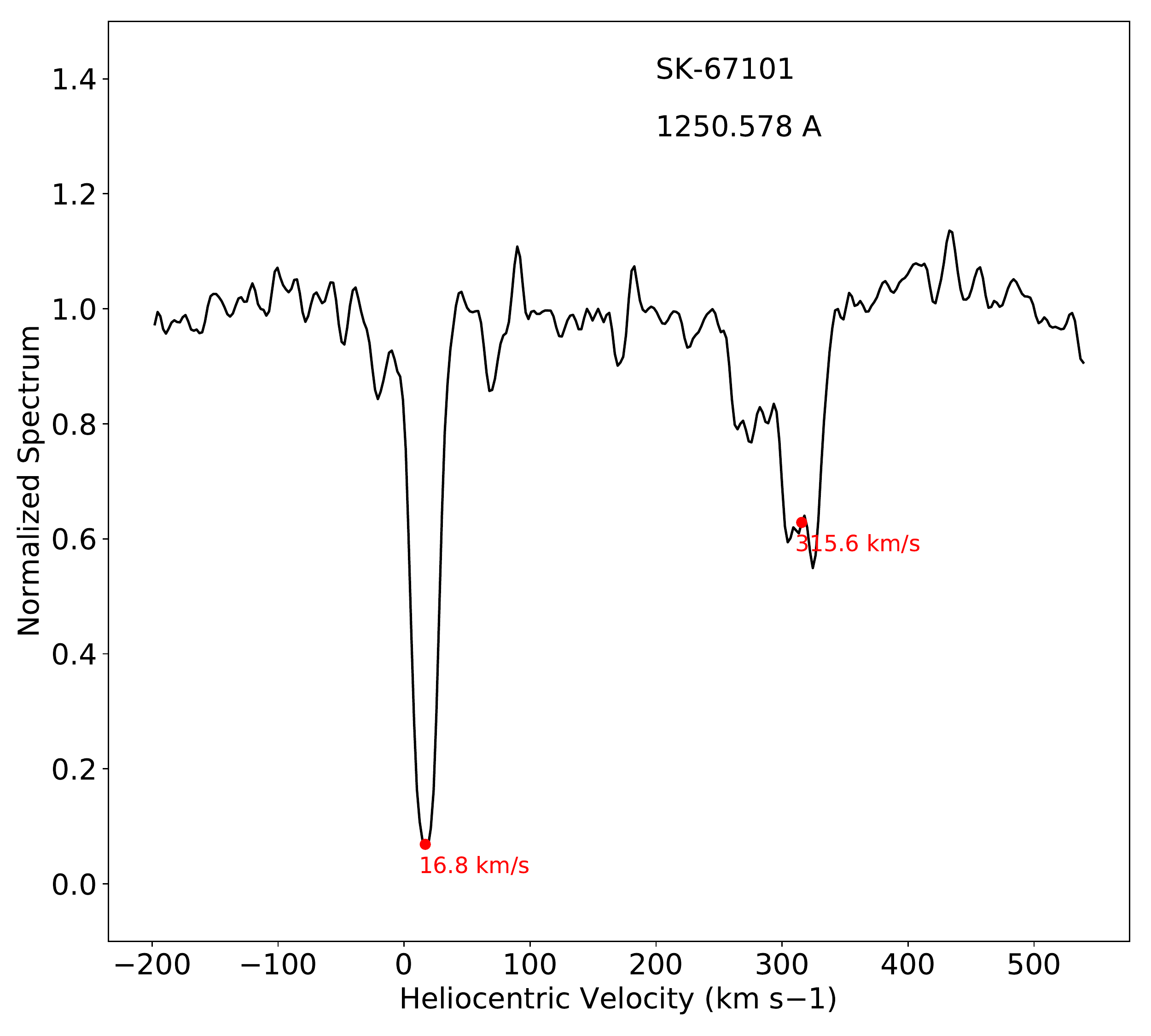}
\includegraphics[width=7.7cm]{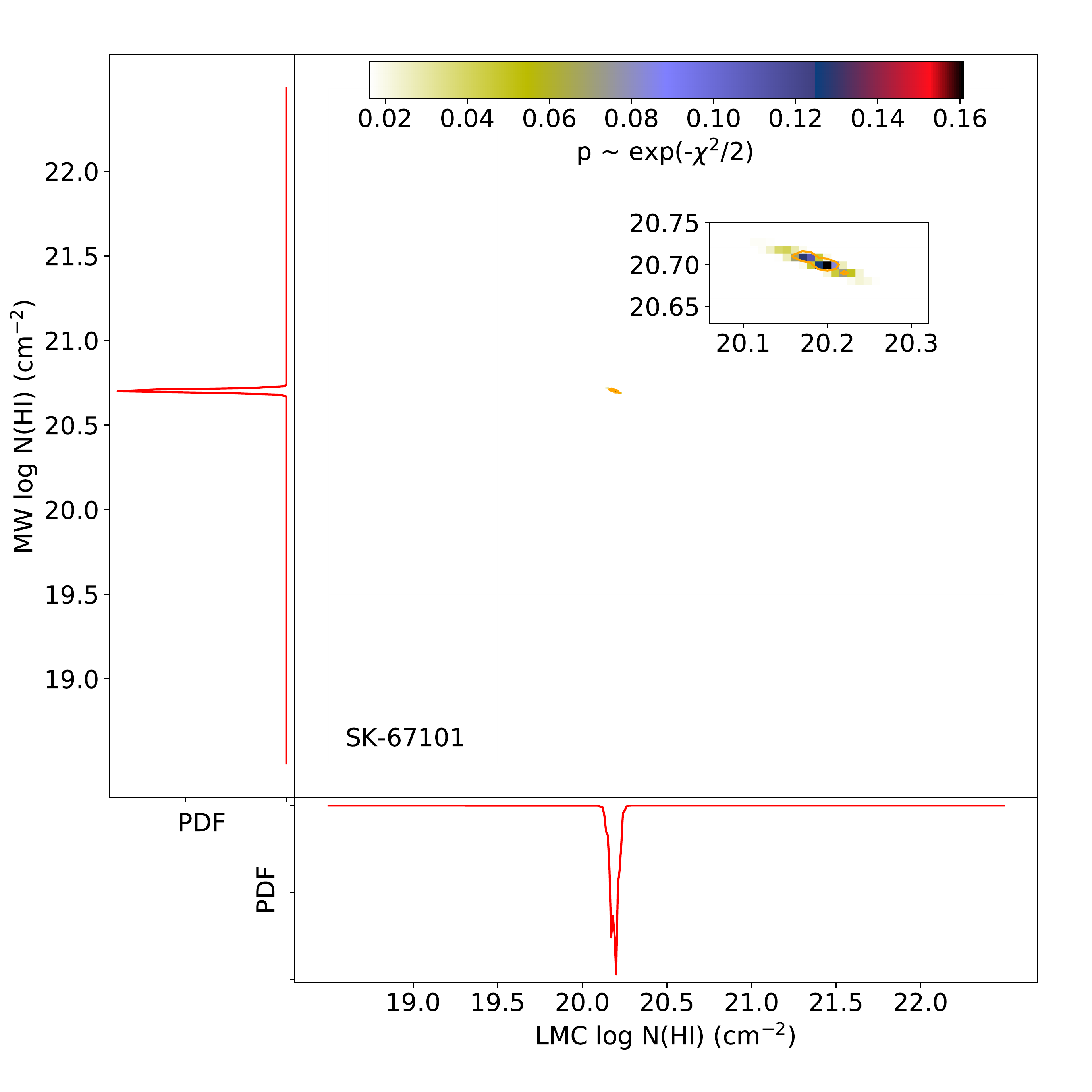}
\includegraphics[width=\textwidth]{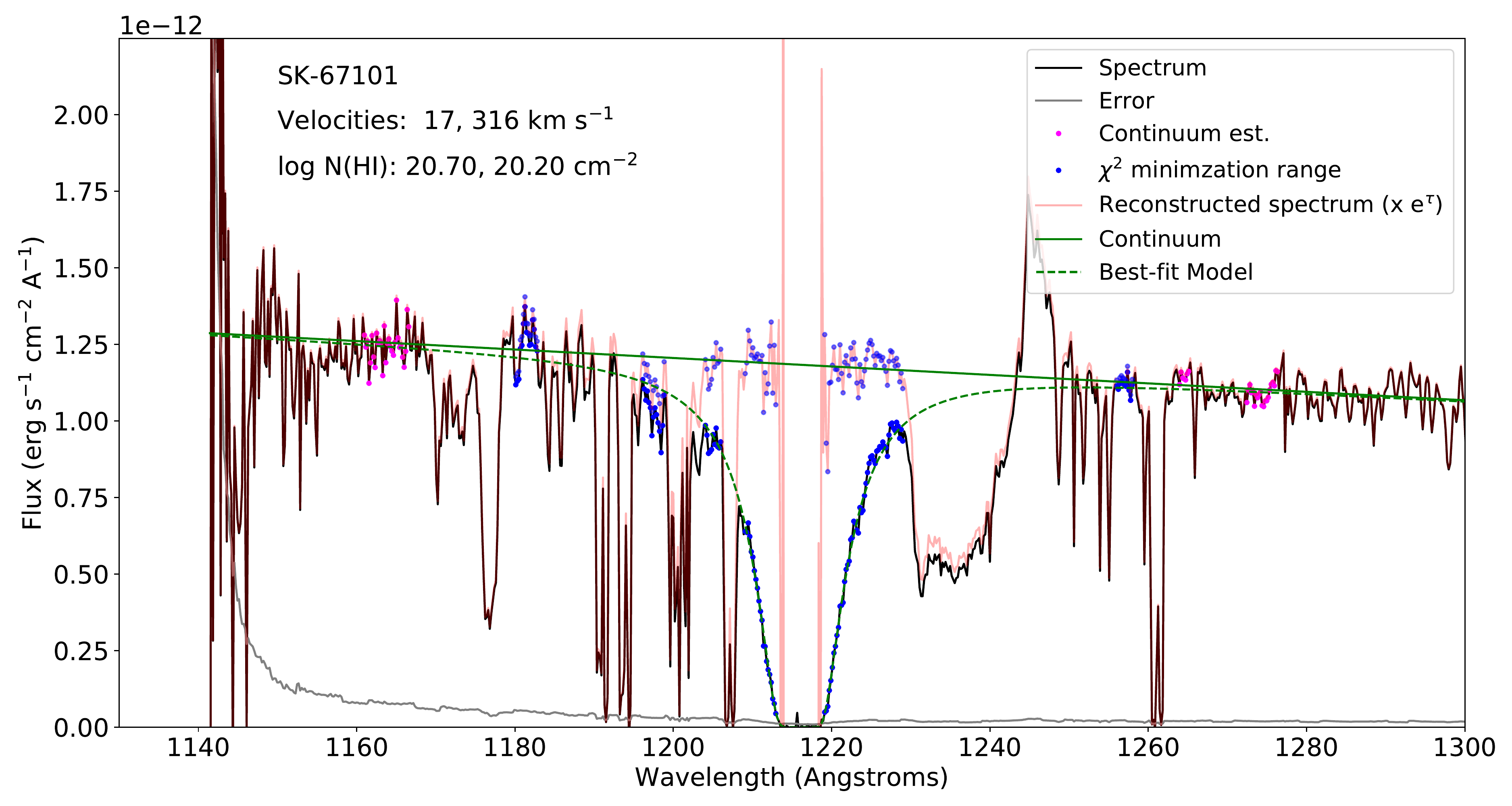}

\caption{Example of MW and LMC \his column density determination (SK-67101). (Top Left) The heliocentric velocities of the main MW and LMC absorption (red dots) are determined from the S II 1250 \mAA absorption line. (Top Right) The 2D joined PDF of the LMC and MW \his column densities is computed, and the best fit pair (N(\hi)$_{\mathrm{MW}}$, N(\hi)$_{\mathrm{LMC}}$) is estimated from the maximum probability (minimum $\chi^2$). Some degeneracy exists between the two components, which is captured by our uncertainties computed from the $\chi^2$ $=$ $\chi^2_{\mathrm{min}}$ $+$ 2.3 contour and the equivalent level ($P$ $=$ 0.32 $\times$ $P_{\mathrm{max}}$) in marginalized PDFs, obtained from taking the maximum probability along each axis. (Bottom) A linear continuum (green solid line) is estimated on either sides of the Lyman-$\alpha$ line in the METAL spectrum (black), using spectral windows pre-defined by eye to be free of stellar or interstellar metal absorption (magenta dots).  The $\chi^2$ between the model (green dashed-line), obtained from damping the continuum with a double Lorentzian profile at the MW and LMC velocities obtained from the S II 1250 \mAA line (top left panel), is estimated in pre-defined windows (blue dots). For this example of SK-67 101, the best-fit estimated from the minimum $\chi^2$ is $\log{}$ (N(\hi)$_{\mathrm{MW}}$, N(\hi)$_{\mathrm{LMC}}$) $=$ (20.70$\pm$0.01, 20.20$\pm$0.03) cm$^{-2}$. The reconstructed spectrum (red) is shown for reference. }
\label{plot_nhi_fit}
\end{figure*}

\begin{figure*}
\centering
\includegraphics[width=6.8cm]{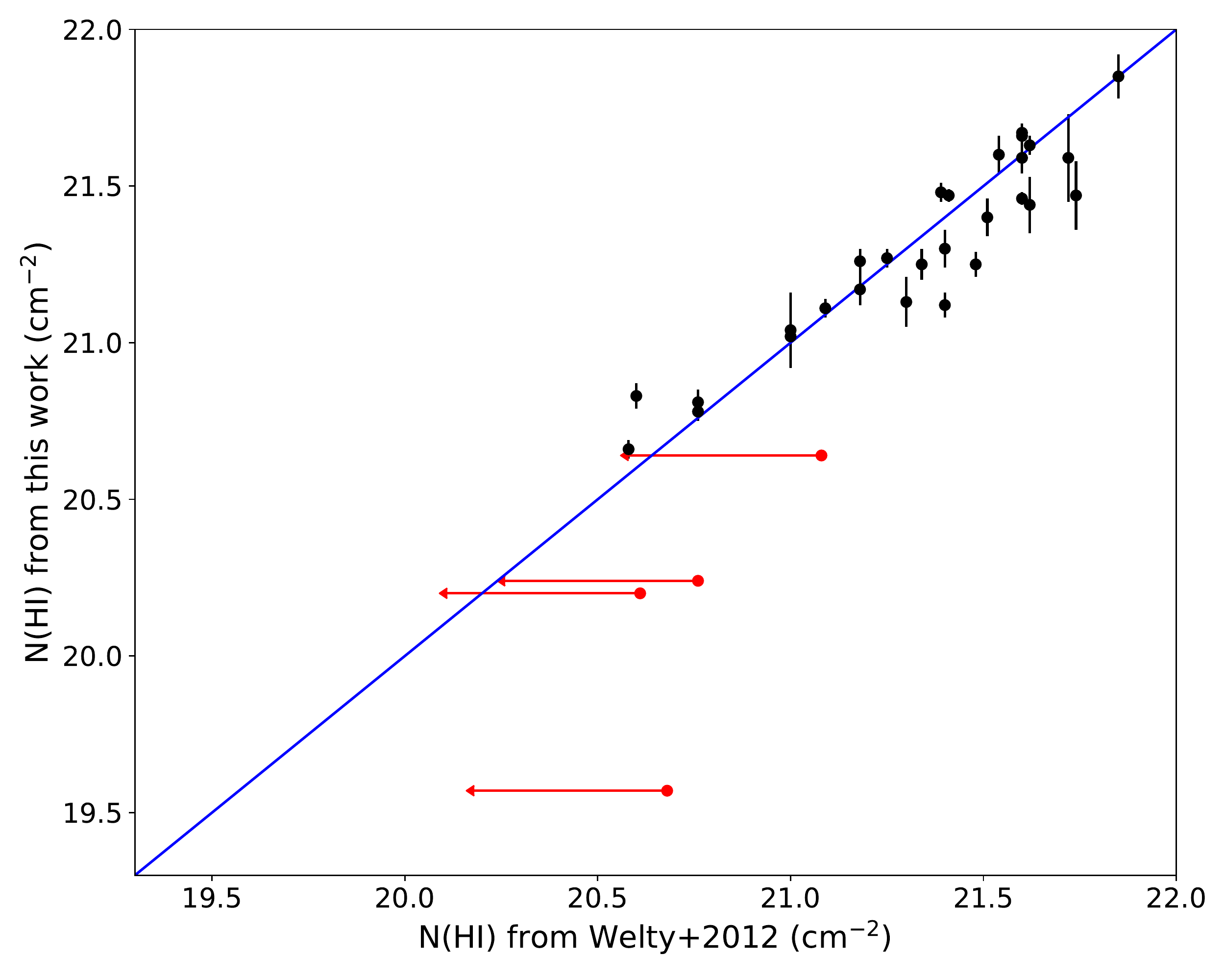}
\includegraphics[width=10.2cm]{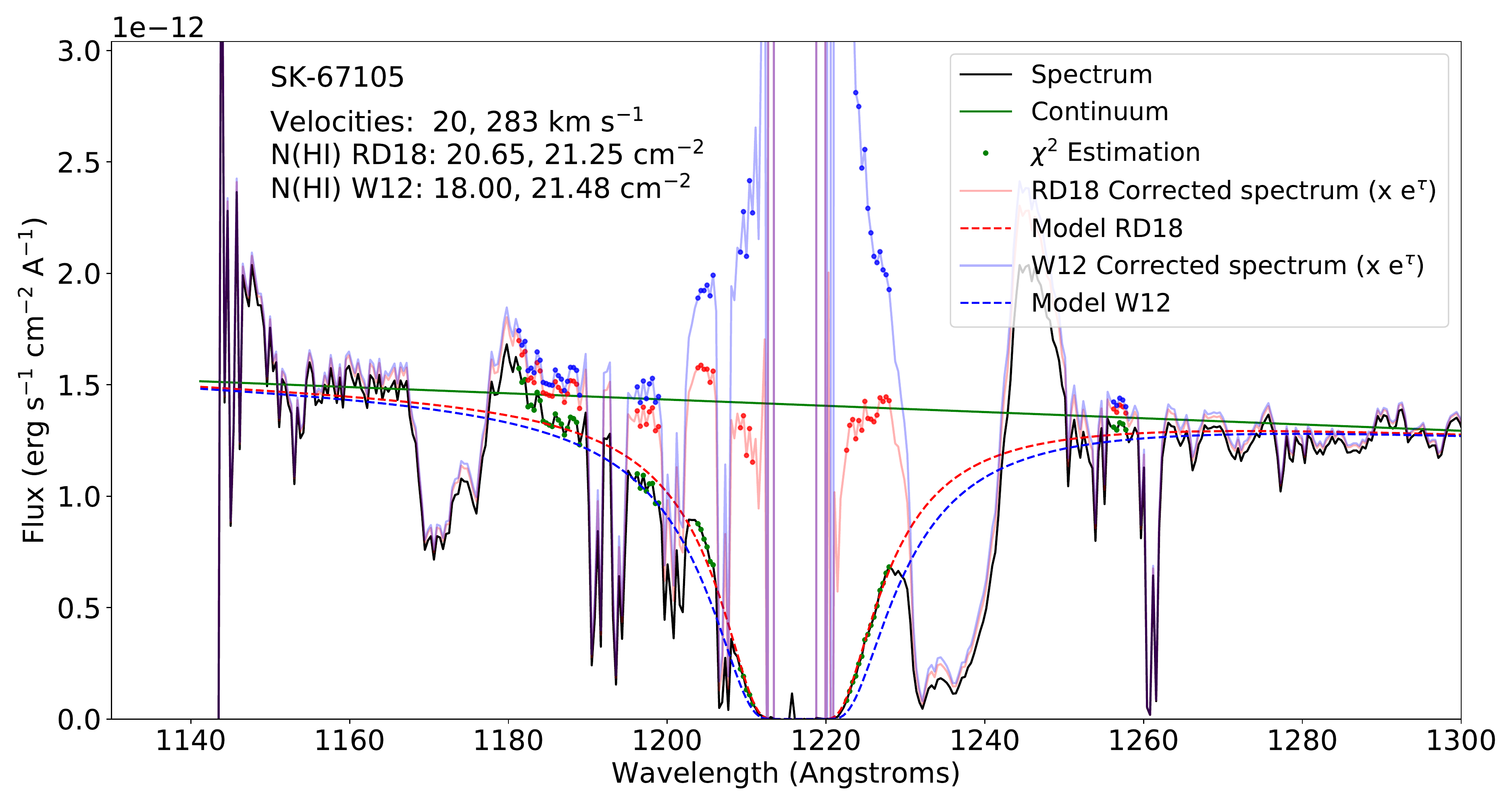}

\caption{(Left) Comparison between LMC \his column densities derived in \citet[][and references therein]{welty2012}, and in this work. (Right) Best-fit models (dashed lines) and reconstructed spectra (solid lines) for SK-67105, for which \citet{welty2012} give $\log{}$ N(\hi) $=$ 21.48 cm$^{-2}$ based on a Lyman-$\beta$ continuum reconstruction with FUSE (blue), while our determination is $\log{}$ N(\hi) $=$ 21.25 $\pm$ 0.04 cm$^{-2}$ based on the Lyman-$\alpha$ line in METAL spectra (red). The dots indicate the spectral range used for the $\chi^2$ estimation. The red wing of the Lyman-$\alpha$ line is incompatible with the larger LMC column density determined from Lyman-$\beta$.  }
\label{plot_comp_jrd_welty}
\end{figure*}

\indent Neutral hydrogen column densities for our sample of targets had previously been derived by \citet{welty2012} using the continuum reconstruction method \citep{bohlin1978, diplas1994} on the Lyman-$\alpha$ and Lyman-$\beta$ lines observed in archival FUV HST (STIS, COS, FOS, GHRS), IUE, and FUSE (Lyman-$\beta$) observations of massive stars in the LMC. The resulting sample of \his column densities has some limitations. First, it is heterogenous since it was based on a variety of literature studies, using a range of instruments, methods, and spectral resolutions (high, medium, and low resolution gratings on HST for instance). Second, a number of the METAL targets only has upper limits on their \his column densities from the \citet{welty2012} sample, Third, \citet{welty2012} did not provide uncertainties for the LMC \his column densities. \\
\indent With our uniform METAL spectroscopic sample of HST STIS E140M and COS G130M spectra, we re-derive \his column densities toward all of our targets along with the associated uncertainties (except for the LBV SK-69 220, for which we use the \his column density estimate from \citet{welty2012}). We use a modified version (detailed in the following) of the continuum reconstruction method outlined in \citet{diplas1994} on the Lyman-$\alpha$ line. The MW and LMC \his column densities can be estimated in a non-degenerate way because of the relatively large velocity offset between the two galaxies, which creates an asymmetry in the Lyman-$\alpha$ absorption profile. \\
\indent In order to determine the \his column densities originating in the MW and LMC toward each METAL sight-line, we first determine the heliocentric velocities of the main MW and LMC absorption using the S II 1250 \mAA line (see an example in the top left panel of Figure \ref{plot_nhi_fit}). Second, we fit the continuum with a linear function (degree 1 polynomial) using pre-defined spectral windows on either side of the Lyman-$\alpha$ line. The spectral windows for fitting the continuum are determined by eye for each target, so as to be free from ISM or stellar metal absorption features. An example is given in the bottom panel of Figure \ref{plot_nhi_fit}. Third, we fit the MW and LMC \his column densities by minimizing the $\chi^2$ between the observed Lyman-$\alpha$ spectrum re-binned by 21 pixels and a model spectrum, $f_m$, obtained by damping the linear continuum, $f_c$, with two Lorentzian profiles at the MW and LMC velocities. The binning of 21 pixels (corresponding to approximately 50 km s$^{-1}$) was chosen so that samples are relatively independent. It was empirically determined that fix pattern noise and weak stellar features are correlated on this scale. The model spectrum is given by $f_m$ $=$ $f_c \exp{\left (- (\tau_{\lambda}(L\alpha)_{\mathrm{MW}}  + \tau_{\lambda}(L\alpha)_{\mathrm{LMC}})\right )}$. $\tau_{\lambda}(L\alpha)_{gal}$ is the optical depth of the Lyman-$\alpha$ line, given by:

\begin{equation}
\tau_{\lambda}(L\alpha)_{gal} = \frac{4.26\times10^{-20} \mathrm{cm}^2 N(\mathrm{H I})_{gal}}{6.04\times10^{-10} + (\lambda - \lambda_{0, gal})^2}
\end{equation}

\noindent where $\lambda_{0, gal}$ $=$ 1215.67 \mAA $\left (1 + v_{0,gal}/c \right)$ and $v_{0, gal}$ are the MW and LMC heliocentric velocities determined from the S II 1250 \mAA absorption lines. We estimate the $\chi^2$ for a 2D grid of models covering $\log$ N(\hi)$_{\mathrm{MW}}$ and $\log$ N(\hi)$_{\mathrm{LMC}}$ in the range 18.5---22.5 cm$^{-2}$. The $\chi^2$ is computed in pre-defined spectral windows in the wings of the Lyman-$\alpha$ line that are free from stellar and interstellar metal absorption. We exclude the very low count regions in the core of the Lyman-$\alpha$ line where errors might not behave in a Gaussian way. In the computation of the $\chi^2$, we include photometric errors on the re-binned spectra, as well as random spectral features (weak stellar lines and fix pattern noise). The latter source of error is estimated from the standard deviation between the linear continuum and the observed re-binned spectrum in the windows used for the continuum estimation. Our resulting best-estimate of (N(\hi)$_{\mathrm{MW}}$, N(\hi)$_{\mathrm{LMC}}$) is the pair that minimizes the 2D $\chi^2$ map. The bottom panel of Figure \ref{plot_nhi_fit} shows an example of \his column density estimation, with the model Lorentzian profile and the corresponding reconstructed spectrum ($f_r$ $=$ $f_{\mathrm{obs}} \exp{(\tau_{\lambda}(L\alpha)_{\mathrm{MW}}  + \tau_{\lambda}(L\alpha)_{\mathrm{LMC}})}$, where $f_r$ is the reconstructed spectrum and f$_{\mathrm{obs}}$ is the observed spectrum), along with the placement of the spectral windows used for the estimation of the continuum and $\chi^2$.  \\
\indent To estimate the uncertainties on N(\hi)$_{\mathrm{LMC}}$ (resp. N(\hi)$_{\mathrm{MW}}$), we compute the joined likelihood $L$ $\sim$ $\exp{(-\chi^2/2)}$ for the (N(\hi)$_{\mathrm{MW}}$, N(\hi)$_{\mathrm{LMC}}$) pair. We derive the contour for which $L = L_{max}\times 0.32$ corresponding to $\chi^2 = \chi^2_{\mathrm{min}} + 2.3$. The resulting area enclosed by this contour corresponds to the 1$\sigma$ uncertainty for the two \his column density parameters \citep{lampton1976}. To transform this contour to an uncertainty on each N(\hi), we marginalize the 2D likelihood by taking the maximum probability along each axis, and compute the N(\hi) value for which the likelihood decreases by a factor 0.32. An example of 1D PDF for N(\hi)$_{\mathrm{LMC}}$ and N(\hi)$_{\mathrm{MW}}$ is shown in the top right panel of Figure \ref{plot_nhi_fit}.  Addtionnally, the $\chi^2$ and joined likelihood can only be estimated in spectral windows free of metal lines. Therefore, it underestimates the true uncertainty, particularly for some targets with high column densities where the bottom of the Lyman-$\alpha$ wings is not well constrained due to the presence of numerous metal lines. To account for the sparseness of the window placement and obtain a more realistic estimate of the uncertainties, we add in quadrature to the $\chi^2$ based uncertainty a systematic uncertainty of 0.02 dex or 0.05 dex depending on the spectral coverage of the windows used for $\chi^2$ estimation. The targets with a 0.05 dex systematic uncertainty on N(\hi)$_{\mathrm{LMC}}$ are SK-68 129, SK-68 140, SK-68 26, Sk-66 19, SK-68 52, SK-68 155, SK-69 279, SK-67 2. Those values were obtained from visually examining reconstructed spectra using difference spectral windows. The LMC and MW \his column densities and their uncertainties are reported in Table 2. \\
\indent To validate the method presented here, co-author Jenkins independently derived \his column densities using the same METAL data, but a slightly different method, selecting the \his column densities for the MW and LMC that led to the flattest reconstructed spectrum. The resulting column densities are within our estimated errors, indicating that both our revised \his column densities and their uncertainties are reasonably estimated. Additionally, Figure \ref{plot_comp_jrd_welty} compares the \his column density determination from this work and \citet{welty2012}. For the most part, differences are within uncertainties. Our new N(\hi) determinations supersede upper limits from \citet{welty2012} for SK-69 104, SK-67 101, SK-69 175, and SK-67 14. A few cases stand out as presenting significant differences: SK-68 135 (21.46$\pm$0.02 vs 21.6 cm$^{-2}$, 7$\sigma$, based on Lyman-$\beta$); BI 184 (21.12$\pm$0.04 vs 21.4 cm$^{-2}$, 7$\sigma$, based on Lyman-$\beta$ in FUSE);  SK-67 105 (21.25$\pm$0.04 vs 21.48 cm$^{-2}$, 7$\sigma$, based on Lyman-$\beta$ in FUSE).\\
\indent In the case of SK-68 135, SK-67 105 and BI 184, our determination of N(\hi) was based on Lyman-$\alpha$ while the determination from \citet{welty2012} was based on Lyman-$\beta$ since only FUSE spectra were available at the time of publication. In the bottom panel of Figure \ref{plot_comp_jrd_welty}, we compare the reconstructed spectra of SK-67 105, from our \his column density determination based on Lyman-$\alpha$ ($\log{}$ N(\hi) $=$ 21.25 cm$^{-2}$) and the determination from \citet{welty2012} based on Lyman-$\beta$  ($\log{}$ N(\hi) $=$ 21.48 cm$^{-2}$). For the latter, we do not know their estimate of the MW column density, so we assume a conservatively low value of $\log{}$ N(\hi) $=$ 18 cm$^{-2}$. Even with the lowest possible MW contribution, the red wing of the Lyman-$\alpha$, which includes the largest contribution from the LMC absorption, is incompatible with the amount of hydrogen in the LMC derived from Lyman-$\beta$ for this target. As acknowledged by \citet{welty2012}, this example highlights the difficulty and large uncertainties associated with determining \his column densities from the Lyman-$\beta$ reconstruction technique, due in part to the large amount of H$_2$ lines present in the Lyman-$\beta$ trough. For such cases, the use of Lyman-$\alpha$ from METAL medium-resolution spectra should therefore lead to a substantial improvement in accuracy.\\

\subsection{Silicon Depletions}\label{depletion_section}

\begin{deluxetable*}{cccccc}
\centering
\tabletypesize{\scriptsize}
\tablecolumns{6}
\tablewidth{\textwidth}
\tablecaption{Silicon gas-phase column densities and depletions from METAL}
\tablenum{4}
 
 \tablehead{Target & N(H)\tablenotemark{a}& Velocity range (heliocentric) &  $W_{\lambda}$ & $log$ N(Si II) & $\delta$(Si)\tablenotemark{b}}\\
 \startdata
 &     cm$^{-2}$ & km s$^{-1}$ & m\AA& cm$^{-2}$ & \\
\hline
&&&&&\\
SK-67 2 & 21.46 $\pm$ 0.12 & 220---310 & 148.0 $\pm$ 17.8 & 15.63 $\pm$ 0.13 & -1.18 $\pm$ 0.18 \\
SK-67 5 (E230H) & 21.04 $\pm$ 0.04 & 240---320 &  156.5$\pm$6.2 & 15.79$\pm$0.03 & ...\\
SK-67 5 (E230M)  & 21.04 $\pm$ 0.04 &  240---320 & 161.2$\pm$12.7 & 15.64$\pm$0.03 & ...\\
SK-67 5 (Adopted)  & 21.04 $\pm$ 0.04 & .. & ... & 15.79 $\pm$ 0.03 & -0.60 $\pm$ 0.06 \\
SK-69 279 & 21.63 $\pm$ 0.05 & 210---335 & 319.8 $\pm$ 24.8 & 15.96 $\pm$ 0.13 & -1.02 $\pm$ 0.14 \\
SK-67 14 & 20.24 $\pm$ 0.06 & 230---335 & 83.9 $\pm$ 17.7 & 15.23 $\pm$ 0.1 & -0.36 $\pm$ 0.12 \\
SK-66 19 & 21.87 $\pm$ 0.07 & 225---330 & 254.8 $\pm$ 36.7 & 15.91 $\pm$ 0.18 & -1.31 $\pm$ 0.19 \\
PGMW 3120 & 21.48 $\pm$ 0.03 & 220---330 & 254.3 $\pm$ 21.9 & 15.91 $\pm$ 0.21 & -0.92 $\pm$ 0.21 \\
PGMW 3223 & 21.40 $\pm$ 0.06 & 220---330 & 217.7 $\pm$ 20.6 & 15.91 $\pm$ 0.165 & -0.84$\pm$ 0.18 \\
SK-66 35 & 20.85 $\pm$ 0.04 & 235---320 & 209.0 $\pm$ 12.0 & 15.75 $\pm$ 0.125 & -0.45 $\pm$ 0.13 \\
SK-65 22 & 20.66 $\pm$ 0.03 & 240---340 & 146.8 $\pm$ 9.1 & 15.58 $\pm$ 0.125 & -0.43 $\pm$ 0.13 \\
SK-68 26 & 21.65 $\pm$ 0.06 & 225---320 & 190.2 $\pm$ 25.4 & 15.73 $\pm$ 0.135 & -1.27 $\pm$ 0.15 \\
SK-70 79 & 21.34 $\pm$ 0.04 & 180---270 & 207.7 $\pm$ 18.2 & 15.75 $\pm$ 0.13 & -0.94 $\pm$ 0.14 \\
SK-68 52 & 21.31 $\pm$ 0.06 & 195---350 & 239.6 $\pm$ 26.9 & 15.76 $\pm$ 0.13 & -0.90 $\pm$ 0.14 \\
SK-69 104 & 19.57 $\pm$ 0.68 & 200---290 & 63.9 $\pm$ 16.0 & 15.07 $\pm$ 0.125 & 0.15 $\pm$ 0.69 \\
SK-68 73 & 21.68 $\pm$ 0.02 & 215---330 & 234.5 $\pm$ 12.9 & $>$ 15.92 & $>$ -1.11  \\
SK-67 101 & 20.2 $\pm$ 0.04 & 230---360 & 110.2 $\pm$ 19.8 & 15.33 $\pm$ 0.09 & -0.22 $\pm$ 0.10 \\
SK-67 105 & 21.26 $\pm$ 0.04 & 235---350 & 98.6 $\pm$ 9.7 & 15.36 $\pm$ 0.06 & -1.25 $\pm$ 0.07 \\
BI 173 & 21.25 $\pm$ 0.05 & 195---280 & 236.6 $\pm$ 11.7 & 15.78 $\pm$ 0.05 & -0.82 $\pm$ 0.07 \\
BI 184 & 21.15 $\pm$ 0.04 & 200---330 & 250.0 $\pm$ 24.7 & 15.8 $\pm$ 0.11 & -0.70 $\pm$ 0.12\\
SK-71 45 & 21.11 $\pm$ 0.03 & 200---345 & 373.9 $\pm$ 10.3 & 16.01 $\pm$ 0.10 & -0.45 $\pm$ 0.10 \\
SK-69 175 & 20.64 $\pm$ 0.03 & 190---370 & 126.6 $\pm$ 57.5 & 15.43 $\pm$ 0.19 & -0.56 $\pm$ 0.19 \\
SK-67 191 & 20.78 $\pm$ 0.03 & 235---340 & 179.0 $\pm$ 11.0 & 15.61 $\pm$ 0.05 & -0.52 $\pm$ 0.06 \\
SK-67 211 & 20.81 $\pm$ 0.04 & 240---355 & 209.8 $\pm$ 9.2 & 15.74 $\pm$ 0.12 & -0.42 $\pm$ 0.13 \\
BI 237 & 21.65 $\pm$ 0.03 & 220---350 & 269.6 $\pm$ 21.0 & 15.91 $\pm$ 0.14 & -1.09 $\pm$ 0.14 \\
SK-68 129 & 21.62 $\pm$ 0.14 & 230---330 & 265.0 $\pm$ 16.8 & $>$15.98  & $>$ -1.00  \\
SK-66 172 & 21.27 $\pm$ 0.03 & 240---320 & 188.4 $\pm$ 11.7 & 15.71 $\pm$ 0.13 & -0.91 $\pm$ 0.13 \\
BI 253 & 21.68 $\pm$ 0.03 & 200---320 & 327.8 $\pm$ 20.6 & $>$16.15 & $>$-0.88  \\
SK-68 135 & 21.48$\pm$ 0.02 & 225---340 & 273.6 $\pm$ 27.3 & $>$ 16.1 & $>$ -0.73  \\
SK-69 246 & 21.48 $\pm$ 0.02 & 215---320 & 326.7 $\pm$ 7.0 & 16.03 $\pm$ 0.145 & -0.80 $\pm$ 0.15 \\
SK-68 140 & 21.51 $\pm$ 0.11 & 220---340 & 345.8 $\pm$ 35.7 & 16.07 $\pm$ 0.185 & -0.79 $\pm$ 0.22 \\
SK-71 50 & 21.24 $\pm$ 0.05 & 190---320 & 319.7 $\pm$ 15.4 & 15.94 $\pm$ 0.105 & -0.65 $\pm$ 0.12 \\
SK-68 155 & 21.47 $\pm$ 0.09 & 220---360 & 387.1 $\pm$ 22.8 & 16.09 $\pm$ 0.13 & -0.73 $\pm$ 0.16 \\
SK-70 115 (E230H)&21.18 $\pm$ 0.08 & 195---360 & 322.4$\pm$11.9 & 15.96$\pm$0.02 & ... \\
SK-70 115 (E230M) & 21.18 $\pm$ 0.08 &  195---360  & 326.0$\pm$17.1 & 15.90$\pm$0.12 & ...\\
SK-70 115 (Adopted) & 21.18 $\pm$ 0.08 & ... & ... &  15.96 $\pm$ 0.02 & $-$0.57 $\pm$ 0.09\\

\enddata
\tablecomments{For SK-67 5 and SK-70 115, Si II column densities were also derived from archival E230H/1913 spectra (program GO-9757). The adopted column density is from the high-resolution measurement.}
\tablenotetext{a}{N(H) $=$ N(\hi) + 2N(H$_2$), where N(\hi) is derived from the METAL spectra in Section 4.1 and N(H$_2$) is from \citet{welty2012}}
\tablenotetext{b}{Depletions are derived assuming a reference abundance of Silicon of 12 + log(Si/H) $=$ 7.35. The random error on depletions combines in quadrature the error on Si and H column densities.}
\end{deluxetable*}

\indent A primary goal of METAL was to perform a detailed and complete census of metals in the gas and dust phases for all targets. The quality of the spectra allows us to achieve this for Fe, Mg, Si, Ni, Cu, Cr, Zn, S, and all targets except SK-69 220. SK-69 220 is an LBV and has a very complex stellar spectrum, making accurate measurements of interstellar features impossible.  The full sample of depletions and the description of the methodology and results will be published in Roman-Duval, in prep (Paper II). In this overview of the METAL program, we derive gas-phase abundances of silicon using the 1808 \mAA line.  \\
\indent The upper portions of the velocity ranges over which we perform absorption line measurements are well defined by the sharp, long wavelength edges of strong transitions, such as the Fe II 2344.2 \mAA line (see Figure \ref{show_zooms}). However, at low velocities, the boundary for gas associated with the LMC is poorly defined, owing to the presence of additional features that are at velocities between those associated local Milky Way and those arising clearly from within the LMC (see Section \ref{ivg_section}). Observations of \his 21 cm emission in the GASS III survey \citep{mccluregriffiths2009, kalberla2010, kalberla2015} in the directions of our target stars reveal a well defined feature arising from gas in the LMC, and only rarely is there any evidence for components at intermediate velocities (and when they do appear, they have very low amplitudes). For this reason, we define the lower limit for velocity at the edge of the \his 21 cm emission, where the brightness temperature rises above 0.2 K ($\sim$4$\sigma$). This usually corresponds to a heliocentric radial velocity of about 200 km s$^{-1}$. The velocity ranges used for deriving the Si II column densities are listed in Table 4 (as heliocentric velocities) and examples are shown in Figure \ref{show_zooms}.\\
\indent We define continuum levels from best-fitting Legendre polynomials to fluxes on either side of the absorption profiles. We determine the Si II column densities by integrating the apparent optical depth \citep[AOD, ][]{savage1991, jenkins1996} over the velocity intervals defined in Table 4, assuming an oscillator strength of $\log{\lambda f_{\lambda}}$ = 0.575 \mAA \citep{morton2003}. Some lines appeared to be saturated; in such cases the AOD method can underestimate the true column density. When saturation appeared to be evident, we either raised the upper uncertainty bound or declared the measurement simply as a lower limit, depending on our perceived severity of the saturation.\\
\indent We estimate errors for the column densities from the effects of three different sources: (1) noise in the absorption profile, (2) errors in defining the continuum level, and (3) uncertainties in the transition f-values, all of which were combined in quadrature. We evaluate the expected deviations produced by such errors by remeasuring the AODs at the lower and upper bounds for the continua, which are derived from the expected formal uncertainties in the polynomial coefficients of the fits as described by \citet{sembach1992}. We multiply these coefficient uncertainties by 2 in order to make approximate allowances for additional deviations that might arise from some freedom in assigning the most appropriate order for the polynomial.\\
\indent Ionization corrections for Si II are negligible in the column density range of our sight-lines \citep{tchernyshyov2015, jenkins2017}, and we therefore assume that all the silicon is in the form of Si II. Silicon gas-phase abundances are derived by taking the ratio of the measured Si II column densities to the total hydrogen column density, N(H) = N(\hi) + 2N(H$_2$), where N(H) is listed in Table 4. The \his column densities are determined from the METAL spectra in Section \ref{nhi_fitting}, while the H$_2$ column densities are from \citet{welty2012}. The silicon depletion (fraction of Si in the gas-phase) is then calculated assuming that the total (gas and dust) Si ISM abundance is equal to the photospheric abundance of Si in young stars, $12 + log(Si/H)$ $=$ 7.35 \citep[][and references therein]{tchernyshyov2015}. 

\begin{equation}
\delta(\mathrm{Si}) = \log{ \left ( \frac{\mathrm{N(Si)}}{\mathrm{N(H)}}\right)} - \log{\left (  \frac{\mathrm{Si}}{\mathrm{H}}\right )}_{\mathrm{LMC,tot}}
\end{equation}

\begin{figure*}
\centering
\includegraphics[width=12cm]{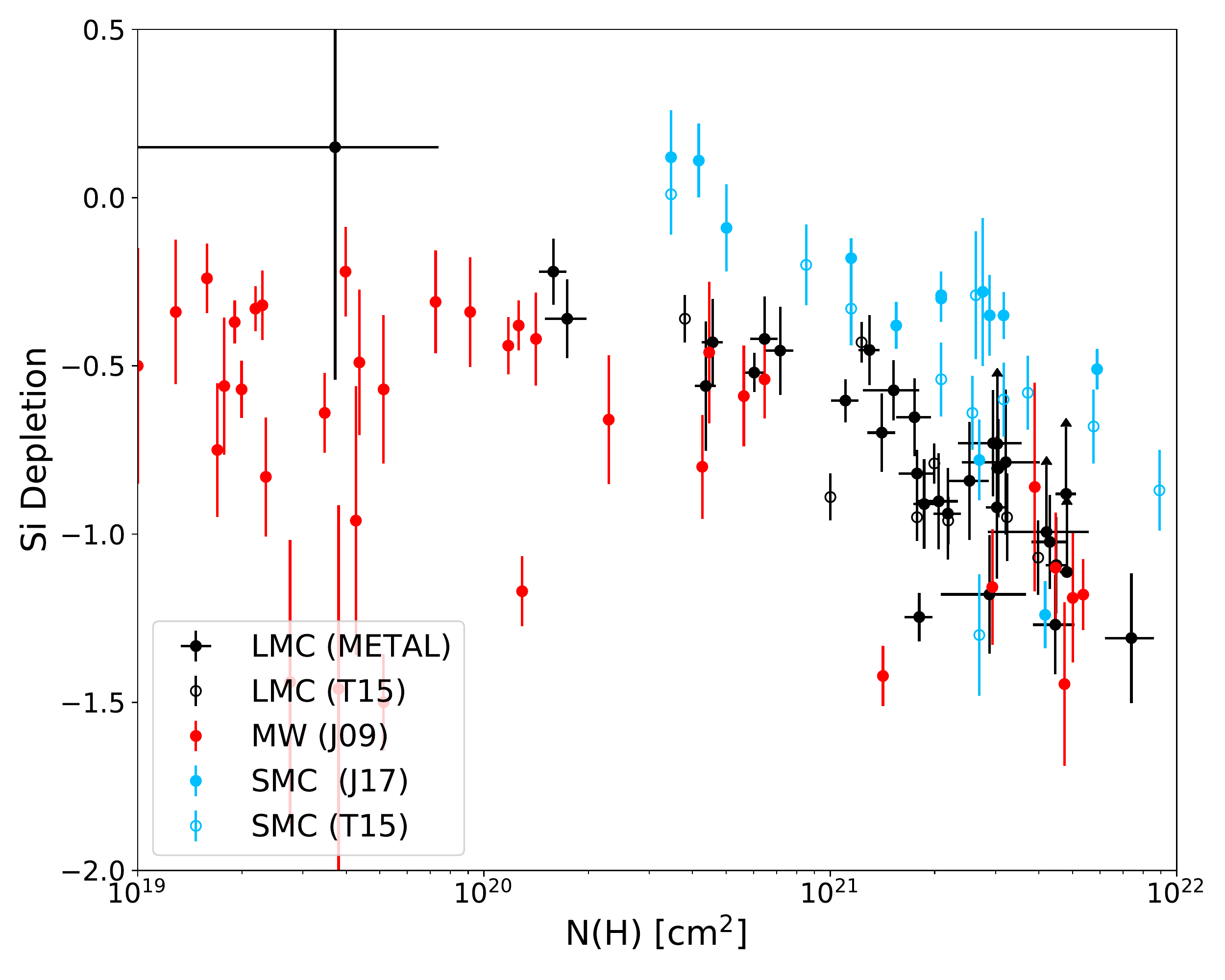}
\includegraphics[width=12cm]{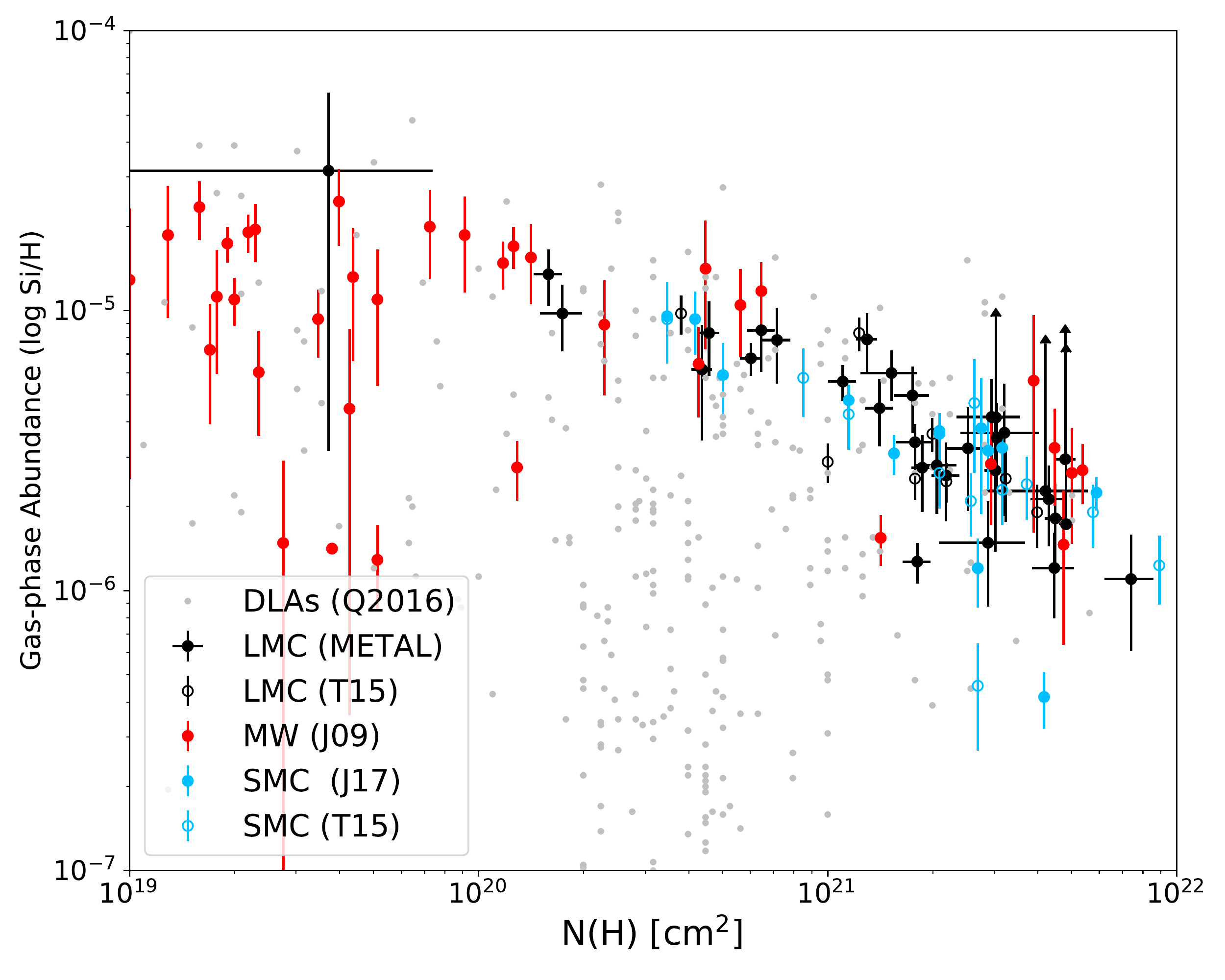}
\caption{Logarithm of the fraction of Silicon in the gas-phase (i.e., depletion of Si, top) and gas-phase abundance of Si (bottom) as a function of total hydrogen column density $N($H$) = N($\hi$) + 2 N($H$_2)$ in the LMC (black), MW (red), and SMC (blue). The LMC measurements include METAL (filled circles) and \citet{tchernyshyov2015} (open circles). The SMC measurements combine results by \citet{tchernyshyov2015} and \citet{jenkins2017}. in the bottom panel, gray points represent silicon gas-phase abundances for a large sample of DLAs and sub-DLAs from \citet{quiret2016}. }
\label{plot_si_dep}
\end{figure*}

\indent The errors on the depletions are obtained from summing the errors on the logarithms of the silicon and hydrogen column densities in quadrature. The Si 1808 \mAA equivalent widths, column densities, depletions, and their errors are listed in Table 4 for all targets (except SK-69 220, which is not usable due to its complex stellar spectrum). We note that SK-67 5 and SK-70 115 have archival high-resolution E230H spectra. These can use to benchmark the E230M-based column density estimates, which may underestimate column densities slightly due to the combined effects of lower spectral resolution and saturation in individual components. N(Si) determinations are within errors between the two gratings for SK-70 115 (0.5$\sigma$). For SK-67 5 however, the high resolution measurement is 5$\sigma$ higher than the column density measured from the medium-resolution spectrum. This is likely due to unresolved saturation, and were therefore adopt the high-resolution measurement toward both stars. \\
 \indent Figure \ref{plot_si_dep} compares the silicon depletions in METAL (LMC) with silicon depletions in other similar studies performed in the Milky Way \citep{jenkins2009}, LMC \citep{tchernyshyov2015}, and SMC \citep{tchernyshyov2015, jenkins2017}, as a function of total hydrogen column density ($N($H$)=N($\hi$)+2N($H$_2)$). In all three galaxies, the fraction of silicon in the gas-phase decreases with increasing hydrogen column density, but the trends are different between the MW, LMC, and SMC. The fraction of silicon in the gas-phase clearly increases from the MW, to the LMC, to the SMC: it is about 0.4 dex higher in the SMC than in the LMC, and the fraction of Si in the gas in the LMC is about 0.2 dex higher than in the MW. \\
 \indent The fraction of silicon in the dust (or gas) phase in the LMC and SMC appears to scale with inverse metallicity. As a result, the gas-phase abundances of silicon in the MW, LMC, and SMC are similar and follow similar trends with total hydrogen column density (Figure \ref{plot_si_dep}). The bottom panel of Figure \ref{plot_si_dep} also shows that the silicon gas-phase abundances in a large sample of DLAs and sub-DLAs from \citet{quiret2016} follow very similar trends. These trends will be examined for other elements in Paper II.
\indent If other elements follow the same trend (which will be determined in Paper II), then the dust-to-metal ratio (D/M) in the SMC would be four times lower than in the MW, and D/M would be 1.5 times lower in the LMC than in the MW.  The variation of D/M with metallicity implies that the gas-to-dust ratio (G/D = 1/(metallicity $\times$ D/M)) should vary non-linearly with metallicity (as Z$^{-2}$): G/D would be a factor 4 (resp. 1.5) higher in the SMC (resp. LMC) compared to a linear scaling with metallicity, which is consistent with previous studies based on FIR, \hi, and CO emission \citep{remyruyer2014, RD2014, RD2017} as well as recent chemical evolution models \citep{feldmann2015}. The \citet{asano2013} and \citet{feldmann2015} models propose as an explanation for these observations that above a certain critical metallicity ($Z_{\mathrm{crit}}$ $=$ 0.1---0.2 $Z_{\odot}$ for typical star-formation timescales of 0.5---5Gyr), dust formation in the ISM is efficient and can balance the effects of dilution by galactic inflows and outflows and destruction by shocks. Below this critical metallicity, the dust input is dominated by the (non-efficient) formation process in evolved stars (mainly AGB and SN), resulting in a low dust-to-metal ratio and thus a low dust abundance.\\

\subsection{Intermediate and high velocity gas}\label{ivg_section}

\begin{deluxetable*}{ccccccc}
\centering
\tabletypesize{\scriptsize}
\tablecolumns{7}
\tablewidth{\textwidth}
\tablecaption{LSR Velocities of the IVC and HVC}
\tablenum{5}
 
 \tablehead{Target & R.A. & DEC & $v_{\mathrm{LSR}}$(MW) & $v_{\mathrm{LSR}}$(IVC) & $v_{\mathrm{LSR}}$(HVC) & $v_{\mathrm{LSR}}$(LMC)}\\ 
 \startdata
 &   Deg & Deg &  km s$^{-1}$ & km s$^{-1}$ & km s$^{-1}$& km s$^{-1}$ \\
\hline
&&&&&&\\

SK-67 2 & 71.76854706 & -67.11475372 & 0 & 50 & 184 & 260  \\
SK-67 5 & 72.5788269 & -67.6605835 & 3 & 70 & 135 & 275  \\
SK-69 279 & 73.55940247 & -69.25370789 & 5 & 60 & 115 & 260  \\
SK-67 14 & 73.63287354 & -67.25682831 & 5 & 92 & 153 & 262  \\
SK-66 19 & 73.97479248 & -66.41648865 & 0 & 80 & 140 & 275  \\
PGMW 3120 & 74.1950531 & -66.41297913 & 5 & 65 & 150 & 265 \\
PGMW 3223 & 74.25357819 & -66.40697479 & 0 & 75 & 145 & 270  \\
SK-66 35 & 74.26850128 & -66.5773468 & 3 & 95 & 140 & 260  \\
SK-65 22 & 75.34612274 & -65.87594604 & 5 & 80 & 140 & 275  \\
SK-68 26 & 75.3843689 & -68.17858887 & 3 & 80 & 155 & 252  \\
SK-70 79 & 76.65525818 & -70.49004364 & 1 & 60 & 90 & 220 \\
SK-68 52 & 76.83509827 & -68.53572083 & 0 & 90 & 160 & 230  \\
SK-69 104 & 79.74791718 & -69.21522522 & 3 & 50 & 135 & 232  \\
SK-68 73 & 80.74908447 & -68.02961731 & 0 & 50 & 128 & 280  \\
SK-67 101 & 81.48425293 & -67.50796509 & 3 & 55 & 108 & 305  \\
SK-67 105 & 81.52580261 & -67.18244171 & 5 & 55 & 132 & 290  \\
BI 173 & 81.79141998 & -69.13234711 & 3 & 56 & 112 & 235 \\
BI 184 & 82.62773895 & -71.04211426 & 5 & 75 & 115 & 250  \\
SK-71 45 & 82.81522369 & -71.06935883 & 5 & 63 & 113 & 235  \\
SK-69 175 & 82.85633087 & -69.09405518 & 3 & 52 & 117 & 270  \\
SK-67 191 & 83.39178467 & -67.50547791 & 3 & 50 & 140 & 280  \\
SK-67 211 & 83.80793762 & -67.55764008 & 3 & 50 & 110 & 280  \\
BI 237 & 84.06095123 & -67.65532684 & 4 & 50 & 97 & 275  \\
SK-68 129 & 84.11153412 & -68.9588623 & 3 & 55 & 110 & 260  \\
SK-66 172 & 84.2724762 & -66.35977173 & 3 & 70 & 105 & 275  \\
BI 253 & 84.39358521 & -69.01950073 & 2 & 62 & 170 & 264 \\
SK-68 135 & 84.45463562 & -68.91713715 & 5 & 55 & 153 & 255  \\
SK-69 246 & 84.722435 & -69.03359222 & 3 & 55 & 110 & 265  \\
SK-68 140 & 84.73825073 & -68.94808197 & 5 & 55 & 168 & 270  \\
SK-71 50 & 85.17996979 & -71.48351288 & 5 & 91 & 145 & 250  \\
SK-68 155 & 85.72887421 & -68.94847107 & 5 & 55 & 93 & 280  \\
SK-70 115 & 87.20689392 & -70.06606293 & 3 & 50 & 0 & 290  \\
\enddata
\end{deluxetable*}

\begin{figure*}
\centering
\includegraphics[width=8cm]{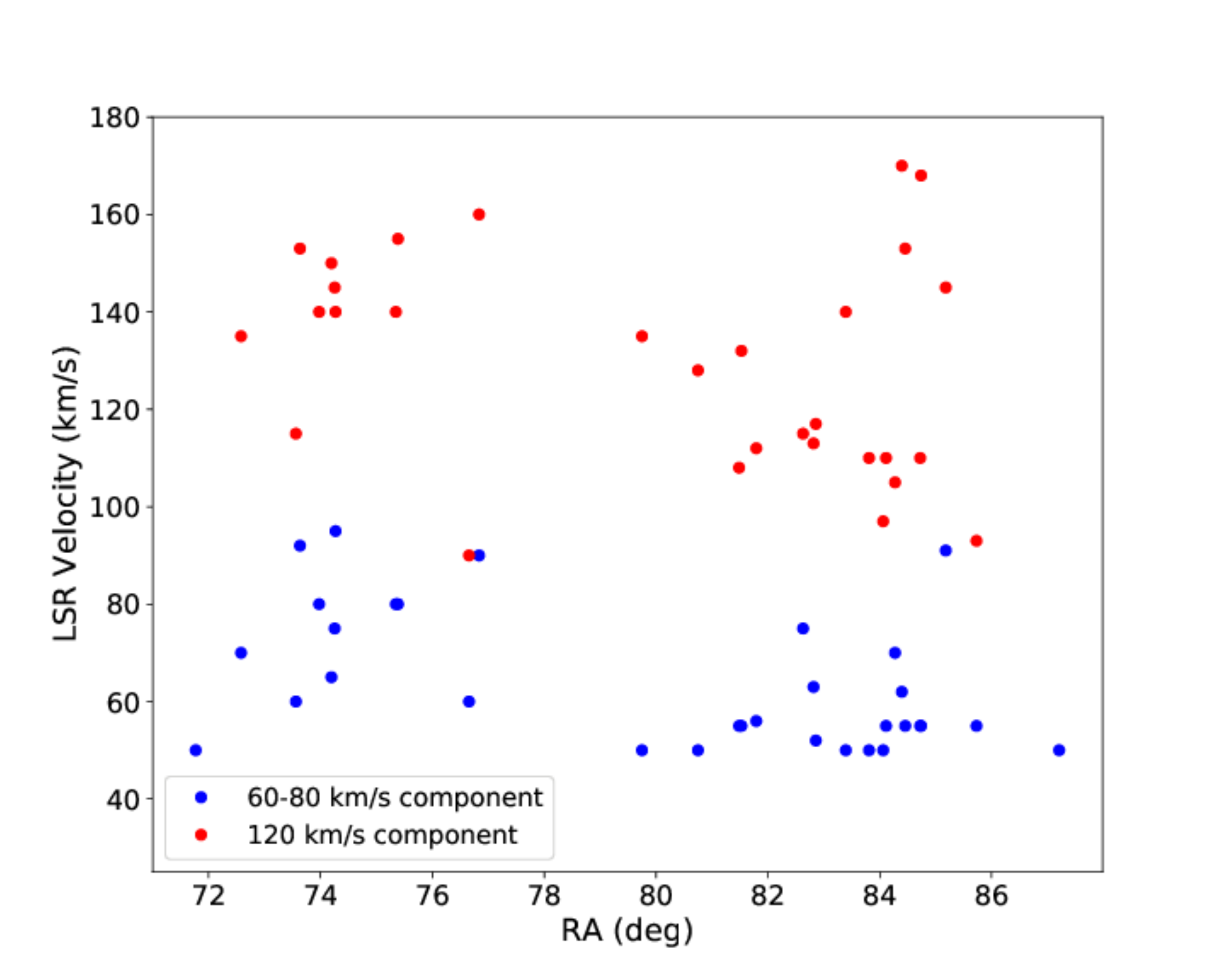}
\includegraphics[width=8cm]{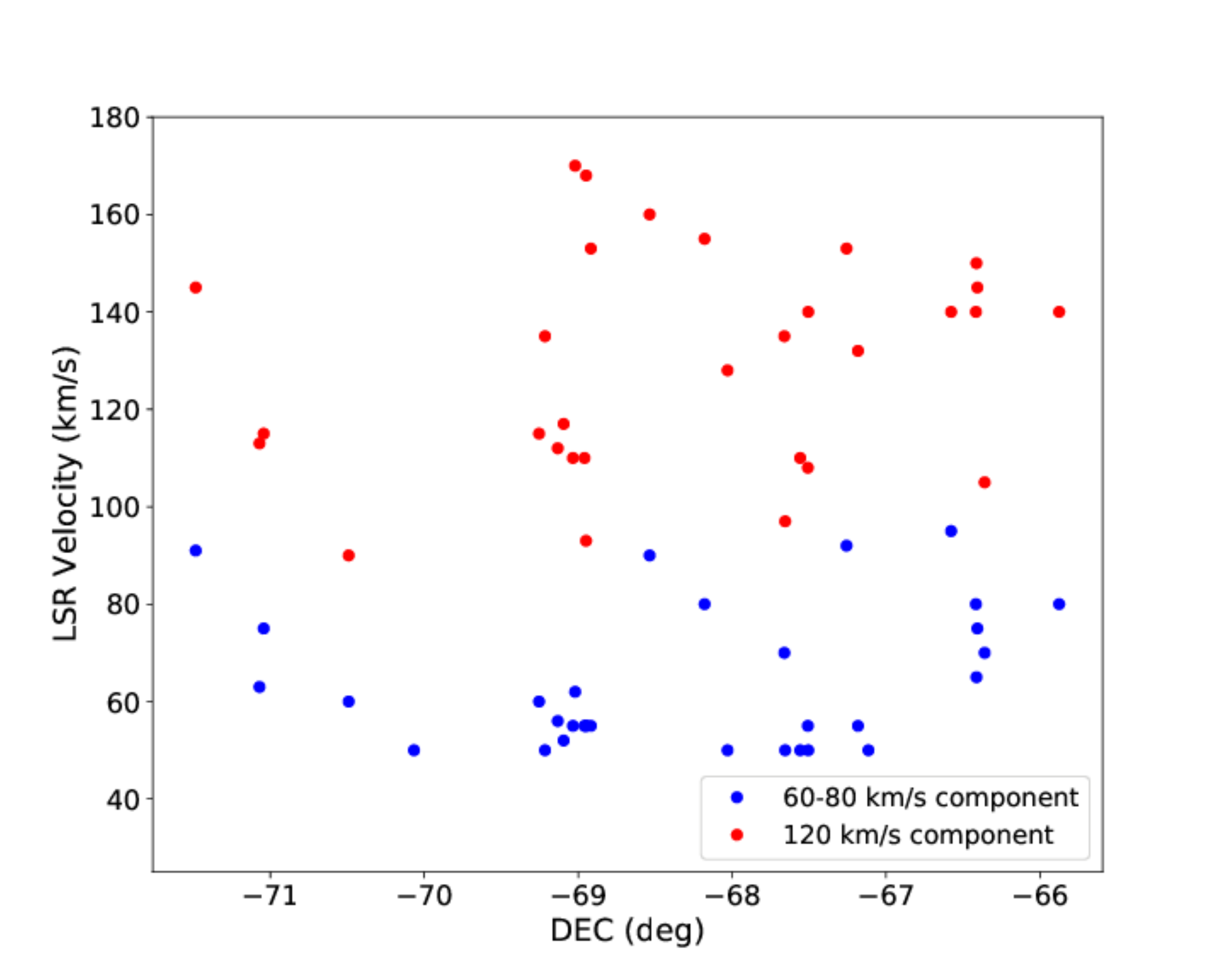}

\caption{LSR velocities of the intermediate (blue, 60 km s$^{-1}$ component) and high (red, 120 km s$^{-1}$ component) velocity gas toward METAL targets (all except SK-69 220), as a function of R.A. (left) and DEC (right).}
\label{show_ivg_ra_dec}
\end{figure*}

\indent We detect intermediate and/or high-velocity gas (i.e., at velocities higher than those associated with the Milky Way and lower than those arising from within the LMC, as listed in Table 4) in all but one target (SK-69 220, due to its complex stellar spectrum). The two main components are observed at LSR velocities of roughly 60 km s$^{-1}$ (typical intermediate velocity cloud, IVC) and 120 km s$^{-1}$ (typical high-velocity cloud, HVC). Table 5 lists the LSR velocities of each IVC and HVC component toward each target (except for SK-69 220), determined from the Fe II 2344 \mAA and O I 1302 \mAA lines (see Figure \ref{show_zooms}). The velocities of the IVC and HVC are also plotted as a function of R.A. and DEC in Figure \ref{show_ivg_ra_dec}. \\
\indent  These components have previously been detected in ''down-the-barrel'' studies with {\it Far Ultraviolet Spectroscopic Explorer} (FUSE) observations of massive stars \citep{danforth2002, lehner2007, lehner2009}, but also UV absorption line studies of a QSO \citep{barger2016}. As shown in Figure \ref{show_zooms}, the IVC and HVC absorption are detected in strong lines such as Fe II 2344 \mAA and O I 1302 \AA, and in some cases in weaker lines, such as Si II 1808 \mAA or even Mg II 1239 \AA. In about 1/3 of the sight-lines, the intermediate/high velocity gas metal absorption is associated with weak \his 21 cm emission in the GASS survey \citep{mccluregriffiths2009, kalberla2010, kalberla2015}, consistent with the results by \citet{lehner2009}. In another 1/3 of the sample, the IVC and HVC gas components are also detected in the Si IV 1393 \mAA and/or C IV 1550 \mAA lines. The HVC has previously been observed in even higher ions accessible in FUSE data, such as O VI \citep{lehner2007, lehner2009}. Based on the ratio of singly ionized silicon (Si II) to neutral oxygen (O I), \citet{lehner2009} determined that the HVC gas is dominantly ionized. With the additional Si IV 1393 \mAA and C IV 1550 \mAA spectral lines present in the METAL spectra, the survey will enable further ''down-the-barrel'' studies of the ionization state of the IVC and HVC gas.\\
\indent Three pieces of evidence strongly suggest that the HVC originates from star-formation driven outflows in the LMC, and not from a galactic fountain in the MW halo or tidally stripped material from the LMC or SMC. First, \citet{staveley-smith2003} observed that the \his column density peaks of the HVC spatially coincide with voids in the underlying \his distribution of the LMC disk and are linked in position-velocity place to the LMC disk. Second, the metallicity of the HVC complex is consistent with that of the LMC disk, and its ionization state is consistent with that of energetic outflows  \citep{lehner2009}. Third, the blue-shifted and red-shifted absorption profiles on the respective near and far sides of the LMC are symmetric \citep{barger2016}, indicating an outflow origin. In this context, the gradient in the velocities of the 120 km s$^{-1}$ component with R.A., which has previously been observed by \citet{lehner2009}, can be attributed to the larger concentration of supergiant shells on the left side of the LMC compared to the right side \citep{kim1999}, implying stronger outflows in the high R.A. regions \citep{lehner2009}. We also note that the IVC and HVC velocities are incompatible with the gas originating from Narrow Absorption Components \citep[NAC, see][]{prinja2002}, often seen in outflows very near luminous, hot stars.  Such components can have velocities as high as $-$1000 to $-$2000 km s$^{-1}$ and are time variable. Since no components at negative velocities are seen in the METAL spectra, such an origin can be ruled out.\\
\indent While the origin of the HVC toward the LMC is already well constrained, the METAL data will enable further studies of the metallicity of this gas (accounting for its ionization state) by extending the UV coverage of a large sample of massive star spectra across the LMC, to include additional spectral lines, such as S II 1250 \AA, Mg II 1239 \AA, Ni II 1370 \AA, Si IV1393 \AA, and C IV 1550 \AA.

\subsection{Extinction Curves}

\begin{figure}
\centering
\includegraphics[width=8cm]{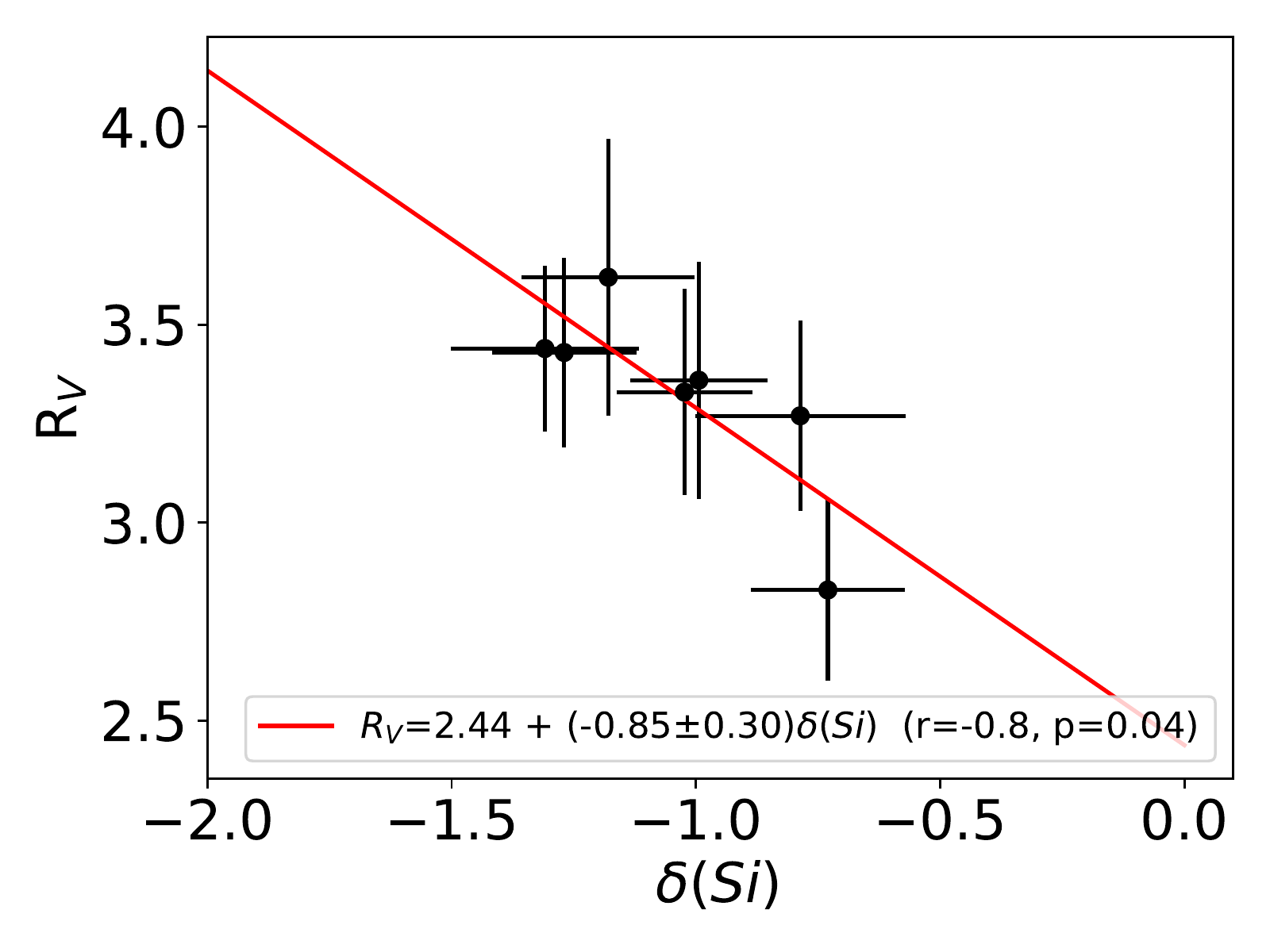}
\caption{Absolute to selective extinction $R_V$ as a function of silicon depletion (fraction of silicon in the gas phase) for the 7 sight-lines in common between the METAL program and \citet{gordon2003}.}
\label{plot_rv_fstar}
\end{figure}

\indent Abundance and depletion measurements from medium-resolution spectroscopy provide the detailed composition of dust. Dust extinction curves, which describe the variations of the optical properties of dust as a function of wavelength, provide important complementary information on the nature of dust grains. In particular, extinction curves have been observed to vary dramatically in the UV. While some of the variation can be attributed to changes in $R_V$ and therefore in the dust grain size distribution,  the physical processes that drive variations in the 2175 \mAA bump strength and FUV rise are still unknown \citep[e.g.,][]{gordon2003}. One of the goals of METAL is to complement the depletion information with UV extinction curves. Out of the 33 METAL targets, 7 sight-lines already have extinction curves published in \citet{gordon2003}. We are in the process of deriving UV extinction curves for the rest of the targets from the archival IUE data, along with the low-resolution STIS spectra acquired as part of METAL (Table A3), using the "standard pair" method \citep{fitzpatrick1985} and the "extinction without standards" method \citep{fitzpatrick2005}.  \\
\indent Each extinction curve in \citet{gordon2003} is described by 6 parameters \citep[$c_1$, $c_2$, $c_3$, $c_4$, $x_0$, $\gamma$, see][]{fitzpatrick1990}. Additionally, the absolute-to-selective extinction, $R_V$, a proxy for average dust grain size, is measured directly in the extinction curves in \citet{gordon2003}. To investigate whether there is a relation between depletions and the shape of the extinction curve with this limited sample, we show in Figure \ref{plot_rv_fstar} $R_V$ from \citet{gordon2003} as a function of the the silicon depletion measured in the METAL data. There is a hint at a correlation between $\delta$(Si) and $R_V$, with slope of $-$0.85$\pm$0.3, an intercept of 2.44, a correlation coefficient of $-$0.8 and a p-value of 0.04 (the correlation is statistically significant). This may indicate that more depleted sight-lines also contain larger grains. Such a trend would be expected given that the level of depletion and average grain-size should both increase with density due to accretion of gas-phase metals onto dust grains and coagulation, respectively. This tentative correlation highlights the need to obtain a larger sample of extinction curves for our depletion sight-lines. Extinction curves will be derived for the entire METAL sample from archival IUE and METAL spectra and the in-depth study of extinction curves in the LMC will be published in a future paper (Hagen et al., in prep).\\

\subsection{Extinction mapping with the BEAST}\label{ext_map_section}

\begin{figure*}
\centering
\includegraphics[width=\textwidth]{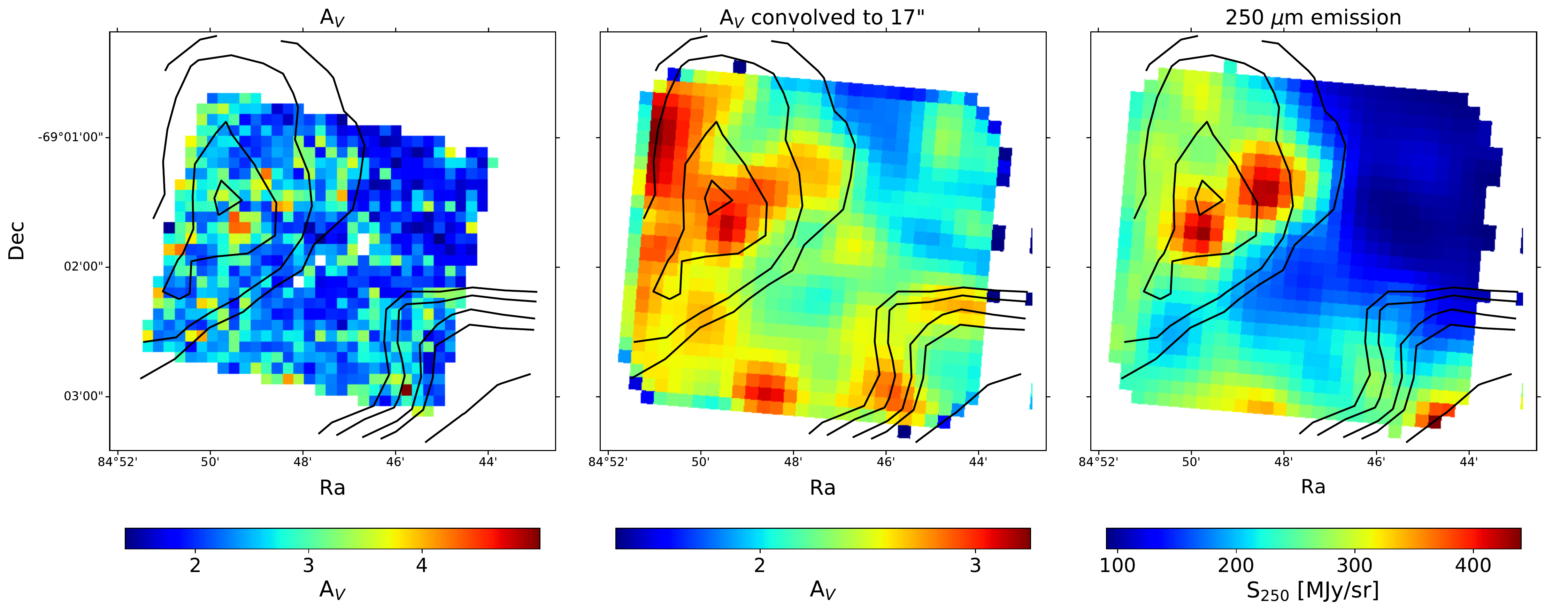}
\caption{Maps of $A_V$ at 5'' resolution (left), $A_V$ at 17.6'' resolution (middle), and 250 $\mu$m dust emission in Herschel SPIRE (right) in the field obtained in parallel with the SK-68140 observations. The black contours indicate CO 1-0 integrated intensity from the MAGMA data (the levels are 0.4, 0.8, 1.5 and 2.4 K km/s). }
\label{show_av_em}
\end{figure*}

\indent Dust extinction is clearly visible in the color-magnitude diagrams from Figure \ref{plot_cmds}. In particular, the red clump stretches out to fainter magnitudes and redder colors due to dust extinction. Visual inspection of Figure \ref{plot_cmds} suggests that sources detected in three or more filters are affected by extinction up to $A_B$ $\simeq$ 3 and $A_V$ $\simeq$ 2. In the future, we will perform a study similar to \citet{yanchulova2017} to model the CMD of the red clump and the red giant branch, and verify the NUV-NIR extinction curve in the LMC derived from spectroscopy.  \\
\indent Furthermore, we can use the NUV-NIR photometry to individual stars to determine their stellar and dust parameters, and model the distribution of those stars in the dust disk to derive extinction maps, similar to \citet{dalcanton2015}. The point source photometry and associated bias described in Section \ref{photometry_section} are being run through the BEAST \citep[Bayesian Extinction and Stellar Tool,][]{gordon2016} to fit the NUV-NIR SED of each star to a model that includes stellar atmospheres, stellar evolution, and dust extinction. Fits to each star produce 6 parameters: mass, age, metallicity, $A_V$, $R_V$, $f_A$, where $f_A$ is a parameter that describes the mixture of SMC-like dust and MW-like dust \citep{gordon2016}. The $A_V$ values toward each star can be modeled along with the geometry of the stellar distribution with respect to the dust disk and the completeness of the data to produce extinction maps. As a first step, we averaged the $A_V$ values toward each star in 5'' bins on the sky. These "naive" extinction maps are not suitable for a rigorous derivation of the spatial variations of the dust optical properties because they do not account for the completeness of the observations, or the relative geometry of the stellar and dust disk. However, the naive extinction maps showcase the concept of extinction mapping with this type of data. An example is shown in Figure \ref{show_av_em}, which includes the naive $A_V$ map at 5'' resolution, the $A_V$ map convolved to Herschel SPIRE 250 $\mu$m resolution (17.6''), and the dust thermal emission in the SPIRE 250 $\mu$m band \citep{meixner2013}. There is a good spatial correspondence between the convolved $A_V$ map and the dust thermal emission. CO 1-0 emission seen in the MAGMA survey \citep{wong2011} is also co-spatial with the $A_V$ and thermal dust emission peaks. The ratio of the dust FIR emission corrected for temperature effects \citep[known from FIR SED fitting, see][]{gordon2014} to the full extinction maps, derived from the modeling of the $A_V$ distributions, completeness, and geometry of the disks, can be used to estimate the FIR emissivity of dust in each of the fields, as a function of the local environment (e.g., surface density, radiation field).

\section{Summary}\label{conclusion_section}
\indent In this paper, we presented an overview of The METAL large HST program (GO-14675). In 101 prime orbits and 101 parallel orbits, METAL obtained medium-resolution FUV---NUV (1150 \mAA --- 2380 \AA) spectra of 33 massive stars in the LMC, and NUV---NIR WFC3 images of parallel fields in the neighborhood of the massive stars. Interstellar depletions (for Fe, Mg, Si, Ni, Zn, Cr, S), UV dust extinction curves, and extinction maps are being derived from the spectra and images.\\
\indent The silicon depletions in the LMC vary with hydrogen column density, with more Si locked in dust grains at higher column densities. Additionally, the fraction of silicon locked in dust decreases from the Milky Way to the LMC (1.5 times lower than in the MW) to the SMC (4 times lower than in the MW). As a result of the rough scaling of the silicon depletion with inverse metallicity, the gas-phase abundances of silicon observed in METAL are constant between those three galaxies, for a given hydrogen column densities. In other words, we reach the striking conclusion that gas in the SMC contains as much silicon as gas in the MW. \\
\indent So far, seven of the METAL targets have published UV extinction curves. Comparisons between the depletions and the absolute-to-selective extinction ($R_V$) indicate tentative trends in agreement with the physical expectation that higher depletions are associated with larger grains. However, with only 7 points, these trends are marginally significant and require further investigation. A larger sample of extinction curves will be derived from the METAL data to determine whether there is a relation between depletions and extinction curves, and what physical parameters drive the shape of the UV extinction curves.\\
\indent Point-source photometry has been derived for all imaging fields using the same photometry pipeline as the PHAT survey. The photometry is being run through the BEAST SED fitting algorithm in order to derive the stellar (mass, age, metallicity) and dust ($A_V$, $R_V$, $f_A$) parameters for each star. By combining the $A_V$ measurements of individual stars in binned pixels on the sky and modeling the geometry of the stars with respect to the dust disk, we can derive extinction maps at 5'' resolution. The extinction maps will be compared to the FIR emission maps from {\it Herschel} and {\it Spitzer} in order to estimate the FIR emissivity of dust. \\
\indent The METAL data will provide a comprehensive and detailed characterization of dust in the LMC.  Combined with already available similar observations in the SMC, the METAL program will provide the missing measurements to evaluate how dust properties change with metallicity,  and an excellent observational anchor to chemical evolution models.\\
\indent Beyond studies of the dust properties in the LMC, the METAL data will carry an important legacy value by enabling studies of galactic-scale inflow and outflow and stellar winds. Intermediate and high velocity gas is detected in a variety of metals toward all of our targets, and we are working on constraining the origin, metallicity, and ionization state of this gas. Furthermore, studies of massive stars and their stellar winds based on the METAL data are already underway.

\acknowledgments{We thank the referee for insightful comments and suggestions. Thank you to Daniel Welty for useful discussions about \his column density measurements. Julia Roman-Duval acknowledges support from the European Space Agency. Edward B. Jenkins was supported by grant HST-GO-14675.004-A to Princeton University.  Benjamin Williams acknowledges support from grant HST-GO-14675.009-A to the University of Washington. Lea Hagen and Karl Gordon were supported by grant HST-GO-14675.002-A to the Space Telescope Science Institute. Karin Sandstrom and Petia Yanchulova Merica-Jones acknowledge support from grant HST-GO-14675.008-A to the University of California, San Diego. This work is based on observations with the NASA/ESA Hubble Space Telescope obtained at the Space Telescope Science Institute, which is operated by the Associations of Universities for Research in Astronomy, Incorporated, under NASA contract NAS5-26555. These observations are associated with program 14675. Support for Program number 14675  was provided by NASA through a grant from the Space Telescope Science Institute, which is operated by the Association of Universities for Research in Astronomy, Incorporated, under NASA contract NAS5-26555.
}

\bibliographystyle{/users/duval/stsci_research/bibtex/apj11}
\bibliography{/Users/duval/stsci_research/biblio_all}

\appendix

This Appendix contains five tables summarizing the spectroscopic and imaging observations, including their instrumental set-up, exposure times, and S/N. Table A1 lists the spectral observations for the primary targets taken with STIS in both the FUV and NUV. Table A2 lists the observations taken with COS in the FUV, and STIS in the NUV. Table A3 lists the low-resolution observations taken with STIS to complement the IUE  and HST archive and compute extinction curves. Table A4 lists select parallel imaging observations. The entirety of Table A4 is available online as a machine-readable table. Table A5 provides the photometric depths for each field, and is available online as a machine-readable table in its entirety.

\setcounter{table}{0}
\renewcommand{\thetable}{A\Alph{section}\arabic{table}}

\begin{deluxetable*}{ccccccccccc}
\centering
\tabletypesize{\scriptsize}
\tablecolumns{11}
\tablewidth{\textwidth}
\tablecaption{Targets and observations with STIS in the FUV and NUV}
 
 \tablehead{Target &  \multicolumn{5}{c}{STIS FUV Observations} & \multicolumn{5}{c}{STIS NUV Observations}}\\
 \startdata
 Target & Grating & Aperture &  PID & $T_{\mathrm{exp}}$ (s) & S/N\tablenotemark{a}& Grating/Cenwave & Aperture & PID & $T_{\mathrm{exp}}$ (s)& S/N\tablenotemark{a} \\
 \hline
 &&&&&&&&&&\\
SK-67 2 &  E140M& 0.2X0.2 & 14675 & 4970.0 & 11 & E230M/1978  & 0.2X0.2 & 14675 & 5010.0 & 18\\
SK-67 5 &  E140M& 0.2X0.2 & 14675 & 2585.0 & 18 & E230M/1978  & 0.2X0.2 & 14675 & 2240.0 & 27 \\
SK-67 14 &  E140M& 0.2X0.2 & 14675 & 2585.0 & 13 & E230M/1978  & 0.2X0.2 & 14675 & 2240.0 & 20 \\
PGMW 3120 &  E140M & 0.2X0.06 & 8320 & 8595.0 & 8  & E230M/1978 & 0.2X0.2 & 14675 & 3810.0 & 9 \\
PGMW 3223 &  E140M & 0.2X0.2 & 14675 & 7768.0 & 10  & E230M/1978 & 0.2X0.2 & 14675 & 4970.0 & 13 \\
SK-66 35 &  E140M & 0.2X0.2 & 14675 & 2585.0 & 12 & E230M/1978 & 0.2X0.2 & 14675 & 2240.0 & 18 \\
SK-65 22 & E140M & 0.2X0.2 & 9434 & 1200.0 & 11&E230M/1978  & 0.2X0.2 & 14675 & 2240.0 & 24 \\
SK-70 79 &  E140M& 0.2X0.2 & 14675 & 7595.0 & 12  & E230M/1978  & 0.2X0.2 & 14675 & 5040.0 & 15 \\
SK-68 52 &  E140M& 0.2X0.2 & 14675 & 2580.0 & 11  & E230M/1978  & 0.2X0.2 & 14675 & 4950.0 & 28 \\
SK-69 104 &  E140M & 0.2X0.2 & 14675 & 2585.0 & 17 & E230M/1978 & 0.2X0.2 & 14675 & 2240.0 & 24 \\
SK-67 101 &  E140M & 0.2X0.06 & 7299 & 5443.0 & 14& E230M/1978  & 0.2X0.2 & 14675 & 2240.0 & 18 \\
SK-67 105 &  E140M& 0.2X0.2 & 12218 & 2839.0 & 14  & E230M/1978 & 0.2X0.2 & 14675 & 4990.0 & 27 \\
BI 173 &  E140M & 0.2X0.2 & 14675 & 4770.0 & 13  & E230M/1978 & 0.2X0.2 & 14675 & 4300.0 & 18 \\
SK-71 45 &  E140M & 0.2X0.2 & 8662 & 1440.0 & 10  & E230M/1978 & 0.2X0.2 & 14675 & 2240.0 & 19 \\
SK-69 175 &  E140M & 0.2X0.2 & 14675 & 2260.0 & 11 & E230M/1978  & 0.2X0.2 & 14675 & 2585.0 & 20 \\
SK-67 191 &  E140M & 0.2X0.2 & 14675 & 5310.0 & 13  & E230M/1978  & 0.2X0.2 & 14675 & 4990.0 & 18 \\
SK-67 211 &  E140M& 0.2X0.2 & 9434 & 1200.0 & 12& E230M/1978 & 0.2X0.2 & 14675 & 2240.0 & 23 \\
SK-69 220 &  E140M & 0.2X0.2 & 14675 & 2585.0 & 10  & E230M/1978  & 0.2X0.2 & 14675 & 2240.0 & 18 \\
SK-66 172 &  E140M & 0.2X0.2 & 14675 & 4960.0 & 11  & E230M/1978  & 0.2X0.2 & 14675 & 4950.0 & 15 \\
SK-68 135 &  E140M & 0.2X0.2 & 14675 & 4950.0 & 15 & E230M/1978 & 0.2X0.2 & 14675 & 2580.0 & 21  \\
SK-69 246 &  E140M & 0.2X0.2 & 9434 & 1200.0 & 12  & E230M/1978 & 0.2X0.2 & 14675 & 2240.0 & 27 \\
SK-71 50 &  E140M & 0.2X0.2 & 14675 & 5275.0 & 9  & E230M/1978 & 0.2X0.2 & 14675 & 6640.0 & 15\\
SK-70 115 &  E140M & 0.2X0.2 & 12218 & 2885.0 & 13  &E230M/1978  & 0.2X0.2 & 14675 & 4990.0 & 26 \\
\enddata
\tablenotetext{a}{The S/N column corresponds to the median S/N per pixel over the wavelength range covered by each setting. } 
\end{deluxetable*}

\begin{deluxetable*}{cccccccccccc}
\centering
\tabletypesize{\scriptsize}
\tablecolumns{12}
\tablewidth{\textwidth}
\tablecaption{Targets and observations with COS in the FUV and STIS in the NUV}
 
\tablehead{Target &  \multicolumn{6}{c}{COS FUV Observations} & \multicolumn{5}{c}{STIS NUV Observations}}\\
 \startdata
 &   \multicolumn{2}{c}{PID} &  \multicolumn{2}{c}{$T_{\mathrm{exp}}$ (s)} &  \multicolumn{2}{c}{S/N\tablenotemark{a}}& Grating/Cenwave & Aperture &  PID & $T_{\mathrm{exp}}$ (s)& S/N\tablenotemark{a} \\
&&&&&&&&&\\
\hline
&&&&&&&&&\\
 & G130M & G160M  &  G130M & G160M &  G130M & G160M & &  & & &  \\
 &&&&&&&&&&&\\
\hline
&&&&&&&&&&&\\
SK-69 279 & 12581 & 12581 & 318.0 & 3769.8 & 7 & 17 &E230M/1978  & 0.2X0.2 & 14675 & 9910.0 & 16 \\
SK-66 19  & 14675 & 14675 & 650.0 & 600.1 & 5 & 3 & E230M/1978  & 0.2X0.2 & 14675 & 9954.0 & 9 \\
SK-68 26  &  14675 & 14675 & 440.1 & 600.1 & 8  & 8 & E230M/1978  & 0.2X0.2 & 14675 & 7700.0 & 17 \\
SK-68 73 &  14675 & 12581 & 1850.1 & 1254.1 & 17  & 13 &E230M/1978 & 0.2X0.2 & 12978 & 16114.0 & 31 \\
BI 184 &  14675 & 14675 & 2220.1 & 1800.1 & 20 & 10  & E230M/1978  & 0.2X0.2 & 14675 & 7150.0 & 11  \\
BI 237  &  14675 & 14675 & 540.1 & 600.1 & 9 & 6 & E230M/1978 & 0.2X0.2 & 14675 & 7120.0 & 12 \\
SK-68 129  & 12581 & 12581 & 1207.6 & 3740.8 & 12 & 15 & E230M/1978 & 0.2X0.2 & 14675 & 13200.0 & 16 \\
BI 253 & 14675 & 14675 & 540.1 & 600.1 & 8 & 6 & E230M/1978  & 0.2X0.2 & 14675 & 9895.0 & 14 \\
SK-68 140  &  12581 & 12581 & 330.0 & 3328.1 & 6 & 14 & E230M/1978 & 0.2X0.2 & 14675 & 10068.7 & 13  \\
SK-68 155 & 12581 & 12581 & 330.0 & 2446.8 & 7 & 14 & E230M/1978  & 0.2X0.2 & 14675 & 9980.0 & 18 \\
\enddata
\tablecomments{GO-14675 (METAL) used the G130M/1291 and G160M/1589 cenwaves, while program 12581 acquired G130M/1327, G160M/1577, 1589, and 1600. For consistency with the METAL program, we only use G160M/1589 of GO-12581. }
\tablenotetext{a}{The S/N column corresponds to the median S/N per pixel over the wavelength range covered by each setting. }
\end{deluxetable*}

\begin{deluxetable*}{cccccccc}
\centering
\tabletypesize{\scriptsize}
\tablecolumns{8}
\tablewidth{\textwidth}
\tablecaption{:Low-resolution spectroscopic observations}
\tablehead{Target & Grating & Cenwave\tablenotemark{a} &Aperture &$T_{\mathrm{exp}}$ (s) & S/N\tablenotemark{b}   & PID & Root}\\
\startdata
PGMW 3120 & G230L & 2376 & 52X2 & 100.0 & 36.4851 & 14675 & oda912030 \\
PGMW 3120 & G140L & 1425 & 52X2 & 100.0 & 25.483 & 14675 & oda912040 \\
BI 173 & G230L & 2376 & 52X2 & 100.0 & 42.4721 & 14675 & oda914030 \\
BI 173 & G140L & 1425 & 52X2 & 100.0 & 30.6142 & 14675 & oda914060 \\
BI 184 & G230L & 2376 & 0.2X0.2 & 100.0 & 23.7564 & 14675 & oda918040 \\
BI 237 & G230L & 2376 & 0.2X0.2 & 100.0 & 22.981 & 14675 & oda926040 \\
BI 253 & G230L & 2376 & 0.2X0.2 & 100.0 & 25.3981 & 14675 & oda929050 \\
SK-71 50 & G230L & 2376 & 52X2 & 100.0 & 33.1934 & 14675 & oda919040 \\
SK-71 50 & G140L & 1425 & 52X2 & 100.0 & 21.57 & 14675 & oda919050 \\
\enddata
\tablenotetext{a}{Cenwave indicate central wavelength}
\tablenotetext{b}{The median S/N is listed per pixel} 
\end{deluxetable*}

\begin{deluxetable*}{ccccccccc}
\centering
\tabletypesize{\scriptsize}
\tablecolumns{9}
\tablewidth{\textwidth}
\tablecaption{Parameters of the parallel WFC3 imaging}
\tablehead{Primary Target & Root & Field Center Ra & Field Center Dec & PA & Detector & Filter & $T_{\mathrm{exp}}$ & Post-flash  }\\
\startdata
 & & $Deg$ & $Deg$ & $Deg$ && & $s$&$s$\\
 \hline
 &&&&&&&&\\
PGMW 3223 & ida974030 & 74.3437517075 & -66.3260553982 & 339.740509 & UVIS & F814W & 674.0 & 4.3 \\
PGMW 3223 & ida974040 & 74.3437517075 & -66.3260553982 & 339.740509 & UVIS & F275W & 2045.0 & 3.9 \\
PGMW 3223 & ida974bqq & 74.3437517075 & -66.3260553982 & 339.740509 & UVIS & F225W & 700.0 & 3.9 \\
PGMW 3223 & ida974btq & 74.3437517075 & -66.3260553982 & 339.740509 & UVIS & F336W & 400.0 & 3.9 \\
PGMW 3223 & ida973040 & 74.3395689304 & -66.3278345963 & 339.740509 & IR & F160W & 1396.9 & --- \\
PGMW 3223 & ida973050 & 74.3395689304 & -66.3278345963 & 339.740509 & IR & F110W & 698.5 & --- \\
PGMW 3223 & ida973060 & 74.3437517075 & -66.3260553982 & 339.740509 & UVIS & F475W & 1205.0 & 4.3 \\
PGMW 3223 & ida973070 & 74.3437517075 & -66.3260553982 & 339.740509 & UVIS & F336W & 1310.0 & 3.9 \\
PGMW 3223 & ida973080 & 74.3437517075 & -66.3260553982 & 339.740509 & UVIS & F225W & 1515.0 & 3.9 \\
SK-65 22 & ida901020 & 75.379223874 & -65.7883197464 & 324.494904 & UVIS & F275W & 696.0 & 4.3 \\
SK-65 22 & ida901030 & 75.379223874 & -65.7883197464 & 324.494904 & UVIS & F336W & 696.0 & 3.9 \\
SK-65 22 & ida901dfq & 75.379223874 & -65.7883197464 & 324.494904 & UVIS & F475W & 155.0 & 4.3 \\
SK-65 22 & ida901dpq & 75.379223874 & -65.7883197464 & 324.494904 & UVIS & F814W & 120.0 & 3.5 \\
BI 173 & ida914070 & 81.9311142278 & -69.0618760415 & 351.467712 & IR & F160W & 1396.9 & --- \\
BI 173 & ida914080 & 81.9311142278 & -69.0618760415 & 351.467712 & IR & F110W & 698.5 & --- \\
BI 173 & ida914090 & 81.936727458 & -69.060475305 & 351.467712 & UVIS & F475W & 725.0 & 4.3 \\
BI 173 & ida9140a0 & 81.936727458 & -69.060475305 & 351.467712 & UVIS & F225W & 1860.0 & 3.9 \\
BI 173 & ida9140b0 & 81.936727458 & -69.060475305 & 351.467712 & UVIS & F336W & 1310.0 & 4.3 \\
BI 173 & ida9140c0 & 81.936727458 & -69.060475305 & 351.467712 & UVIS & F275W & 1245.0 & 4.3 \\
BI 173 & ida9140d0 & 81.936727458 & -69.060475305 & 351.467712 & UVIS & F814W & 710.0 & 4.3 \\
BI 184 & ida920hsq & 82.7731346471 & -71.1203563187 & 191.284698 & UVIS & F475W & 350.0 & 2.0 \\
BI 184 & ida920hvq & 82.7731346471 & -71.1203563187 & 191.284698 & UVIS & F275W & 450.0 & 3.9 \\
BI 184 & ida920ibq & 82.7731346471 & -71.1203563187 & 191.284698 & UVIS & F814W & 300.0 & 2.0 \\
BI 184 & ida920ixq & 82.7793692232 & -71.122224837 & 191.284698 & IR & F160W & 499.2 & --- \\
BI 184 & ida918050 & 82.7371294562 & -70.9634803121 & 340.557404 & IR & F160W & 1396.9& --- \\
BI 184 & ida918060 & 82.7371294562 & -70.9634803121 & 340.557404 & IR & F110W & 698.4\ & --- \\
BI 184 & ida918070 & 82.7423554106 & -70.9617252167 & 340.557404 & UVIS & F475W & 709.0 & 4.3 \\
BI 184 & ida918080 & 82.7423554106 & -70.9617252167 & 340.557404 & UVIS & F275W & 800.0 & 4.3 \\
BI 184 & ida918090 & 82.7423554106 & -70.9617252167 & 340.557404 & UVIS & F336W & 800.0 & 3.9 \\
BI 184 & ida9180a0 & 82.7423554106 & -70.9617252167 & 340.557404 & UVIS & F225W & 800.0 & 3.9 \\
BI 184 & ida9180b0 & 82.7423554106 & -70.9617252167 & 340.557404 & UVIS & F814W & 820.0 & 4.3 \\
\enddata
\tablecomments{The entirety of this table is available online as a machine-readable table}
\end{deluxetable*}

\begin{deluxetable*}{ccccccc}
\centering
\tabletypesize{\scriptsize}
\tablecolumns{7}
\tablewidth{\textwidth}
\tablecaption{Photometric Depth for the WFC3 Observations}
\tablehead{Primary Target & Field Name & Field Center Ra & Field Center Dec & Filter & Total $T_{\mathrm{exp}}$ (s) & Depth\tablenotemark{a}   }\\
\startdata
 & & $Deg$ & $Deg$ & & $s$ & Vega Mag\\
 \hline
 &&&&&&\\
PGMW 3223 & LMC-15562nw-33974 & 74.3395689304 & -66.3278345963 & F110W & 698.5 & 25.2 \\
PGMW 3223 & LMC-15562nw-33974 & 74.3395689304 & -66.3278345963 & F160W & 1396.9 & 23.7 \\
PGMW 3223 & LMC-15562nw-33974 & 74.3437517075 & -66.3260553982 & F225W & 2215.0 & 24.7 \\
PGMW 3223 & LMC-15562nw-33974 & 74.3437517075 & -66.3260553982 & F275W & 2045.0 & 25.0 \\
PGMW 3223 & LMC-15562nw-33974 & 74.3437517075 & -66.3260553982 & F336W & 1710.0 & 25.7 \\
PGMW 3223 & LMC-15562nw-33974 & 74.3437517075 & -66.3260553982 & F475W & 1205.0 & 27.6 \\
PGMW 3223 & LMC-15562nw-33974 & 74.3437517075 & -66.3260553982 & F814W & 674.0 & 25.5 \\
SK-65 22 & LMC-16444nw-32449 & 75.379223874 & -65.7883197464 & F275W & 696.0 & 24.1 \\
SK-65 22 & LMC-16444nw-32449 & 75.379223874 & -65.7883197464 & F336W & 696.0 & 25.0 \\
SK-65 22 & LMC-16444nw-32449 & 75.379223874 & -65.7883197464 & F475W & 155.0 & 25.9 \\
SK-65 22 & LMC-16444nw-32449 & 75.379223874 & -65.7883197464 & F814W & 120.0 & 24.5 \\
BI 173 & LMC-2833ne-35147 & 81.9311142278 & -69.0618760415 & F110W & 698.5 & 25.0 \\
BI 173 & LMC-2833ne-35147 & 81.9311142278 & -69.0618760415 & F160W & 1396.9 & 24.0 \\
BI 173 & LMC-2833ne-35147 & 81.936727458 & -69.060475305 & F225W & 1860.0 & 24.5 \\
BI 173 & LMC-2833ne-35147 & 81.936727458 & -69.060475305 & F275W & 1245.0 & 24.5 \\
BI 173 & LMC-2833ne-35147 & 81.936727458 & -69.060475305 & F336W & 1310.0 & 25.5 \\
BI 173 & LMC-2833ne-35147 & 81.936727458 & -69.060475305 & F475W & 725.0 & 27.2 \\
BI 173 & LMC-2833ne-35147 & 81.936727458 & -69.060475305 & F814W & 710.0 & 25.7 \\
BI 184 & LMC-5378se-19128 & 82.7793692232 & -71.122224837 & F160W & 1896.2 & 23.4 \\
BI 184 & LMC-5378se-19128 & 82.7731346471 & -71.1203563187 & F275W & 1250.0 & 23.9 \\
BI 184 & LMC-5378se-19128 & 82.7731346471 & -71.1203563187 & F475W & 1059.0 & 26.9 \\
BI 184 & LMC-5378se-19128 & 82.7731346471 & -71.1203563187 & F814W & 1120.0 & 25.2 \\
\enddata
\tablenotetext{a}{The depth is given as the Vega mag corresponding to the 50\% completeness limit in the artificial star tests.}
\tablecomments{The entirety of this table is available online as a machine-readable table}
\end{deluxetable*}

\end{document}